\newcommand{\beq}{\begin{equation}}
\newcommand{\eeq}{\end{equation}}
\newcommand{\rc}{\nonumber\\}
\newcommand{\bear}{\begin{eqnarray}}
\newcommand{\eear}{\end{eqnarray}}

\documentclass{JINST}

\usepackage{amssymb}

\title{A comprehensive study of rate capability in Multi-Wire Proportional
Chambers}

\author{A. Andronic$^a$, C. Garabatos$^a$,D. Gonzalez-Diaz$^a$\thanks{Corresponding author}, 
A. Kalweit$^a$, F. Uhlig$^a$\\
\llap{$^a$} GSI Helmoltzzentrum f\"ur Schwerionen Forschung, Darmstadt, Germany \\
E-mail: \email{D.Gonzalez-Diaz@gsi.de}}

\abstract{Systematic measurements on the rate capability of thin MWPCs operated
in Xenon, Argon and Neon mixtures using CO$_2$ as UV-quencher are presented.
A good agreement between data and existing models has been found, allowing us to 
present the rate capability of MWPCs in a comprehensive way and
ultimately connect it with the mobilities of the drifting ions.}


\begin{document}

\section{Introduction}
\label{intro}

Due to their good position resolution, low material budget and low cost, multi-wire 
proportional chambers (MWPCs) are instrumental in high-energy physics. 
They are currently employed in experiments needing high resolution tracking  
(ALICE-TPC \cite{AliceTPC}), muon spectrometers (LHCb \cite{LHCb}) and 
photo-sensitive detectors (HADES-RICH \cite{HadesRich} or ALICE-TRD \cite{AliceTRD}), 
to mention some examples.
In particular, a new generation of fast MWPC counters aimed at TR detection has been 
envisaged for the CBM experiment at FAIR \cite{CBM}. The system's conceptual design 
(presented in \cite{Anton}) is based on a configuration aiming at
 a pion suppression better than 100 for momenta higher than $p\!=\!1.5$ GeV/c with position 
resolution of the order of 200-300 $\mu$m \cite{melanie} at incident fluxes up to 
$\phi\! =\! 100$ kHz/cm$^2$. These requirements are driven by the identification of $J/\Psi$ and $\Psi'$ in the
di-electron decay channel. Populations of $J/\Psi$ and $\Psi'$ 
(in particular the ratio of them) are considered as the most promising signatures of 
Quark Gluon Plasma and its
evolution  \cite{PBM,AntonPsi}. However, the very small production yields (specially
in the case of  $\Psi'$) demand very high interaction rates (up to $10^7$ collisions 
per second), posing challenges for the detector technologies \cite{CBM}.

In this work the rate capability of classical MWPCs with a gap of $h=3$ mm and $s=3, 4$ mm 
anode wire pitch has been explored, for binary gas mixtures Xe-CO$_2$, 
Ar-CO$_2$ and Ne-CO$_2$. CO$_2$ has been chosen as the UV-quencher due to its low 
chemical reactivity, non-flammability and very good ageing properties 
while we focused on Xe due to its high cross-section for X-ray absorption. The data is
interpreted within the theoretical framework developed by 
Mathieson and Smith (\cite{Mat1}, \cite{Mat2}).
 
This document is structured as follows: the setup is described in section \ref{intro};
once the characteristic gain vs voltage curve has been measured (section \ref{Gain_sec}), 
the behavior of the gain as a function of the rate (section \ref{rate_capa}) can be described
within the Mathieson model (subsection \ref{Mat_sec}) after beam-size corrections
are accounted for (subsection \ref{finite_area}), allowing for a determination of the 
mobilities of the drifting ions (subsection \ref{results}). 
An extrapolation to the most common case of uniform illumination by minimum ionizing 
particles (mips), together with a discussion of the results is presented in subsection 
\ref{mips} and section \ref{discussion}, respectively. 

\section{Experimental setup and description of the measurements}
\label{intro}

\begin{figure}[ht!!!]
\begin{center}
\includegraphics[width = 13 cm]{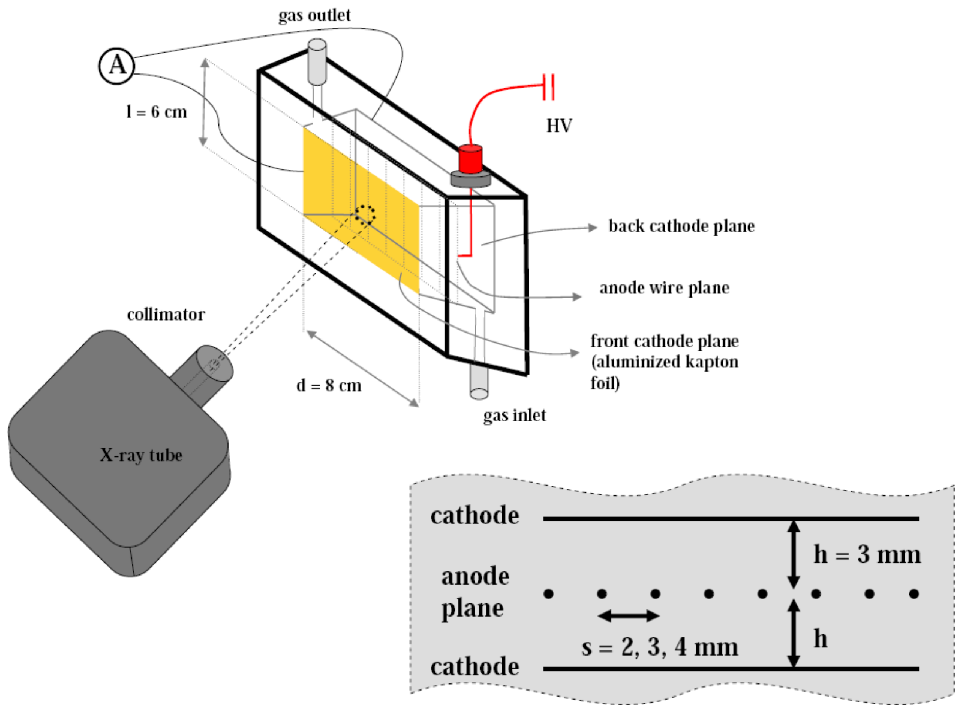}
\caption{\footnotesize Experimental setup. The configuration sketched was
chosen for current measurements, being the amperemeter denoted by -A-. For rate determinations
the amperemeter was replaced by two amplifying stages and a linear Fan-in Fan-out with one
output signal sent to the discriminator and scaler and the other one to the ADC.}
\label{setup} 
\end{center}
\end{figure}

\begin{figure}[ht!!!]
\begin{center}
\includegraphics[width = 11 cm]{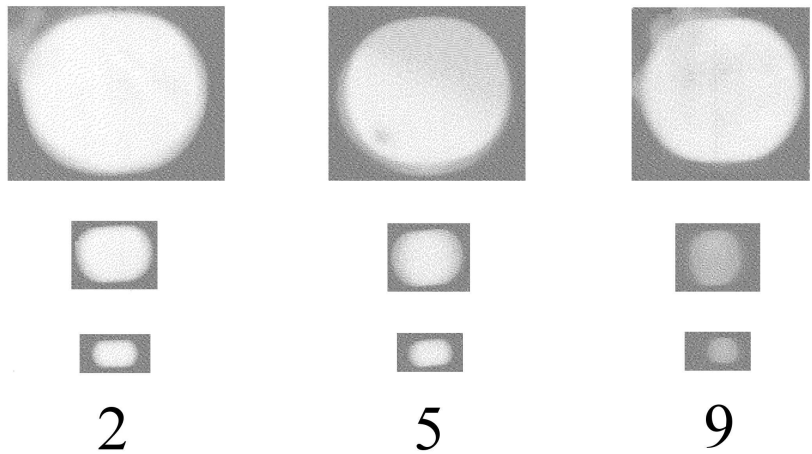}
\caption{\footnotesize Illustration 
of typical beam spots in the Polaroid film stack as a function of the number
of film. From top to bottom: $A=143 \pm 14$ mm$^2$, $A=36 \pm 3.6$ mm$^2$, $A=6 \pm 0.6$ mm$^2$. 
Left-right shows the increasing direction (downstream) of film number. 
A typical choice for the determination of $A$ is the central column (5). Film 2 is over-exposed.
Comparison between 5 and 9 shows a spot fading away in a rather uniform manner, suggesting that the
beam-spot in 5 is highly uniform, as expected.}
\label{beam_spots}
\end{center}
\end{figure}

The experimental setup is shown in Fig. \ref{setup}. The chamber wires are made of gold-coated tungsten 
of 10 $\mu$m radius. The MWPC has a very thin 
entrance foil that acts as one of the cathode planes (25 $\mu$m aluminized kapton). The window
dimensions are 6 cm $\times$ 8 cm but a slightly smaller fiducial area of 5 cm $\times$ 5 cm 
was determined as the region where the gain uniformity was below rms$_{m} \leq 5 \%$.
The chamber was irradiated with an X-ray tube, collimated down to a nominal area 
of approximately $A\simeq\!\!0.5$ cm$^2$, centered
with respect to the afore-mentioned fiducial area. 
The estimation of $A$ was done via simultaneous exposure of 10 Polaroid films attached to the window, 
producing images like the ones shown in Fig. \ref{beam_spots} after developing the film. 
The duration of the exposure was chosen as much as possible such that the first films of the stack were
over-exposed and the last under-exposed. By looking at the beam-profile as a 
function of the film number the beam-quality
can be then assesed. The small presence of tails in some cases
(Fig. \ref{beam_spots}, compare 2-up and 5-up) was taken into account as a 10\% uncertainty 
in the area determined by this procedure.

Unless stated otherwise, all the measurements here presented have been 
performed with an X-ray tube operated at a voltage $V=9$ kV, placed approximately 10 cm
away from the chamber. By using  
different filters (0.25-0.5 mm thick Al foils) most of the low-energy bremsstrahlung and 
characteristic radiation can be suppressed. The feature-less bremsstrahlung
spectra can, by means of this procedure, render a narrow structure close to the
maximum photon energy. 

\begin{figure}[ht!!!]
\begin{center}
\includegraphics[width = 13.8 cm]{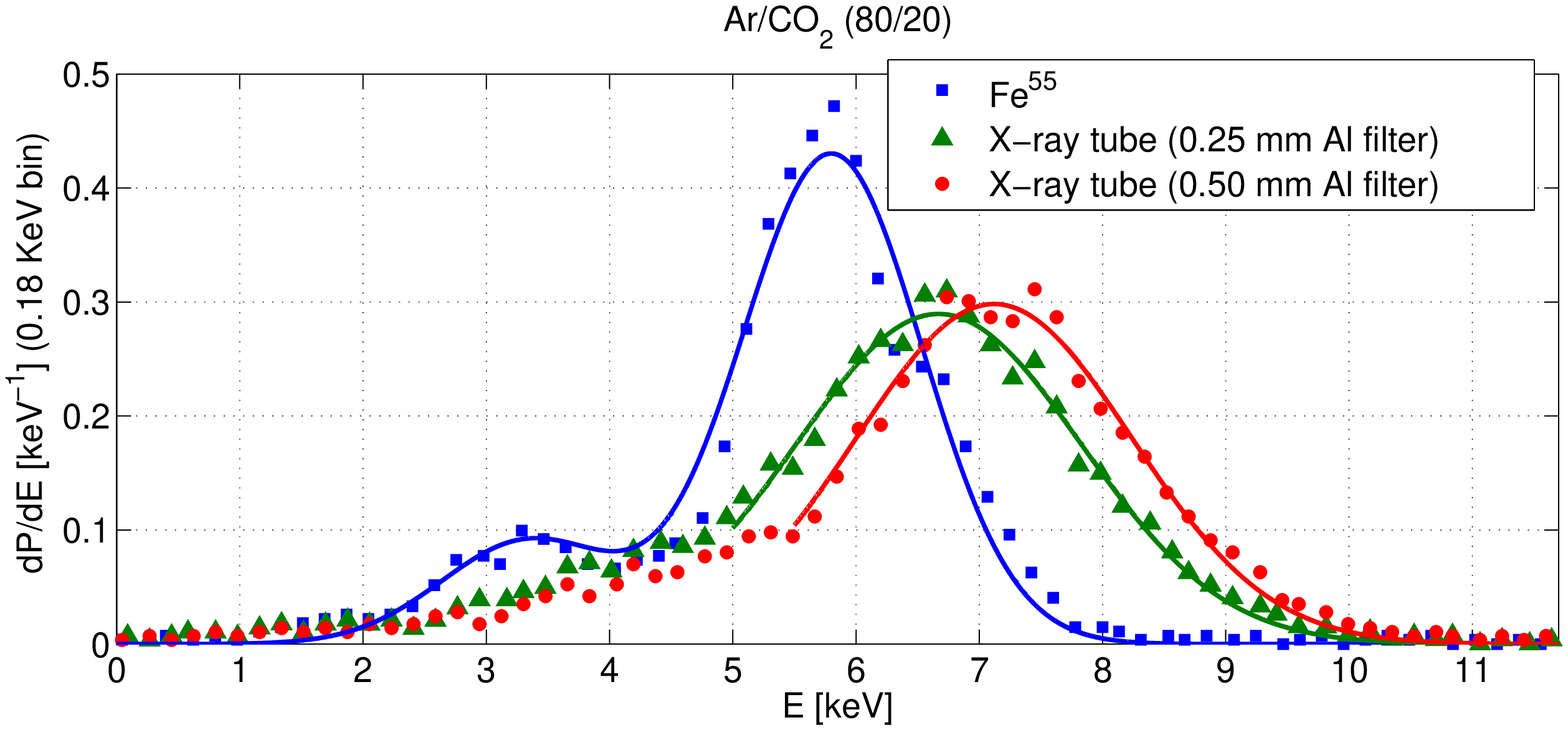}
\caption{\footnotesize Energy loss spectra obtained with the s=4 mm anode-pitch MWPC after calibration. 
The characteristic escape peak from Argon is visible under 
illumination with a Fe$^{55}$ source 
(resolution $\sigma_{E}/\bar{E}=15\%$) but disappears when using the X-ray tube.
The broader energy distribution in such a case stems from the non-monochromatic 
nature of the bremsstrahlung photons. Single (for the tube) and double (for the source) Gaussian distributions
have been fitted to data.}
\label{spectra}
\end{center}
\end{figure}

The measured charge was calibrated by using as reference
the internal conversion line of a Fe$^{55}$ source ($E_{X-ray}=5.9$ kV). Typical 
charge spectra measured with the ADC after calibration 
are shown in Fig. \ref{spectra} for a Ar-CO$_2$(80-20) mixture, obtained with the source (squares) 
and with the tube (triangles, circles) for two different filters. 
The energy spectra obtained with a Fe$^{55}$ source showed a peak-resolution 
$\sigma_{E}/\bar{E}=10-20\%$ with a clear Argon escape peak placed at 
approximately one half of the total absorption one. The spectrum measured with the X-ray tube was 
 broader ($\sigma_{E}/\bar{E}=15-30\%$), resulting from the wider distribution
of the bremsstrahlung photons.
  
In order to obtain the distribution of Fig. \ref{spectra} the signal induced in both 
cathode planes was added and amplified by a custom-made charge sensitive pre-amplifier 
followed by an ORTEC 452 spectroscopic amplifier. The analog signal was duplicated with a linear
Fan-in-Fan-out and sent to a LeCroy 0628 leading edge comparator and to a Lecroy 2249W charge ADC. 
The output of the former
was used for gating the ADC and for the rate determination, via a scaler. 
The chamber current was measured with a Keithley 6487 amperemeter.

Once the current and the avalanche rate are measured, the detector gain can be obtained as:
\beq
m = \frac{i}{n_o r} \label{gain_eq}
\eeq
where $i$ is the total current, $r$ the avalanche rate and $n_o$ the average number of electrons
initially released by the X-ray photon.

For studying the behaviour of the gain as a function of rate, the determination of the later can be 
simplified by resorting to its proportionality with the tube current, that was verified 
over the whole range. Only one direct determination of the
rate (obtained at low rates) is therefore needed at the beginning of each `rate scan' 
and higher values can be obtained by proportionality with the tube current.

\begin{figure}[ht!!!]
\begin{center}
\includegraphics[width = 11 cm]{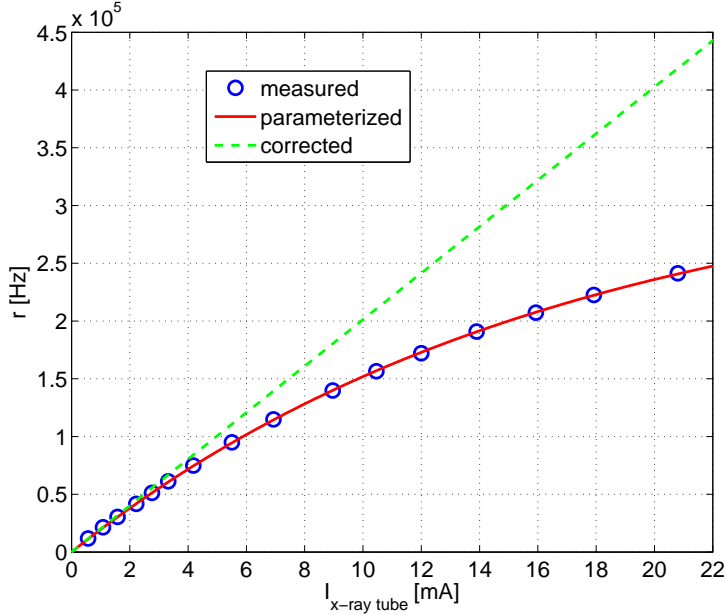}
\caption{\footnotesize Electronic dead-time as measured in pure CO$_2$ at $m \simeq 5\times10^4$.
The points indicate the measured rate as a function of the X-ray tube current. 
The continuous line is a 1-parameter phenomenological fit introduced in the text.}
\label{dead_time}
\end{center}
\end{figure}

In some particular cases like for the determination of the characteristic gain curves, where the nominal
rate was not too high, it was useful to perform direct rate measurements with some regularity 
and care was taken of checking that the electronics dead-time had little impact in the results.
The dead-time of the electronics system was estimated by resorting to the afore-mentioned
proportionallity with the tube current: measurements were performed in pure CO$_2$ 
at $m \simeq 5 \times 10^4$ and a phenomenological model of only one 
parameter was used to fit the data (Fig. \ref{dead_time}), yielding the correction formula needed translate 
 the measured rate $r_{meas}$ to the real rate $r$:
\beq
r = -\frac{1}{\Delta T} \ln(1- \Delta T ~ r_{meas}) \label{dead_time_eq}
\eeq
with $\Delta T$ in the range 2-3 $\mu$s, slightly depending on the gain and mixture.
Due to this dependency, only dead-time corrections not in excess of $20\%$ were allowed 
in practice.

A comparison between these two independent methods (proportionality with the
tube current and dead-time correction) was nevertheless utilized
as a cross-check in a number of cases, showing good agreement up to moderate rates. 
In general, the former method (proportionallity with the tube current) was preferred for 
the measurements of the rate capability and the second (dead-time correction) for the 
determination of the characteristic gain curve. Faster electronics (required for the final 
application in CBM) is currently under development.

All measurements were performed in a monitored (not controlled) atmosphere with the temperature, 
pressure and oxygen content showing values in the range $T\!=$ 22-25$~ ^o$C, 
$P\! =$ 985-1010 mbar, f$_{O_{2}}=$ 8-20 ppm. Fresh gas was injected at 2 detector 
volumes per minute. The voltage was corrected to the standard ambient 
value ($T_o = 20~ ^o$C and $P_o = 1000$ mbar) 
by introducing the reduced potential $V^* = \frac{P_o}{P} \frac{T}{T_o} V$, as is customary.

\section{Gain curve measurements and chamber stability}
\label{Gain_sec}
The characteristic gain curve is a necessary input when attempting a 
quantitative description of the rate capability of MWPCs \cite{Mat1}. On the other hand, the 
maximum operating voltage before glow discharge or breakdown appears in a chamber 
must be also scrutinized critically. The measurements devoted to these aspects
are compiled in this section.

The experimental determination of the gain curve is based on eq. \ref{gain_eq}. The chamber 
was irradiated with the X-ray tube and, after optimization of the filters and tube voltage, 
 a spectrum with a peak energy of $\bar{E}_{X-ray}\simeq=6.6$ keV and $15-30\%$ width
(worsening when increasing the fraction of quencher) was produced. The main advantage of this 
procedure, as compared to irradiation with a Fe$^{55}$ source, is that the very strong decrease in 
absorption cross-section with the CO$_2$ concentration 
(and therefore in counting rate) can be accommodated by adjusting the intensity of the tube with
the additional help of 2 Al filters, so that the measurements of the characteristic 
gain curves can be easily taken at the same avalanche rate. 
This approach largely increased 
the reproducibility of the measurements under different conditions. 

Once the rate and the current are known, the gain can
be calculated from eq. \ref{gain_eq} and the initial number of electron-ion pairs $n_o$:
\bear
&& n_o = \frac{\bar{E}_{X-ray}}{W} \label{ioniz}\\
&& W = \left(\frac{1-f_{CO_2}}{w_{noble}} + \frac{f_{CO_2}}{w_{_{CO_2}}}\right)^{-1} \label{energy_pair}
\eear
being $W$ the effective average energy to produce one electron-ion pair, $w_{noble}$ and 
$w_{_{CO_2}}$ the corresponding values for pure noble gas and CO$_2$, respectively, 
and $f_{CO_2}$ the fraction of CO$_2$.
A slight shift of the peak energy  
$\bar{E}_{X-ray}=6.6-7.1$ keV was observed when using a second filter. This
was taken into account for evaluation of eq. \ref{ioniz}.

In order not to bring the 
amperemeter to the limits of its sensitivity, 
a primary rate $r = 120$ kHz was kept constant for
 the measurements at low gains ($m<5\times 10^4$), 
yielding a minimum current of 5 pA (corresponding to $m\simeq 1$) in pure CO$_2$. 
The chosen rate represents a compromise between a low system dead-time 
(Fig. \ref{dead_time}) and small space-charge, on one hand, and a reasonable 
current at low voltages on the other hand. 
The uncertaintinty of the current was determined through the rms 
of 30 consecutive measurements, resulting in a statistical dispersion of $\simeq 1\%$. 
At $m\simeq 1$ the current stabilized only after several 
minutes (maximum 30 min), showing an uncertainty of $\simeq 10\%$.

Aiming at reducing the effect of space-charge in the characteristic curves, 
gains above $m\simeq 5 \times 10^4$ 
were determined at a reduced rate of $r = 10$ kHz. After
correcting for the system dead-time both the high and low gain regimes were found to be in
good agreement. No significant difference was observed between the gain curve 
obtained with the Fe$^{55}$ source as compared with the tube, for a reference Ar-CO$_2$(80-20) mixture.

\begin{figure}[ht!!!]
\begin{center}
\includegraphics[width = 12.8 cm]{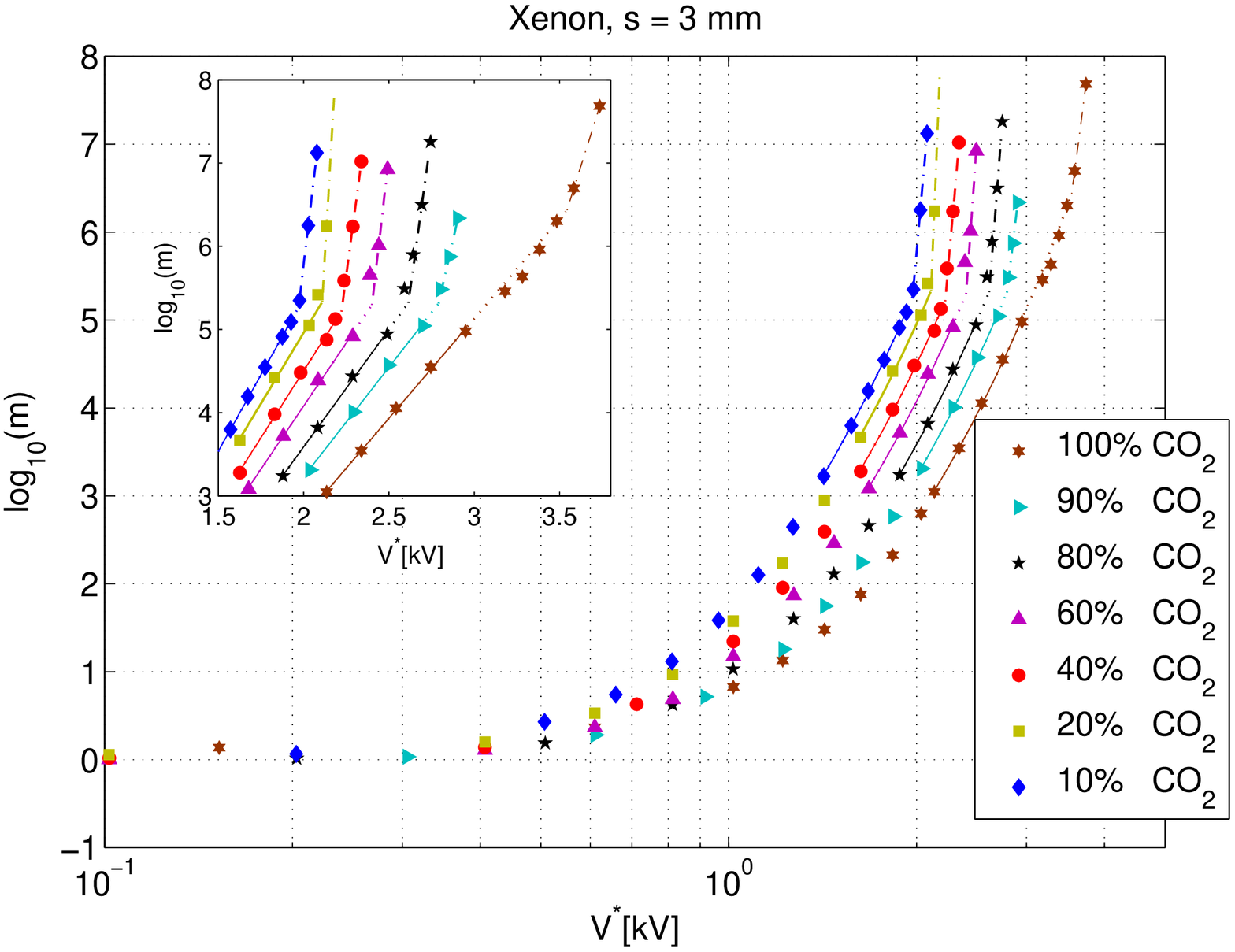}

\includegraphics[width = 12.8 cm]{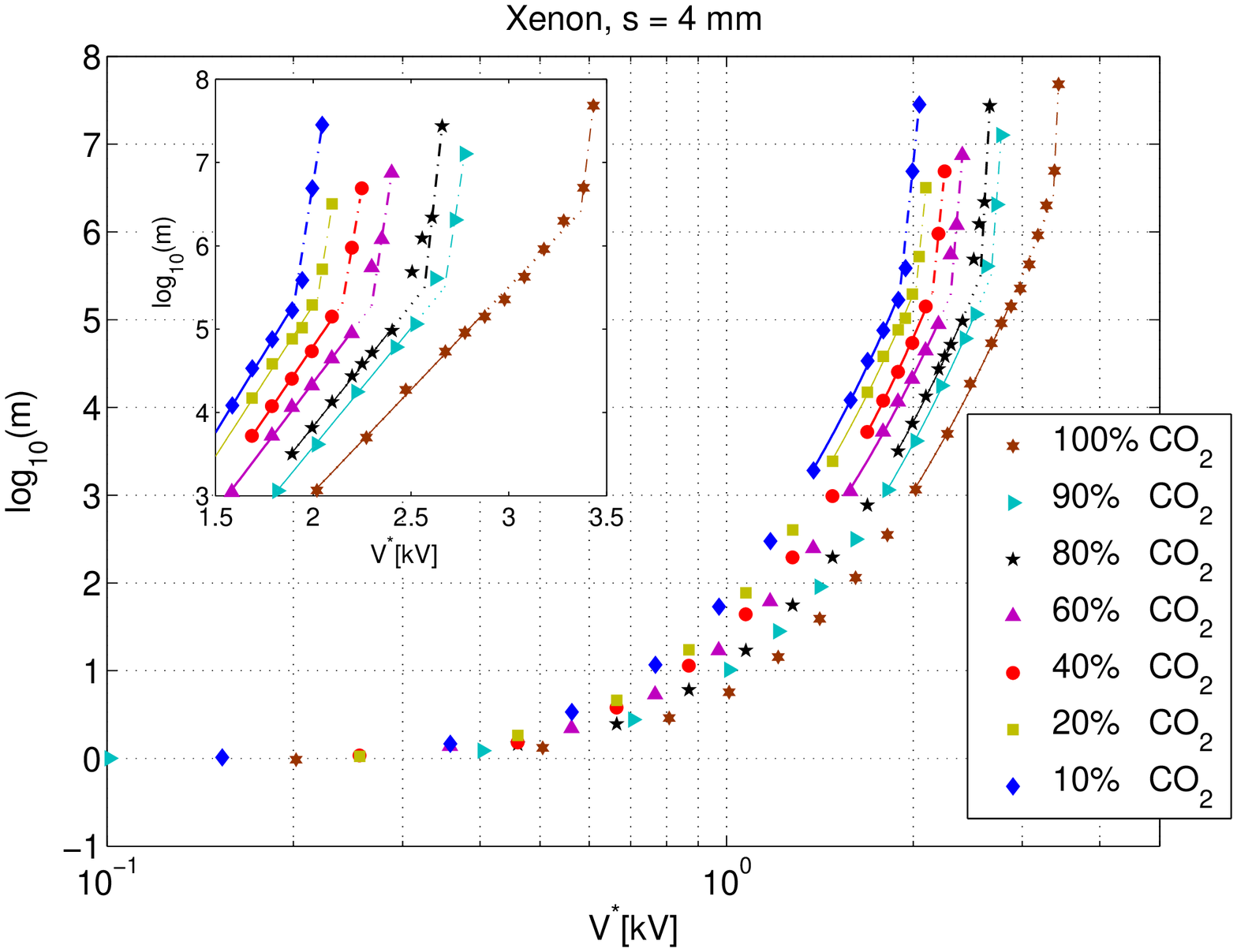}
\caption{\footnotesize Up: gain $m$ vs reduced voltage $V^*$ for the $s=3$ mm -pitch chamber operated in 
Xenon with different fractions of CO$_2$. Down: like up-picture but for the $s=4$ mm -pitch chamber.
The inset zooms in the region where data has been fit to an exponential ($\log_{10}{(m)}=3-5$)
and also the regime of self-sustained discharge ($\log_{10}{(m)}\gtrsim5.5$).}
\label{Xe_gain}
\end{center}
\end{figure}

For illustration, the gain curves as a function of the reduced anode potential $V^*$ 
are shown in Fig. \ref{Xe_gain} for the case of Xenon mixtures in the 
$s = 3$ and 4 mm -pitch chambers. The region $m=1$ is well described when 
obtaining $n_o$ from the tabulated values of Sauli \cite{Sauli}, but an excellent agreement
was also verified for the Ne and Ar mixtures. Overall, the behavior of both chambers 
is very similar, regarding voltage dependence. 
For chambers of pitch $s = 2$ mm (not shown) the gain curves are 
substantially different due to the increasing influence of the neighboring wires in the field 
close to the anode. A systematic study of the gain behavior of these chambers will follow 
to this publication.

Fig. \ref{Xe_gain} shows an exponential fit in the region $m = 10^3$ - $10^5$ 
(solid line), that is used in the next section, together with an 
extrapolation to higher gains (dotted line). The deviation of data from the fit was identified as the 
onset of a self-sustained discharge in the chamber\footnote{The chamber continues to draw the same current
even when the tube is switched off. After little training, the process can be identified by
direct observation in the scope.}.
From the crossing point of the up-extrapolated gain (dotted line) 
and a linear down-extrapolation (dot-dashed line) of the last two data points
one can obtain a crude estimate of the onset of the process. For Xe mixtures the chambers were stable 
up to a maximum gain $m_{max} = 3 \times 10^5$, being little dependent on the fraction of quencher. 
After prolonged operation in the discharge regime, the chamber required a power cycle 
to restore its previous performances. 

\section{Rate capability measurements}
\label{rate_capa}
\subsection{The Mathieson model}
\label{Mat_sec}

The theoretical description used in this work has been introduced
elsewhere \cite{Mat1}. 
For simplicity, an exponential dependence of the gain with the 
applied voltage has been assumed, such that:
\beq
\ln(m) = aV + b \label{gain_curve_eq}
\eeq
that is well satisfied for our chambers in the 
region $m = 10^3-10^5$, and is a convenient parameterization provided
the gain was never reduced below those values, when operated
at high photon fluxes. Following  \cite{Mat1}, 
the gain behaviour as a function of the flux can be expressed as:
\beq
m\!=\!m_o \exp \left(-q_e \frac{ a^2(f)~ n_o(f)}{2 \mu_{\alpha}(f)} \frac{s h^2 ~ \bar{d}_m(s,h)}{C_L(s,h,r_a)}
\frac{m}{\ln{m}-b(f)} \phi\right) 
\label{M_Mat}
\eeq
with $q_e$ being the electron charge, $m_o$ the gain at zero rate, $n_o$ the initial number of electron-ion
pairs, $\mu_{\alpha}$ the mobility of ion-specie $\alpha$, $f$ 
the fraction of quencher, $C_L$ the wire 
capacitance per unit length and $\phi$ the avalanche flux (in [T$^{-1}$L$^{-2}$]). The correction for finite area of 
illumination, here denoted as $\bar{d}_m$, is a function 
defined between 0 and 1, that solely depends on the chamber geometry and the shape of the beam 
spot, under the assumptions of \cite{Mat1}\footnote{In particular ion diffusion is not included
in the model.}. The functional expression of $C_L$ is: 
\beq
C_L=\frac{2\pi\epsilon_o}{\ln(r_c/r_a)}
\eeq
with $r_a$ the anode radius and $r_c$ being well approximated in the limit $s/h<2\pi$ by:
\beq
r_c \simeq \frac{s}{2 \pi} e^{\pi h/s}
\eeq
It becomes apparent that, once
the characteristic `m vs V' curve has been measured ($a$, $b$ are then known), 
the only free parameter for every mixture is $\mu_{\alpha}(f)$. Therefore, within the Mathieson 
model only one curve 
'$m$ vs $\phi$' is needed experimentally for a given mixture (for any $m_o$)
and the behavior for 
different $m_o$ can be derived from eq. \ref{M_Mat}. Moreover, the mobility of the ions is usually 
well described by the Blanc's law \cite{Charpak}, that in our case simply reads:
\beq
\frac{1}{\mu_{\alpha}} = \frac{1-f_{CO_2}}{\mu_{\alpha, noble}} 
+ \frac{f_{CO_2}}{\mu_{\alpha,CO_2}} \label{Blanc}
\eeq 
where $\mu_{\alpha, noble}$ and $\mu_{\alpha,CO_2}$ are the mobilities of the drifting ion $\alpha$ 
 in the corresponding pure gas. In case of pure gases the nature of $\alpha$ must change but, if two rate curves
are obtained at, say, $f_{CO_2}=20\%$ and $f_{CO_2}=80\%$, then eq. \ref{Blanc} together with eq. \ref{M_Mat} 
would allow for a complete characterization of the rate capability of the chamber for all the possible 
admixtures of two given gases (except maybe in the limit of pure gases). 
In general different drifting species may coexist, but on the basis
of a generally accepted fast charge-transfer mechanism \cite{Sauli} only one drifting ion is considered
here for each noble gas-CO$_2$ mixture, its nature not depending 
on the gas concentration (except maybe in the limit of pure gases).
The complex nature of the ion drift and its high sensitivity to the presence of impurities 
\cite{Charpak} advice, nevertheless, 
to verify Blanc's law experimentally for every particular case. 

The main experimental difficulty involved in the precise evaluation of the rate capability of a chamber
can be observed after inverting eq. \ref{M_Mat}: if the rate capability $\phi_{_F}$ 
is defined as the flux needed for causing a certain fractional gain drop $F$, 
the following expression can be obtained:
\beq
\phi_{_F} = q_e \frac{\ln\left((1-F)m_o\right)-b}{(1-F)m_o} \frac{2 \mu_{\alpha}(f)}{ a^2(f)~ n_o(f)} \frac{C_L(s,h,r_a)} {s h^2 \bar{d}_m(s,h)} \ln(1-F) \label{R_Mat}
\eeq
and making use of the fact that the logarithm is a slow varying function of its argument:
\beq
\frac{\delta \phi_{_F}}{\phi_F}\simeq \frac{\delta F}{F} \label{error}
\eeq
Therefore, an experimental uncertainty in the determination of 
$m_o$ by $5\%$ will cause an uncertainty in the determination of
the fractional drop at $F=10\%$ by $50\%$, and this uncertainty will be directly propagated
to the estimated rate capability.

\begin{figure}[ht!!!]
\begin{center}
\includegraphics[width = 12.5 cm]{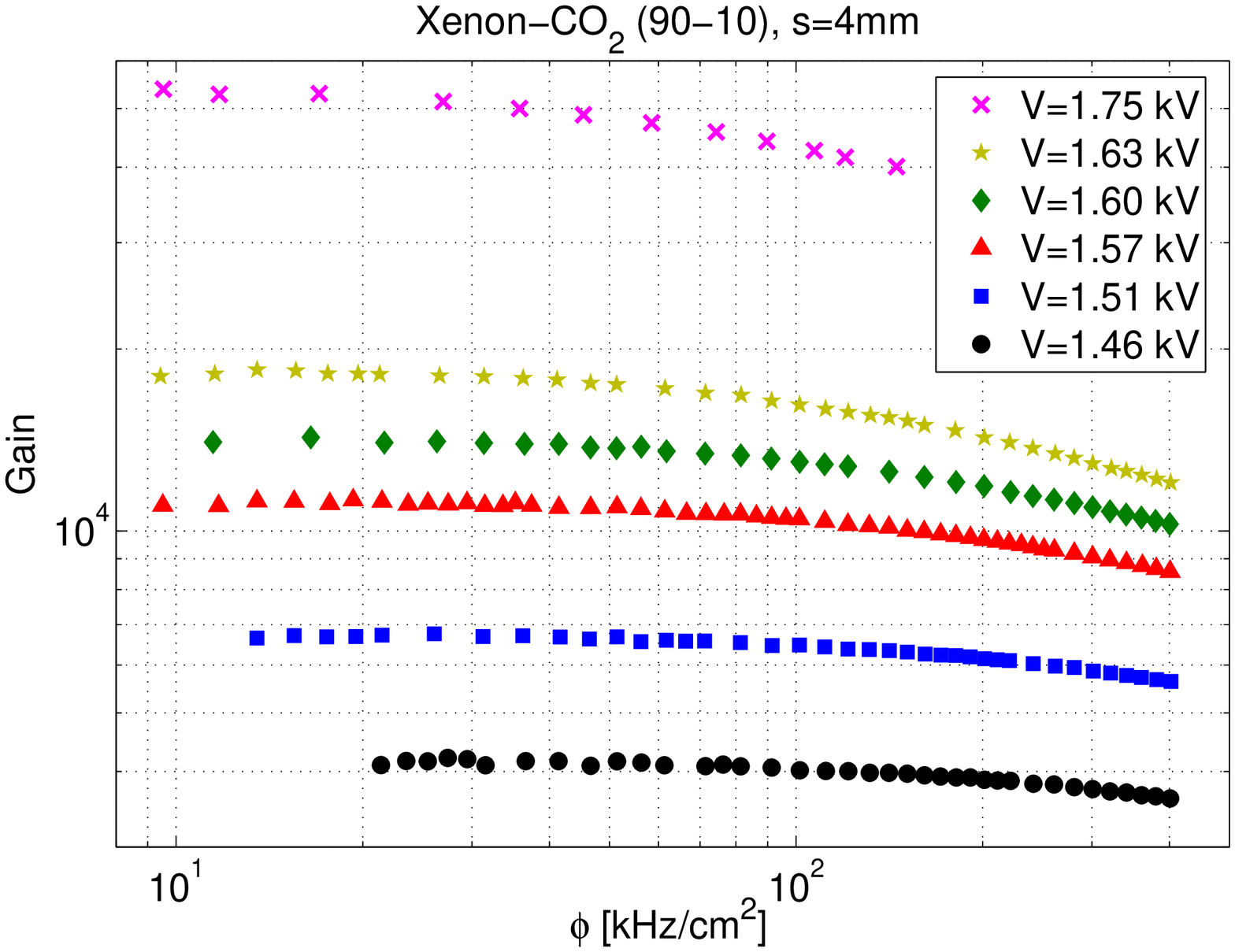}

\includegraphics[width = 12.5 cm]{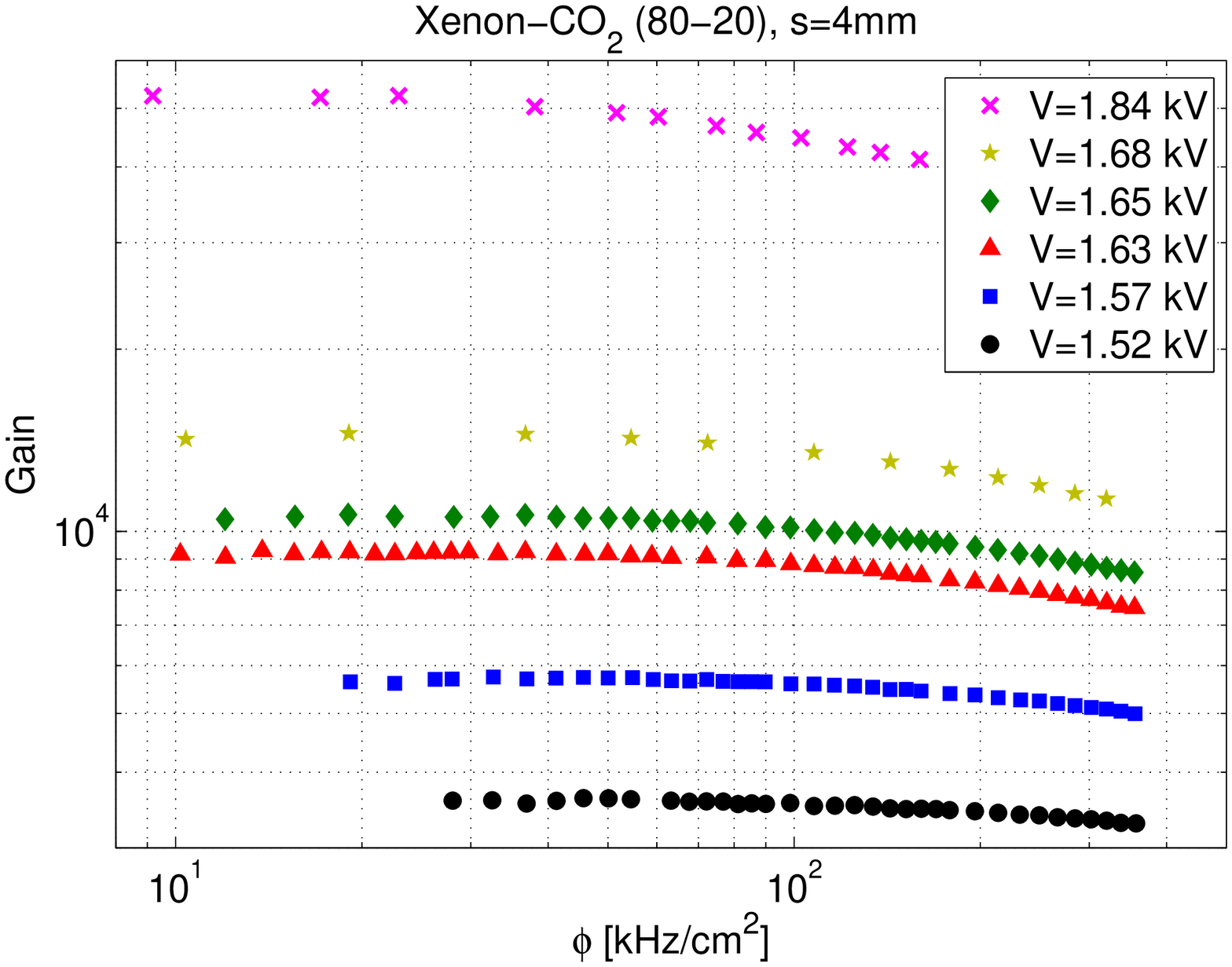}
\caption{\footnotesize Up: gain as a function of the avalanche flux for the $s=4$ mm chamber 
operated in Xenon with 10\% CO$_2$. Down: like up-figure but for 20\% CO$_2$. The flux has been
 extrapolated to the case of uniform irradiation. This correction
implies a scaling-down of the measured flux by a factor 2 for our setup (see text).}
\label{Xe_rate}
\end{center}
\end{figure}

An illustration of the measured gain behavior as a function of the avalanche flux 
is shown in Fig. \ref{Xe_rate} for the case of Xe-CO$_2$(90-10) and 
Xe-CO$_2$(80-10) in the $s=4$ mm chamber. 
The statistical uncertainty is smaller than the size of the data points. 
Indeed, not all the mixtures were equally well characterized: in view of the final
application of this work (TR detection) the measurements in Xenon were performed at 3-4 different values
of $m_o$ for every concentration of quencher $f_{CO_2}$. 
For Argon and Neon -based mixtures almost 2 curves at different $m_o$ 
were measured in average for every value of $f_{CO_2}$.
The flux $\phi$ has been already extrapolated to the case of uniform irradiation as 
$\phi\rightarrow\bar{d}_m \phi$, taking
into account the finiteness of the beam-spot. As shown in the next section, the correction 
factor involved in such an extrapolation $\bar{d}_m$ amounts to $\simeq 0.5$ 
for the bulk of the measurements presented here.

\subsection{Extrapolation to infinite area of illumination}
\label{finite_area}

The `Mathieson model' includes a prescription on how to correct for finite beam size effects 
(\cite{Mat1}, \cite{Mat2}) and even extraordinary experimental agreement has been found for 
the 1-D case \cite{Mat3} (finite irradiation along the direction of the wires, infinite 
in the transverse direction). To the authors' knowledge, no dedicated study exists
for the 2-D case.

Following the original formulation of Mathieson, the voltage drop $\Delta{V}(x,y)$
at the anode 
plane caused by a charge distribution with a certain x-y profile can be obtained
from an electrostatic calculation assuming a constant space-charge density of the
drifting ions in the chamber. A correction factor can be then introduced as:
\beq
d_m(x,y) = \frac{\Delta V(x,y)}{\Delta V_o}
\eeq
where $\Delta V_o$ is the voltage drop at the anode plane 
in case of uniform/infinite irradiation. A calculation of the 
factor $d_m(x,y)$  for a squared beam profile\footnote{The experimental oval-like beam-spot 
will be assimilated to a square, for the sake of simpler theoretical description.} 
can be accomplished by using expressions in \cite{Riegler}, yielding: 
\bear
&& d_m(x,y) = \frac{8}{\pi^2 h^2}\int_o^\infty \int_o^\infty dk_1 dk_2 
\frac{\sin(k_1 H/2)\sin(k_2 L/2)\cos(k_1 y)\cos(k_2 x)}{k_1 k_2 (k_1^2 + k_2^2)} \rc
&& \times \left[ 1 - \frac{1}{\cosh\left( h (k_1^2 + k_2^2)^{1/2} \right)}\right] \label{dmRie}
\eear
Following the notation of \cite{Mat2}, $L$ is the beam size along the wires (x axis) and $H$ across 
them (y axis). 
When $H\rightarrow\infty$ (1-D limit) $d_m(x,y)$
must be averaged in the interval $\pm L/2$ in order to obtain the 
desired correction factor, as has been shown in
\cite{Mat1}. This factor as in eq. \ref{M_Mat} would be naturally interpreted 
as an `effectively' reduced flux $\bar{d}_m \times \phi$ 
or, equivalently, as a certain particle rate over an `effectively' increased area, 
as compared to uniform irradiation.  
Remarkably, according to \cite{Mat1} the correction factor $\bar{d}_m(L,H,s,h)$ depends 
only on the chamber/beam arrangement.

Despite this apparent simplicity, the `effective' voltage distribution at the wire
positions $y_k$ ($\Delta{V}_e(x,y_k)$) seems to be needed in order to generate the exact 2-D
distribution $\Delta{V}(x,y)$ that allows for a precise description of data \cite{Mat3}.
A generalization to the 2-D case by averaging $d_m(x,y)$ over the $H/s$ irradiated wires 
was introduced in \cite{Mat2} based on the magnitude $\Delta{V}_e(x,y)$. The original
approximation will be used in this work, $\Delta{V}(x,y_k) \simeq \Delta{V}_e(x,y_k)$,
for the sake of simplicity, so that eq. \ref{dmRie} can be used directly.

In the continuous case where the density of wires is infinite, 
a direct integration of eq. \ref{dmRie} yields the desired average:
\beq
\bar{d}_{m,c}=\frac{1}{L H}\int_{-L/2}^{L/2} \int_{-H/2}^{H/2} d_m(x,y) dy dx \label{dmInt}
\eeq
while in the discrete case the integral over $y$ must be replaced by a discrete 
average at the position of the wires:
\beq
\bar{d}_m=\frac{1}{N L} \int_{-L/2}^{L/2} \sum_{k=1}^N d_m(x,y_k) dx
\eeq
where $k$ runs up to the $N$ wires within the area of illumination.
 If the discretization of space would influence the value of $\bar{d}_m$ also the 
relative position chamber-beam will. This was evaluated by simulating chamber-beam 
displacements along $y$ in $0.25$mm intervals for different pitches $s=3,4$ mm and a 
beam-spot $L\simeq H\simeq 6$ mm, yielding a deviation from the continuous
value $\bar{d}_{m,c}$ within $\pm 10\%$.

\begin{figure}[ht!!!]
\begin{center}
\includegraphics[width = 12 cm]{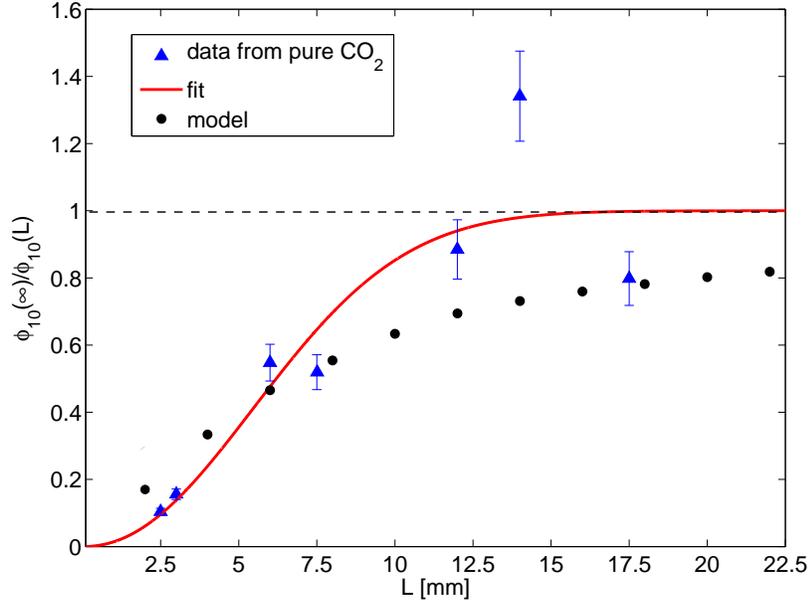}
\caption{\footnotesize Comparison between data (triangles) and model (circles) for the correction 
factor $\bar{d}_m=\phi_{10}(\infty)/\phi_{10}(L)$. $\bar{d}_m$ has been experimentally defined as the 
ratio of the flux at $10\%$ gain
drop for infinite area of irradiation divided by the one measured for a square beam of 
side $L$. $\phi_{10}(\infty)$ is defined as the average of the three higher points.
Data has been taken in pure CO$_2$. A 2-parameter fit is shown also as a continuous red line.}
\label{beam_size_fig}
\end{center}
\end{figure}

  In order to better assess the role of the beam size, a specific set of measurements 
of the rate capability at various beam sizes was accomplished. Taking advantage
of the fact that the correction factor $\bar{d}_m(h,s,L,H)$ should not depend, within the model, 
on the gas mixture, its value was evaluated from measurements of the rate
capability at $10\%$ drop ($\phi_{10}$) in pure CO$_2$ (Fig. \ref{beam_size_fig}).
The correction factor was obtained experimentally after normalizing by 
the rate capability obtained at infinite beam size $\bar{d}_m=\phi_{10}(\infty)/\phi_{10}(L)$. 
The value of $\phi_{10}(\infty)$ was determined as the average of the three 
rate capabilities measured at the largest beam profiles.

Fig. \ref{beam_size_fig} shows the behavior of $\bar{d}_m$ 
both in data (triangles) 
and model (circles) for a square profile, together with a phenomenological 2-parameter fit to a 
function $\bar{d}_m=1-\exp(-{L/L_o}^c)$ (continuous line).
The theoretical value of $\bar{d}_m(L)$ tends asymptotically to 1 although slower than one
may expect (even for $L=20$ mm ($\gg s,h$) 
the required correction factor is still at the level of $20\%$)\footnote{To reduce the correction
below 5\% would require of a square spot with its side being larger than 10 cm.} and, conversely,
it drops to zero very fast when $L\simeq s,h$, as also seen in data.
Since data, model and fit agree reasonably well for our nominal beam profile $L\simeq6$mm, 
 a value of $\bar{d}_m=0.5$ was taken for extrapolating the measured fluxes to the uniform irradiation
case, and a $20\%$ overall uncertainty was estimated for the flux calculated following this procedure. 
This value corresponds approximately to the uncertainty of the mean of the three highest points in 
Fig. \ref{beam_size_fig} that was experimentally defined as $\phi_{10}(\infty)$.

\begin{figure}[ht!!!]
\begin{center}
\includegraphics[width = 7.5 cm]{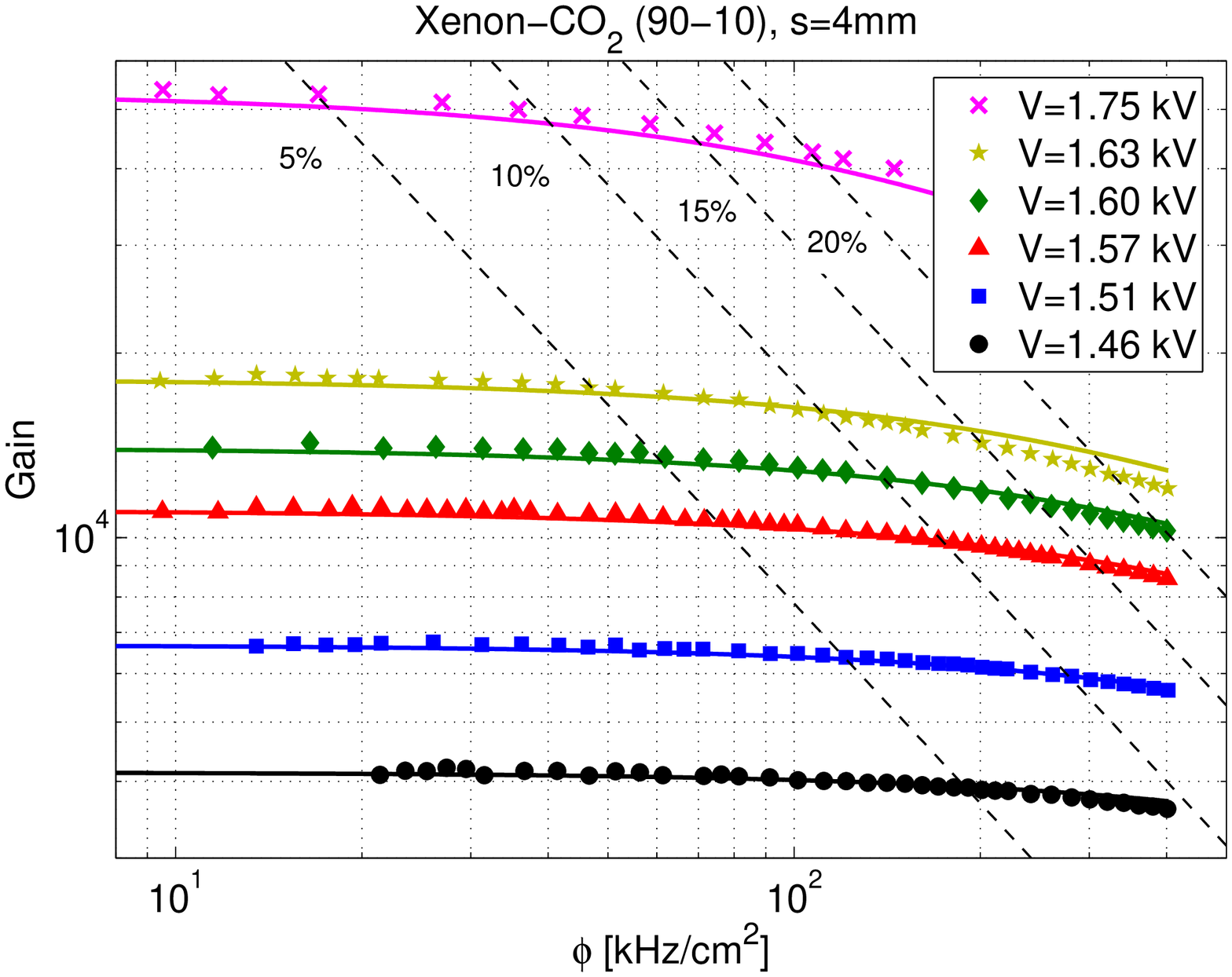}
\includegraphics[width = 7.5 cm]{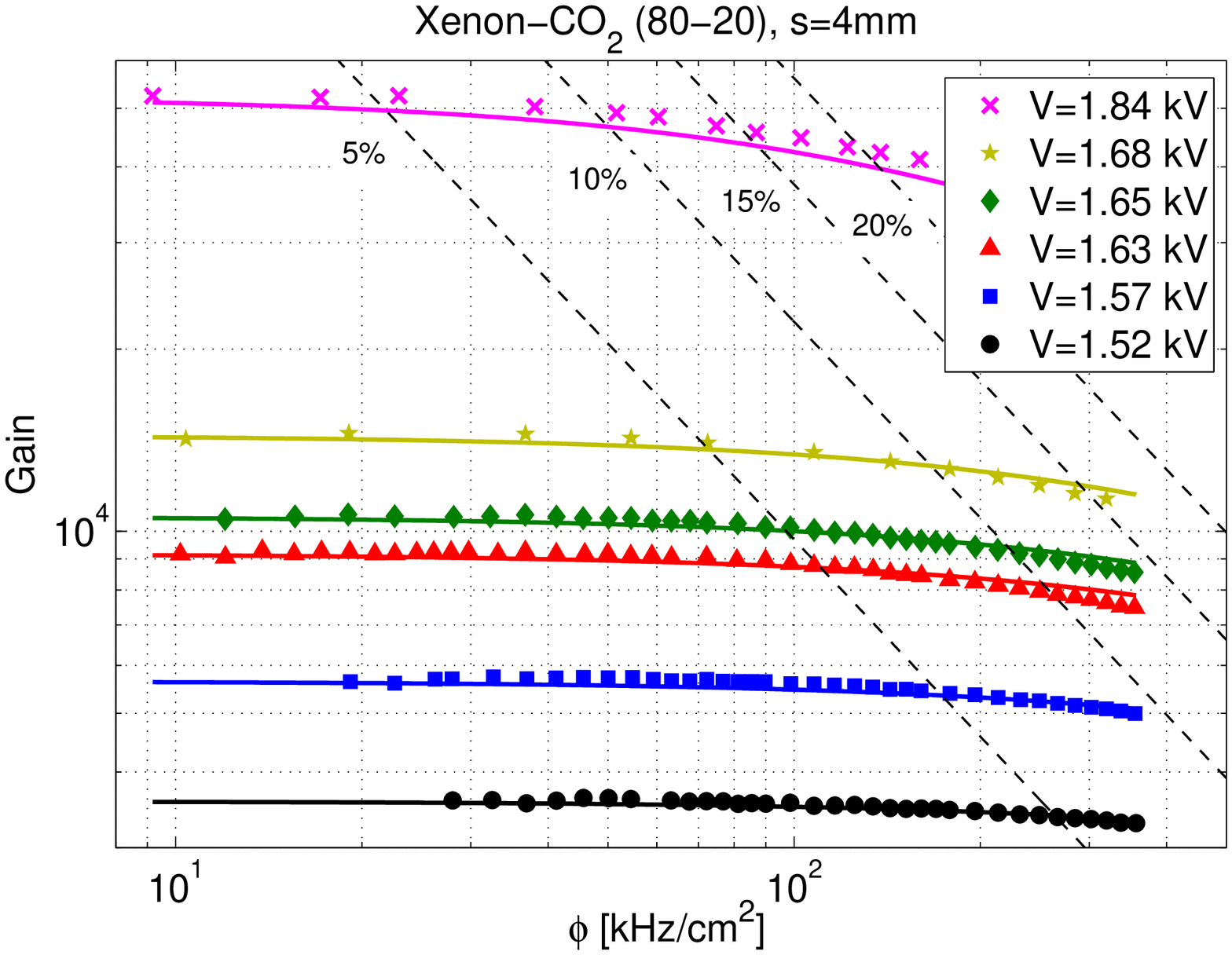}

\includegraphics[width = 7.5 cm]{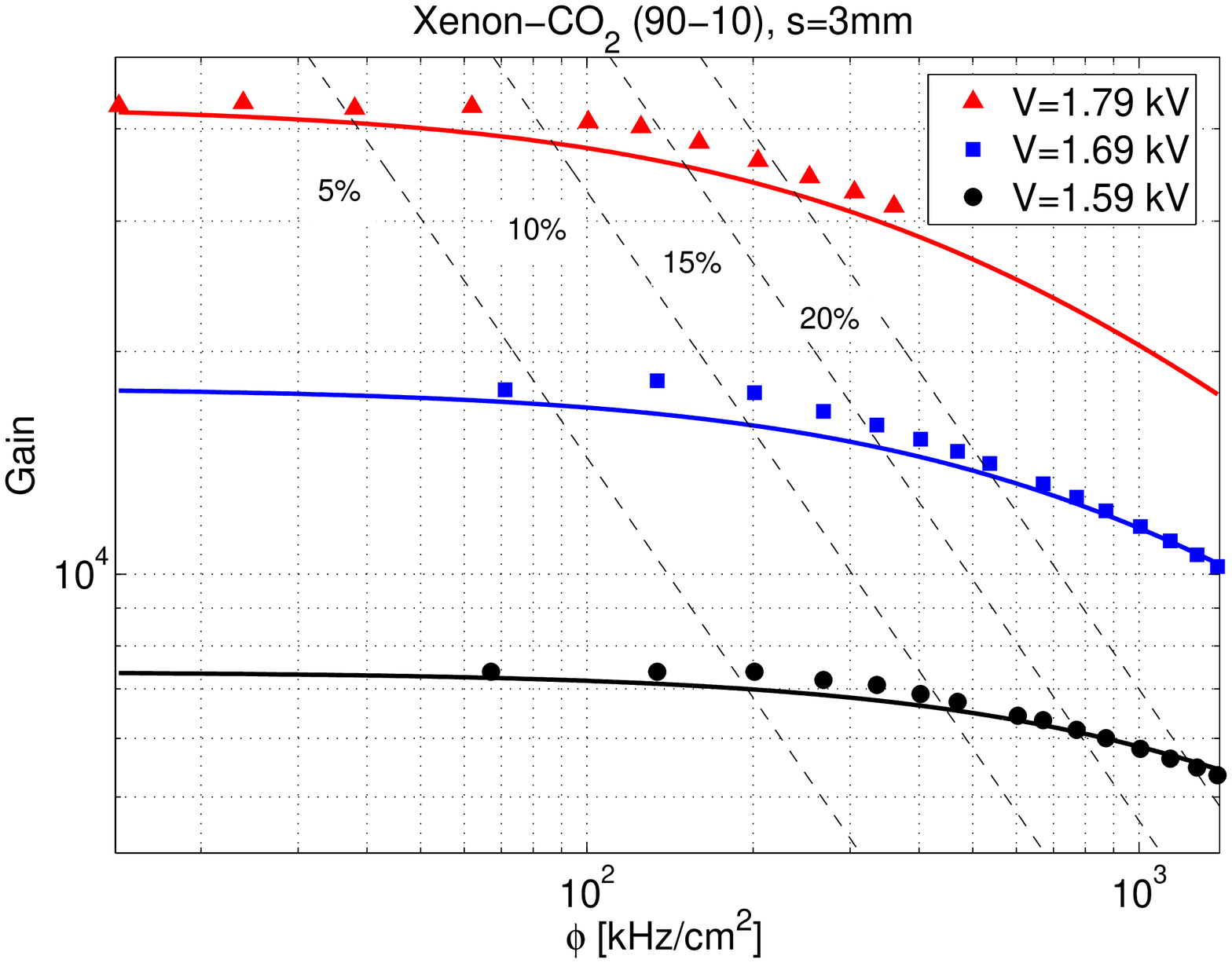}
\includegraphics[width = 7.5 cm]{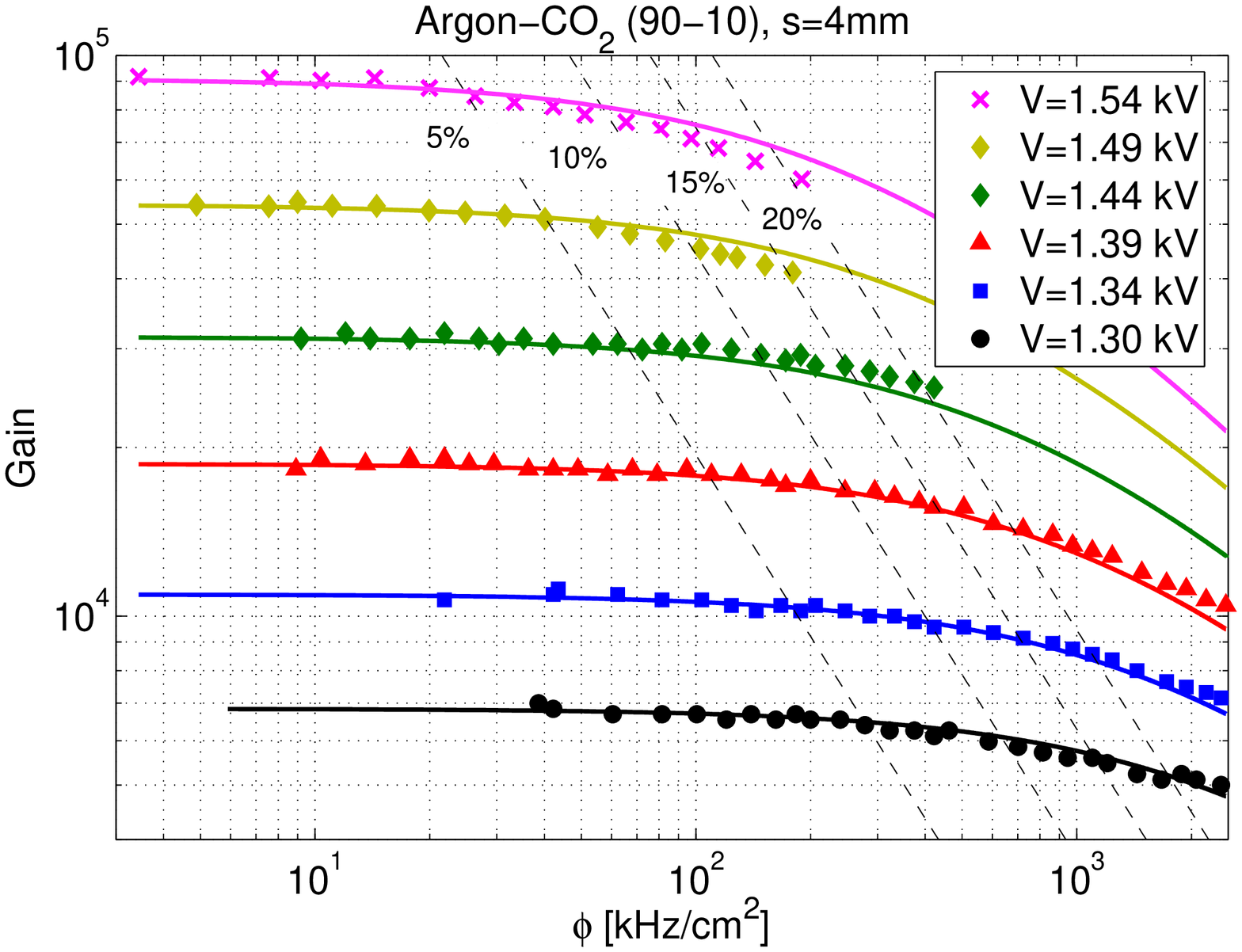}

\includegraphics[width = 7.5 cm]{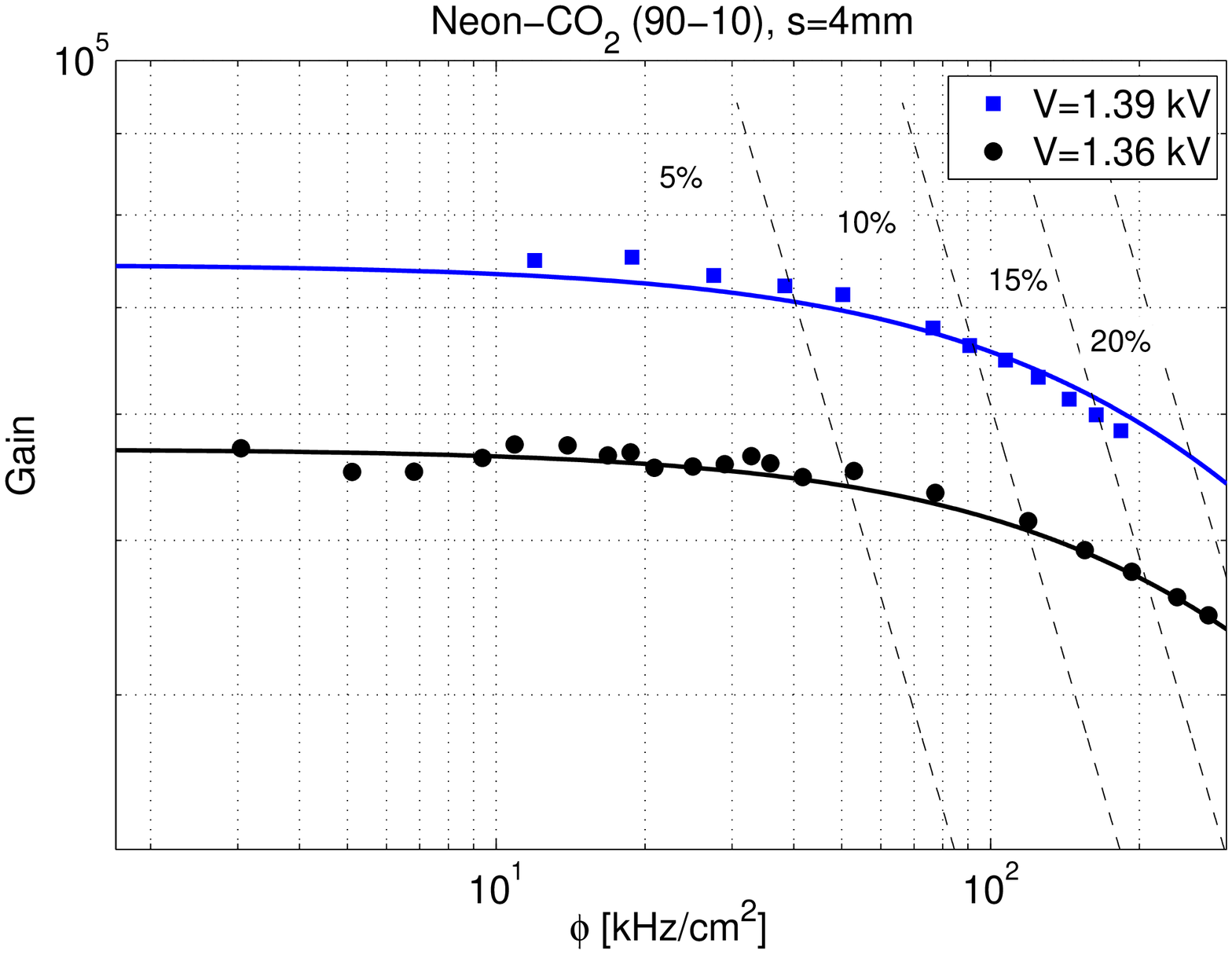}
\includegraphics[width = 7.5 cm]{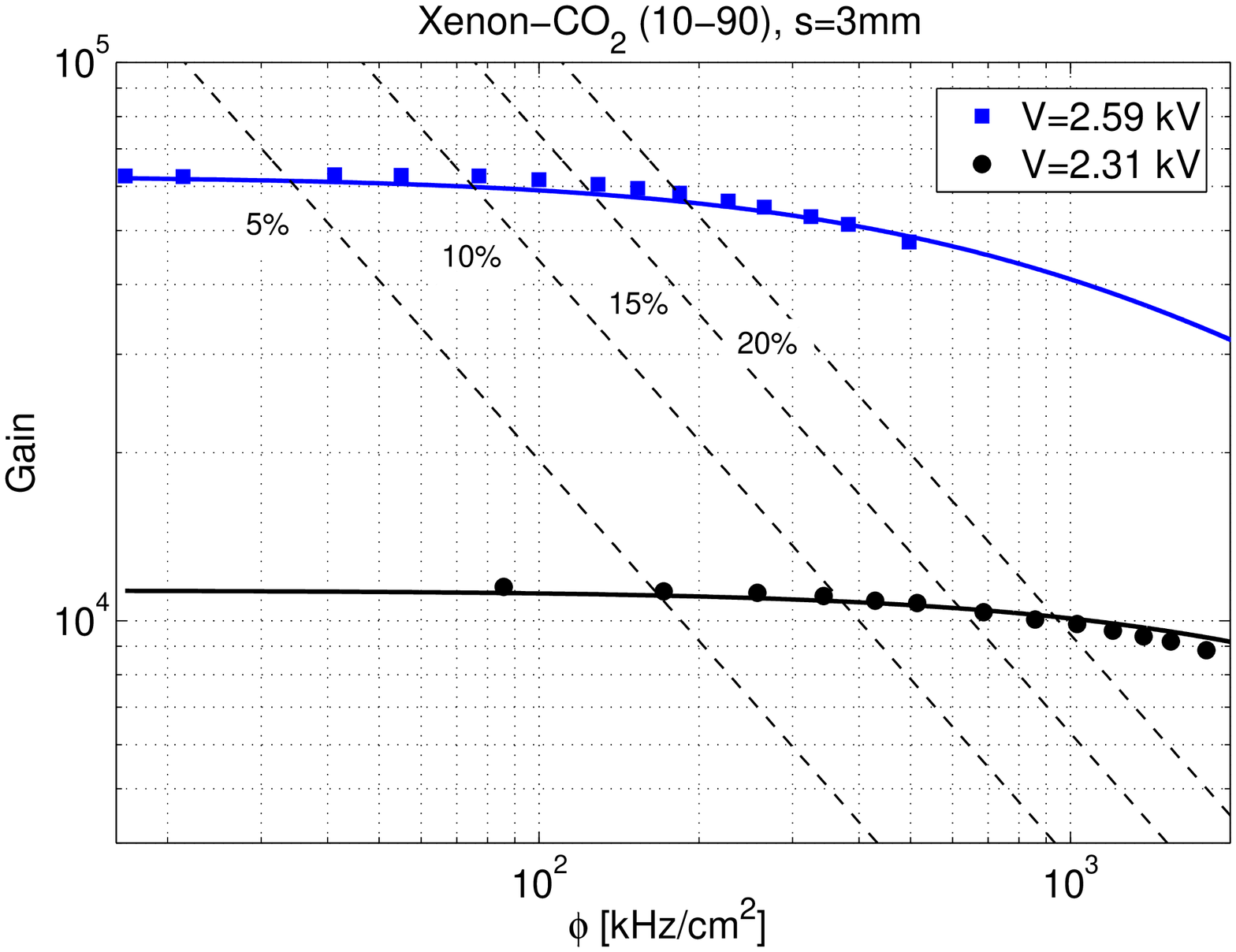}
\caption{\footnotesize Up-left: fit of the rate curves for Xe-CO$_2$ $(90-10)$ obtained for the $s=4$ mm chamber.
Up-right: the same as up-left but for $80-20$ admixture. Center-left: the same as up-left but for $s=3$ mm. 
Center-right: the same as up-left but for Ar-CO$_2$. Down-left: the same as up-left but 
for Ne-CO$_2$. Down-right: fit of the rate curves for Xe-CO$_2$ $(10-90)$ obtained for the $s=3$ mm chamber. 
Dashed lines represent the theoretical flux at given fractional
gain drops from the Mathieson formula with parameters obtained from the fit. 
A total of 88 curves/data sets have been fitted simultaneously.}
\label{Fit_rate}
\end{center}
\end{figure}

\subsection{Fitting procedure}
\label{results}
A simultaneous description of all data based on 
eqs. \ref{M_Mat} and \ref{R_Mat} (Mathieson model including Blanc's law) 
is presented in this section.
Following the spirit of \cite{Charpak} no assumption is done regarding the nature of 
the drifting ion that is denoted by $\alpha$ (in Xe), $\beta$ (in Ar) and $\delta$ (in Ne). There
are therefore 6 free parameters, namely:
$\mu_{\alpha, CO_2}$, $\mu_{\alpha, Xe}$, $\mu_{\beta, CO_2}$, $\mu_{\beta, Ar}$, $\mu_{\delta, CO_2}$, 
$\mu_{\delta, Ne}$.
A global fit of the gain vs flux curves for the $3$ mm -pitch chamber (36 data sets) 
and the $4$ mm -pitch chamber (52 data sets) was therefore attempted by using the $6\times2$ free parameters introduced
above (2 for each noble gas-mixture). The transcendental equation \ref{M_Mat} was evaluated numerically 
by interpolation after being tabulated first.

A $\chi^2$ minimization was performed by assigning 
 equal weight to the data sets for every mixture. This means that each residual is normalized
by the measured value and each data set is additionally weighted by the number of data sets
availables per mixture (for instance, the residuals for each data set for the Xe-CO$_2$(90-10) mixture in s=4 mm
(6 in total) are correspondingly divided by 6, Fig. \ref{Xe_rate}-up). This procedure avoids 
that the fit is dominated by high gains, and compensates for the fact that not all
gas mixtures are equally well represented in terms of data sets (see app. \ref{app_data}).

Two main sources of uncertainties were identified: 
\begin{enumerate}
\item One is related to the estimate of $m_o$ that is obtained for each data set as the average of the 3 
 gains measured at the lowest rates.  This procedure works generally very well but in some cases 
there were not enough points for a good determination. Estimates of $m_o$ wrong by few 
percent are very critical (eq. \ref{error}) and, indeed, a typical disperssion of 1\%
is present in data.
In order to calculate the error introduced in the fit due to this effect, $m_o$ was 
globally scaled up and down for all data sets  in the range $\pm 3\%$ in steps of $1\%$ 
and the fit performed afterwards. The parameters' uncertainties were 
estimated from the rms of the values obtained from these different fits. 
\item A second source of uncertainty comes from the fact that the measurements were 
taken along 3 major campaigns and 3 (slightly) different beam sizes were used. Extrapolations
to uniform irradiation were made according to Fig. \ref{beam_size_fig} for which  
a systematic error of $20\%$ in the re-calculated flux was assigned.
\end{enumerate}

\begin{figure}[ht!!!]
\begin{center}
\includegraphics[width = 11.5 cm]{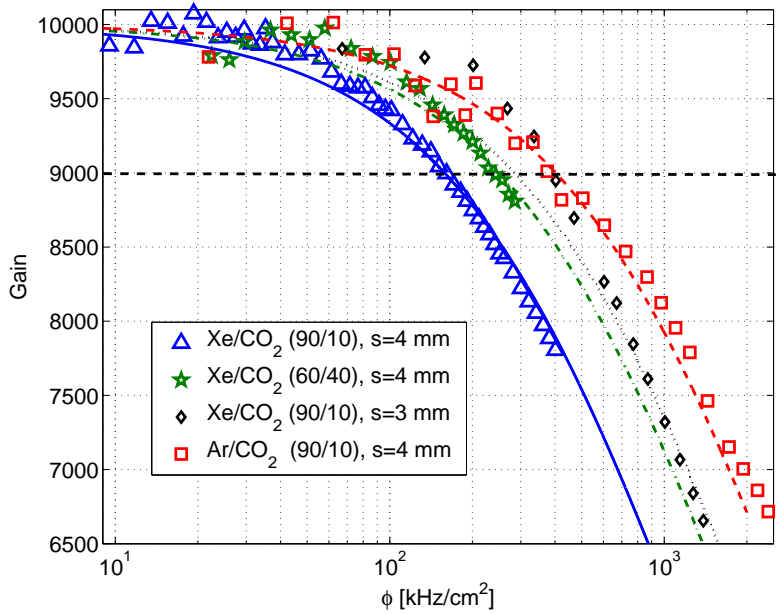}
\caption{\footnotesize Compilation of different curves extrapolated to a 
typical gain $m_o=10^4$. Figure aims at
illustrating the main dependences of the rate capability, together with a detailed comparison with 
the theoretical model (lines).}
\label{Fit_M_mixed}
\end{center}
\end{figure}

The fitted data are compiled in the appendix \ref{app_data} while some relevant cases
are presented in Fig. \ref{Fit_rate}, in order to illustrate 
the basic dependencies of the rate capability. The curves at fixed fractional drop
are shown ($F=5\%, 10\%, 15\%, 20\%$), obtained from evaluation of eq. \ref{R_Mat} 
after substitution of the parameters of the fit.

A different attempt to extract dependencies can be performed by comparing at 
fixed gain. This is largely facilitated if the predicted scaling with $m_o$ 
from eq. \ref{R_Mat} is used: 
\beq
\frac{\phi_{_F}(m_o)}{\phi_{_F}(m_o')} = \frac{\ln\left((1-F)m_o\right)-b}{\ln\left((1-F)m_o'\right)-b}
~~\frac{m_o'}{m_o} \label{extrapol}
\eeq
The second term corrects linearly for the different total charge while the first incorporates the linear 
dependence of the drift velocity with the anode potential. 
Different dependencies can be therefore made more 
apparent, as shown in Fig. \ref{Fit_M_mixed}. 
The data sets have been chosen so that a (relatively small) extrapolation by using eq. \ref{extrapol} 
is done for a ratio $(m_o-m_o')/m_o$ smaller than $50\%$. 

\begin{figure}[ht!!!]
\begin{center}
\includegraphics[width = 7.5 cm]{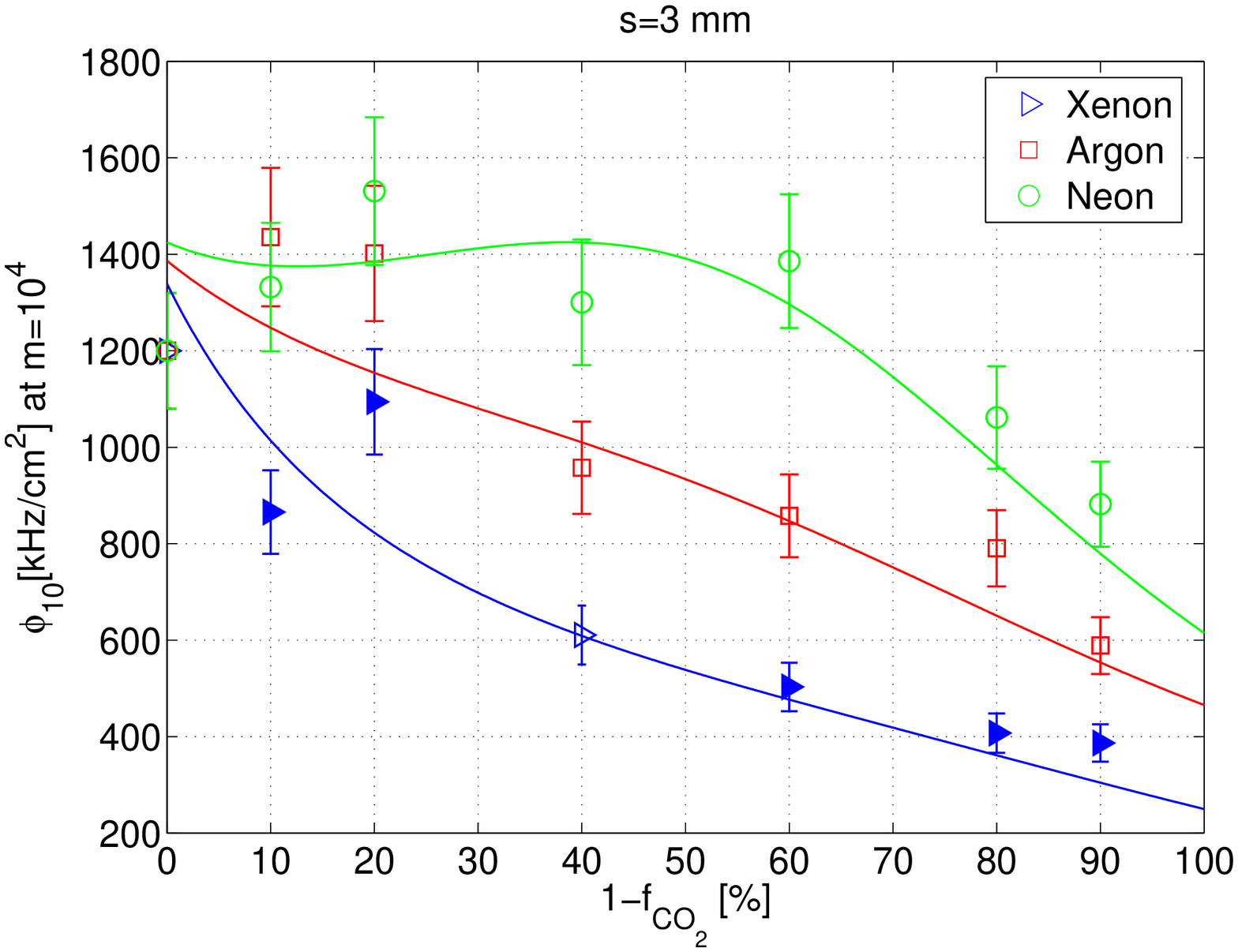}
\includegraphics[width = 7.5 cm]{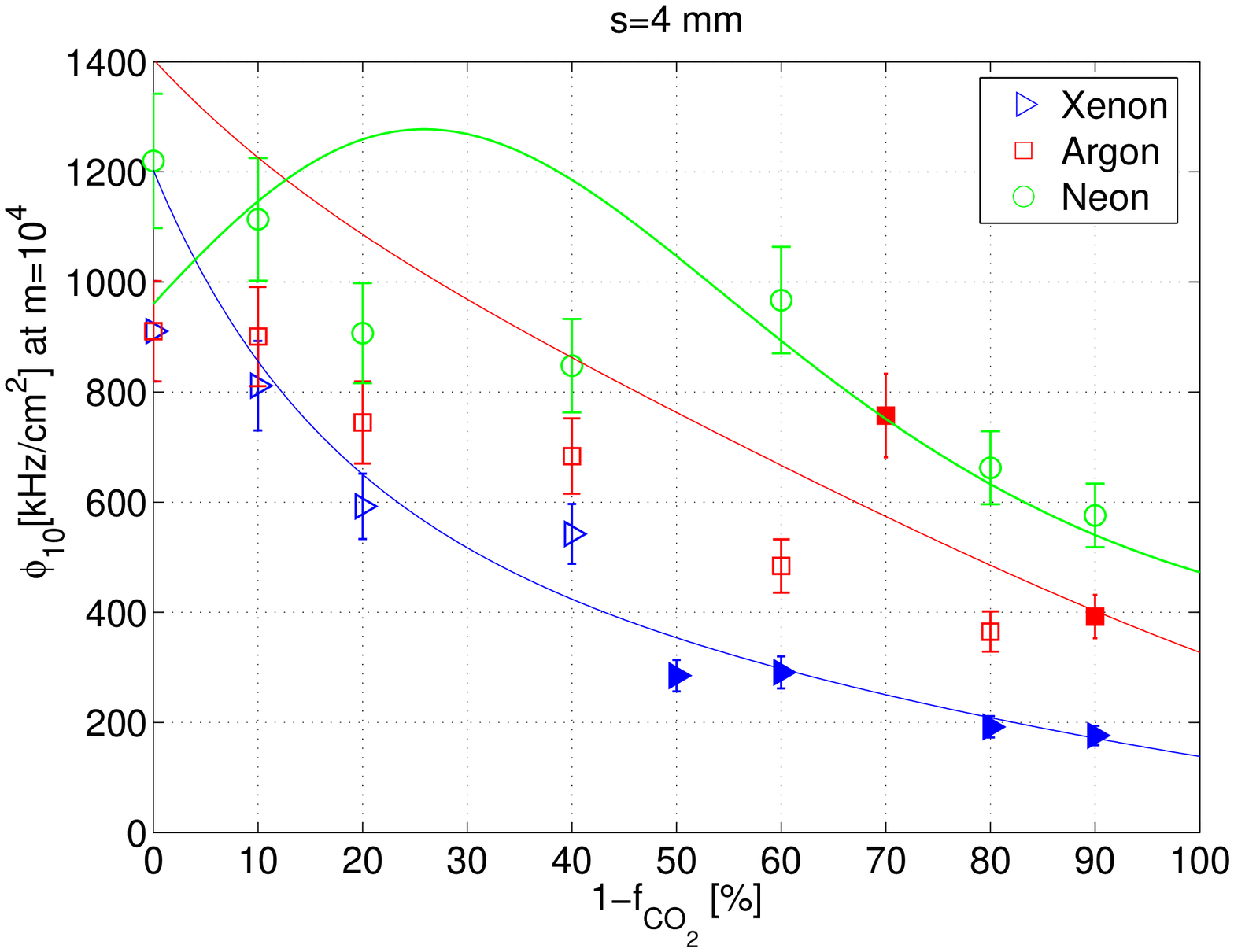}

\includegraphics[width = 7.5 cm]{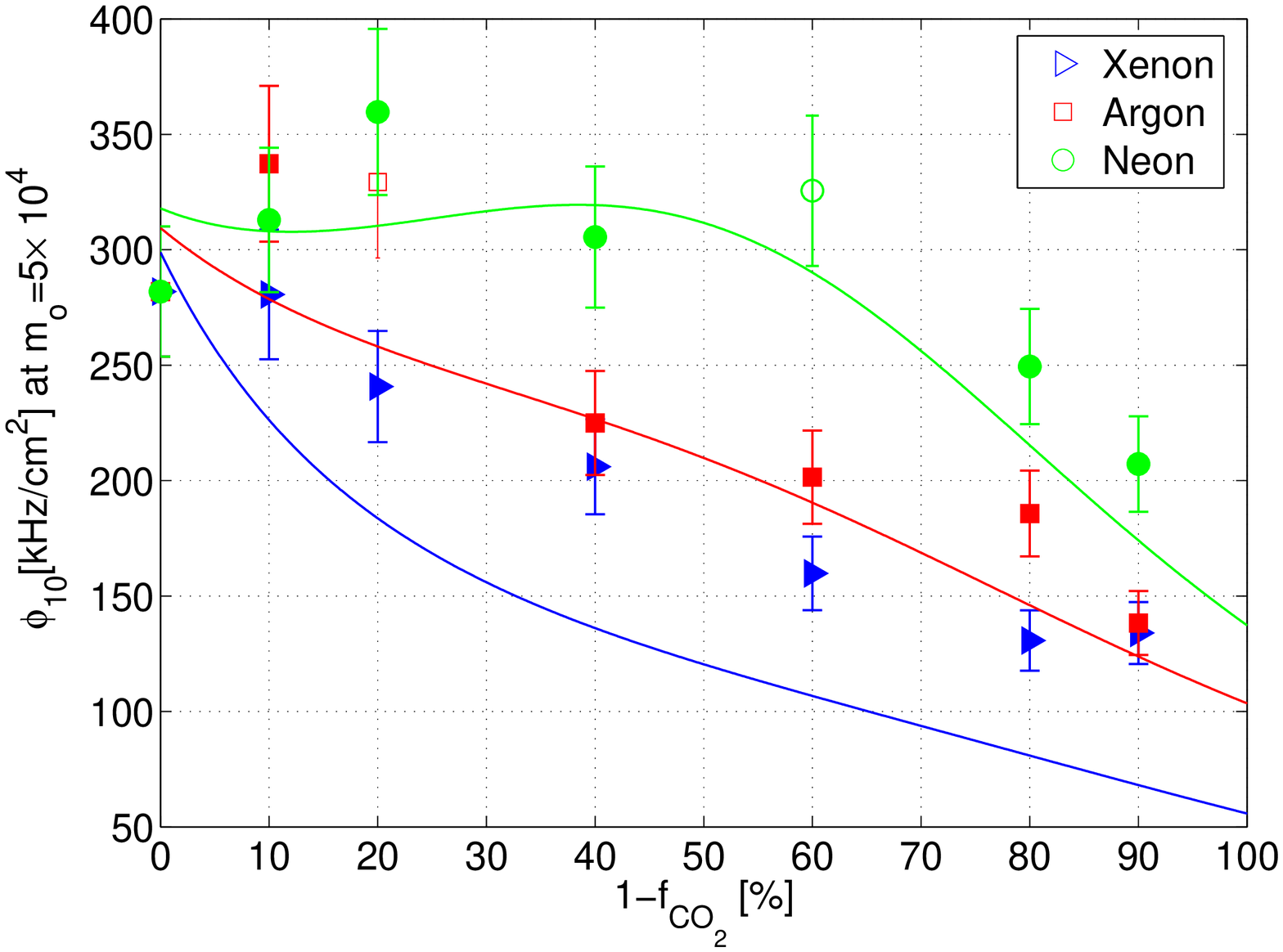}
\includegraphics[width = 7.5 cm]{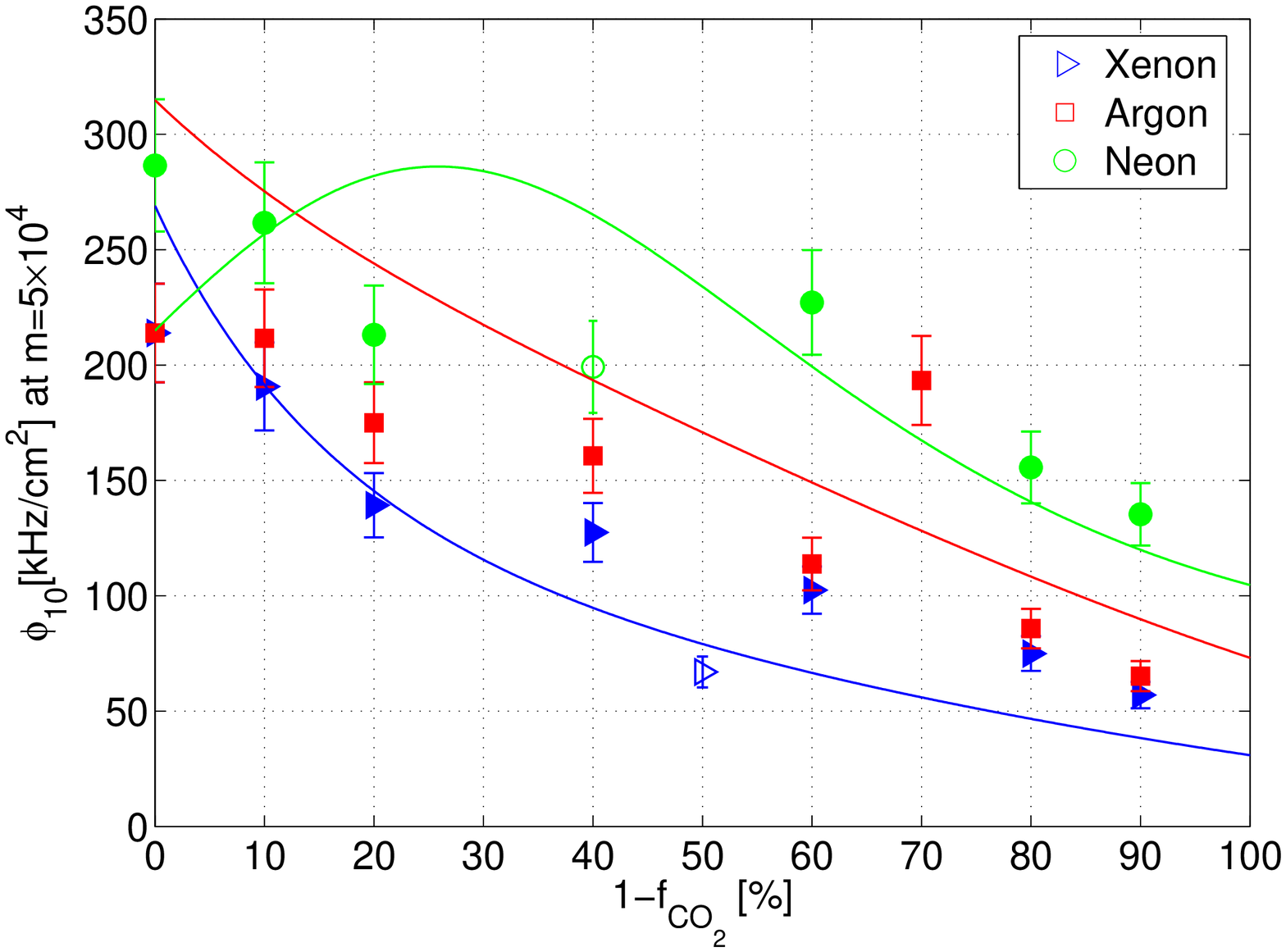}
\caption{\footnotesize Up-row: rate capability at $10\%$ gain drop for 
the $s=3$mm and $s=4$mm pitch chamber at $m_o=10^4$ as a function of
the fraction of noble gas. Data correspond to Xenon (triangles), Argon (squares) 
and Neon (circles) based mixtures. Open symbols indicate that the plotted 
value has been extrapolated over more than $50\%$ difference with respect 
to $m_o$. Lines show the Mathieson model with the mobilities for the drifting 
ions taken from a global fit to 88 data sets. Low-row: as up-row but for $m_o=5\times 10^4$.}
\label{Fit_rate_universal}
\end{center}
\end{figure}
A more systematic approach can be still devised:
the flux for a $10\%$ fractional gain drop from a nominal gain $m_o$ (`$\phi_{10}$ at $m_o$') 
can be experimentally determined by 
extrapolation from the closest value measured at $m_o'$, according to eq. \ref{extrapol}. 
The main uncertainty in the determination of $\phi_{10}$ is the uncertainty in 
the measurement of $m_o$, that has been assumed to be 1\%. By resorting to eq. \ref{error},
an uncertainty of 10\% in $\phi_{10}$ has been estimated.

The resulting $\phi_{10}$ 
points can be plotted as a function of the fraction of
noble gas, as shown in Fig. \ref{Fit_rate_universal} for Xenon (triangles), Argon (squares) and 
Neon (circles) mixtures, for different
 pitches ($s$) and initial gains ($m_o$). 
Full points indicate that the extrapolation has been done 
from a curve measured with a real gain $m_o'$ within less than $50\%$ deviation 
from the extrapolated $m_o$.
The lines represent the fit to the 'Mathieson model with Blanc's formula'. For convenience, 
the dependence of the gain parameters $a$ and $b$ with $f_{CO_2}$ has been parameterized by 
using 3$^{rd}$ order polynomials.

The fitted mobilities used as input for the model are compiled in table \ref{tabla},
with their statistical and systematic uncertainties. The later ($20\%$) arise from 
the uncertainty in the flux determination due to the beam size correction, 
that is reflected linearly in the measured mobilities (eq. \ref{R_Mat}). 
This uncertainty is a correlated one for all the measurements. 
The following features can be observed: i) the values of fitted mobilities in the $s=3$ and $4$ mm
chambers are compatible within $1\sigma$ with exception of  $\mu_{\alpha,Xe}$ and $\mu_{\beta,CO_2}$
($2\sigma$), ii) the mobilities in noble gases differ in a ratio $\simeq 1:3:4$ for Xe:Ar:Ne. 
iii) When the results from the two chambers are combined, ion mobilities in Xe, Ar and Ne differ with
a statistical significance of more than $2\sigma$ while all the mobilities in pure CO$_2$ are 
statistically compatible within $1\sigma$, iv) the values of $\mu_{\alpha,Xe}$ are slightly
higher than expected for Xe$^+$ but slightly smaller than other indirect 
measurements as \cite{Mat3}, v) the mobilities of ion $\beta$ in Ar and CO$_2$ agree well within 
errors with the CO$_2^+$ hypothesis, as has been measured for similar chambers \cite{Charpak}, 
vi) ion $\delta$ shows a mobility much closer  to the value of Ne$^+$ than CO$_2^+$ 
(factor two difference), vii) the average $\chi^2$ is 0.99, providing statistical support
to the approach followed in this analysis. A compilation of these mobilities can be found in \cite{Ellis1}.
 
\begin{table}[htbp]
\begin{center}
\begin{tabular}{|c|c|c|c|c|c|c|c|c|}

\hline
 mobility [mm$^2$V$^{-1}$s$^{-1}$] & $s=3$ mm (fit) & $s=4$ mm (fit)&  mean & $\chi^2$ \\
\hline
$\mu_{\alpha,CO_2}$ & $155 \pm 37 \pm 31$ & $124 \pm 16 \pm 25$ & $129 \pm 15 \pm 26$ & 0.59 \\
\hline
$\mu_{\alpha,Xe}$   & $70 \pm 6 \pm 14$   & $95 \pm 13 \pm 19$  & $74 \pm 5 \pm 15$ & 3 \\
\hline
$\mu_{\beta,CO_2}$  & $173 \pm 18 \pm 35$ & $128 \pm 24 \pm 25$ & $157 \pm 14 \pm 31$ & 2.25 \\
\hline
$\mu_{\beta,Ar}$ & $214 \pm 20 \pm 43$ & $207 \pm 30 \pm 41$ & $212 \pm 17 \pm 42$ & 0.0377 \\
\hline
$\mu_{\delta,CO_2}$ & $126 \pm 23 \pm 25$ & $129 \pm 25 \pm 26$ & $127 \pm 17 \pm 25$ & 0.0081 \\
\hline
$\mu_{\delta,Ne}$ & $298 \pm 21 \pm 60$ & $289 \pm 30 \pm 58$ &  $295 \pm 17 \pm 59$ & 0.0604   \\
\hline

\end{tabular}
\caption{\footnotesize{Ion mobilities from the global fit of the data to the Mathieson model.
The first errors are statistical, the second ones systematic.}}
\label{tabla}
\end{center}
\end{table}

\subsection{Calculations for minimum ionizing particles}
\label{mips}

In the CBM experiment at FAIR, 
most of the particles over the anticipated MWPC-based TRD will be minimum-ionizing particles (mips) \cite{CBM}. 
It is therefore important to estimate the rate
capability in such an environment. A systematic survey of energy loss in Xe-CO$_2$ has been
recently carried out over a broad dynamic range \cite{Anton_data} where a detailed comparison
between data and simulation was presented. A calculation purely based on the Bethe-Bloch formula
overestimates the measured energy loss by some 30\%. Hence, for simplicity, the Bethe-Bloch 
prescription as recommended for mixtures \cite{Seltzer} was used here, 
but with a normalization taken to reproduce 
the value $dE/dx=5$ keV/cm for $p=1$ GeV pions in Xe-CO$_2$(85-15) reported in \cite{Anton_data}. 
When going from pure 
Xenon to pure CO$_2$ such a procedure predicts 
a reduction in energy loss by a factor of 2, due to the smaller density. On the other hand, 
in the above-mentioned reference gas mixture the energy lost by mips is approximately 3 keV in the total
gas thickness $2h=6$ mm, to be compared with our X-ray energy $\bar{E}_{X-ray}= 6.6$ keV. 

\begin{figure}[ht!!!]
\begin{center}
\includegraphics[width = 15 cm]{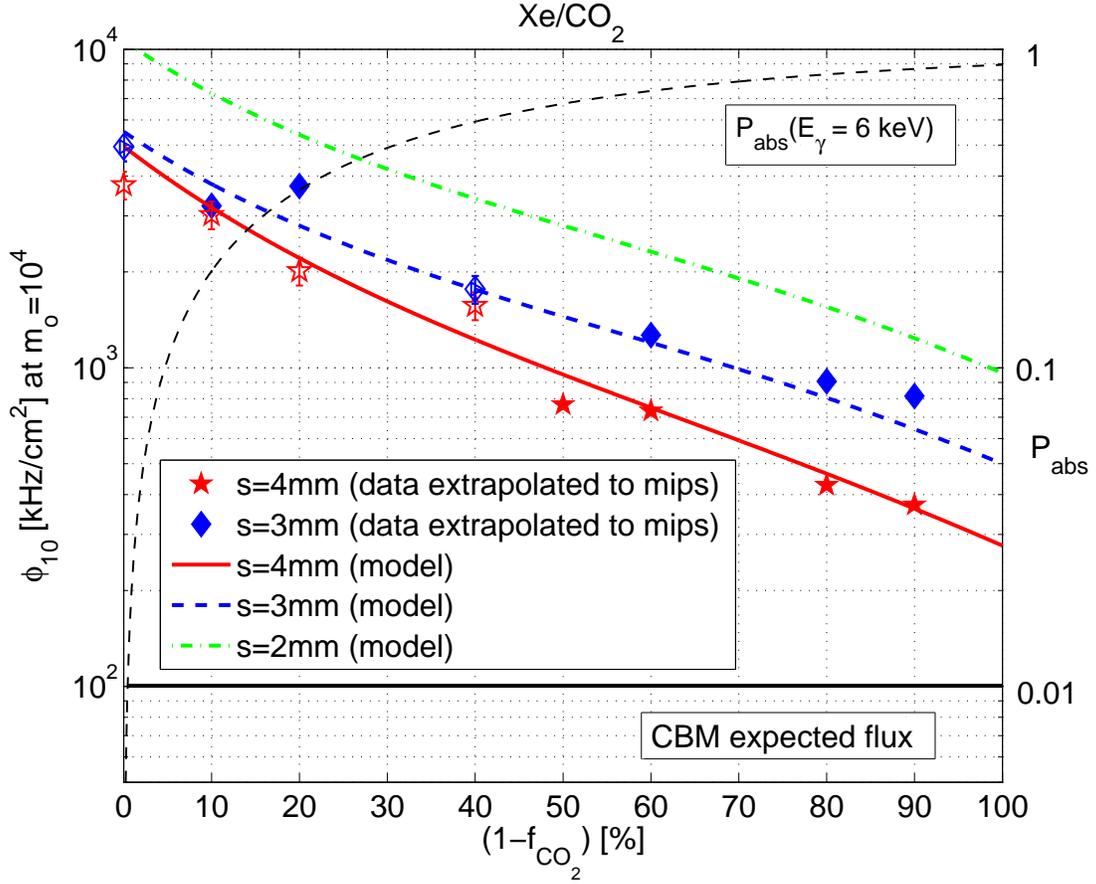}
\caption{\footnotesize Up: rate capability at 10\% gain drop 
for the $s=3$ mm (diamonds) and $s=4$ mm (stars) chambers extrapolated to minimum
ionizing particles. The lines show the Mathieson model without fit. 
The extrapolation to the $s=2$ mm case is also indicated (dot-dashed line),
by using the mobilities of the $s=3$ mm chamber from table in text (the
necessary parameters of the characteristic gain curve are taken from measurements). 
The dotted curves (scale on the right) correspond to the absorption
probability ($P_{abs}$) of X-rays of $E=6$ keV. Open points indicate that
the data has been extrapolated to $m_o$ from a measured value  $m_o'$ different by more than 50\%.}
\label{Mariana}
\end{center}
\end{figure}

The compilation of the measured rate capability in Xenon mixtures for the $s=3$ and $4$ mm pitch chamber, 
 extrapolated to mips
by using the Bethe-Bloch formula obtained as explained above, 
together with the model and an extrapolation to $s=2$ mm is shown in Fig. \ref{Mariana}
for a typical operating gain of $m_o=10^4$ (extrapolation to $s=2$ mm 
indeed requires to use the gain curves measured in such a case, that will be published 
elsewhere).

The improvement of the rate capability 
with decreasing $s$ has two main origins: first, the rate per wire 
is proportionally smaller in such a case; second, the higher wire density implies that higher
fields are required in order to get similar gains, meaning that the coefficient $a$ becomes smaller 
(eq. \ref{gain_curve_eq}). 
The fact that the gain curve is 'softer' in the sense of $a$ being smaller makes the 
chamber more resistant to the space-charge of the ions ($\phi_{F} \simeq 1/a^2$). 
The capacitance per unit length $C_L$ has little impact for normal values of $s$.

On the other hand, the behavior of the rate capability with increasing fraction of CO$_2$ has four
main dependencies: i) The energy required to create a pair in CO$_2$
is higher than in noble gases (exception is Neon, where both are comparable). Still, this 
fact can account for a 30\% increase in Xe-CO$_2$ mixtures due to the lower initial number of 
electron-ion pairs when increasing the concentration of CO$_2$. ii) CO$_2$ is less dense than the 
mixture by a factor amounting by maximum 2 in Xenon-based mixtures, increasing the rate capability
accordingly when going from pure Xe to pure CO$_2$. iii) The curves become 'softer' when the 
fraction of CO$_2$ is increased, resulting in an increase by a factor 2-3 from pure Xe to pure CO$_2$. 
iv) The mobility of the ions can vary also sizably between 
extreme cases: in case of Ne this implies a reduction of up to a factor 3 
when increasing $f_{CO_2}$ while in Xe it increases by a modest $50\%$ when going from pure Xe to pure CO$_2$.
These effects compete in some cases, as in Neon, where the increase in rate capability due to
the much increased mobilities at high Neon concentration is compensated by a much harder gain curve.
This yields a largely flat dependence of the rate capability for X-ray photons 
as a function of CO$_2$ concentration (Fig.\ref{Fit_rate_universal}). In Xe mixtures all effects i-iv 
go in the same direction, and up to one order of magnitude can be gained in rate capability
for minimum ionizing particles, between pure CO$_2$ and pure Xe, as shown in Fig. \ref{Mariana}-up.

The present model has been used to describe recent mips data from M. Petris et al. \cite{Petris},
showing a good agreement. Nevertheless, we believe that the very small deterioration
of the performances observed by the authors
does not add extra support to the message conveyed in the present work.
\section{Discussion}
\label{discussion}

Considerable attention has been devoted in this work to the `beam-size' correction. Nevertheless, if the model 
assumption that `the correction factor $\bar{d}_m$ depends only
on the chamber/beam arrangement' would hold 
true one could eventually resort, in general, to 
the well measured CO$_2^+$ mobility in CO$_2$ to estimate 
the factor $\bar{d}_m$ by imposing agreement with the Mathieson formula \ref{M_Mat} in the limit of pure
CO$_2$ (see \cite{Urquijo}). Instead, if one would resort to the theoretical model to
calculate it, a correction factor that would include i) an accurate knowledge of the 
beam intensity profile and ii) the relative beam/chamber alignment
 would be needed in order to obtain a precise (yet model-dependent) correction. Using 
the known CO$_2^+$ mobility in CO$_2$ as a reference 
or, alternatively, taking one single measurement with defocused beam should 
be probably more sound than attempting a detailed calculation.
Overall, we consider the results of table \ref{tabla} as a strong indication that 
reliable predictions can be done for the rate capability of MWPCs once the proper geometric 
corrections are performed and the nature of the drifting ion is known (and its mobility has been measured
before). Within the precision of the present approach, Blanc's law works well.
There is, nevertheless, a tendency of the Argon and Xenon mixtures to behave similarly at 
high gains ($m_o=5 \times 10^4$), a fact that cannot be accomodated in the model.

With all the above information at hand, we are in position for discussing the more convenient
MWPC geometry for the envisaged application in CBM. 
Due to the nature of TR, hadron identification/suppression worsens at low momenta
where the particle deviates from the minimum ionizing regime. Additionally, the nature
of the CBM spectrometer makes the detection of high energy electrons specially relevant in the context
of $J/\Psi$ and $\Psi'$ detection, while the low momenta regime is well covered by the anticipated
RICH detector and the TOF wall. Based on that we take as a working number the average TR energy released
 by a $p = 2$ GeV electron in a typical CBM-radiator, that is approximately 
$\bar{E}_{_{TR}}\simeq 6$ keV \cite{CBM}. This roughly corresponds to the minimum momentum
for which the CBM TR system will provide sufficient $e/\pi$ suppression, and
that we consider as our benchmark here. From the tabulated X-ray absorption in Xenon a mean free path of
$\lambda_{\gamma}(E= 6 $keV$)=2.7$ mm can be readily obtained, being the probability to absorb the
photon in the gas given by $P_{abs}= 1 - \exp(-2h/\lambda)$. High absorption probability must be balanced
with a moderate rate capability, that has a steep dependence with the gap size (eq. \ref{R_Mat}).

\begin{figure}[ht!!!]
\begin{center}
\includegraphics[width = 15 cm]{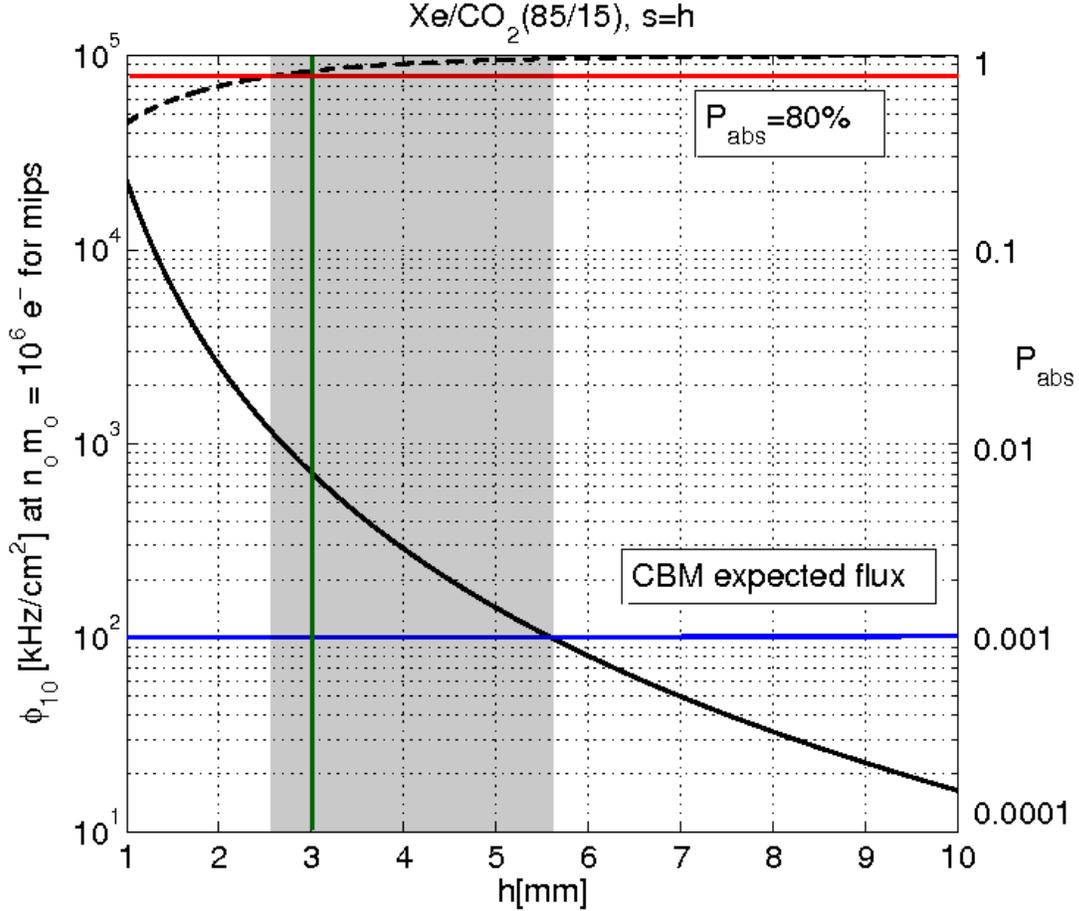}
\caption{\footnotesize Extrapolated rate capability of a MWPC operated in a Xe-CO$_2$(85-15) mixture
as a function of the chamber gap $h$ for mips (axis on left) when the 
average signal amplitude in number of electrons is fixed to 10$^6$ e$^-$ and $s=h$.
The absorption probability of TR photons for the CBM benchmark case is shown with axis on the right
(dashed line). 
The shadowed area represents a comfortable situation in terms of rate capability
and photo-absorption probability. Mobilities have been taken from the weighted average of 
the mobilities in the $s=3$ mm and $s=4$ mm chambers, 
as obtained from the global fit to the Mathieson model.}
\label{compromise}
\end{center}
\end{figure}

 A meaningful determination of the rate capability as a function of $h$ can 
be accomplished by making some further assumptions: first, in order to avoid 
potentially harming operating voltages, the ratio $s/h$ is kept constant. 
Second, the operating conditions are assumed to be such that the 
average number of electrons after multiplication is constant and equal to $n_e\simeq 10^6$e$^-$
for mips (changes in $h$ lead to changes in the initial ionization, that must/can be compensated 
by re-adjusting the gain, so that the induced signal is roughly the same). It is
also possible to directly obtain the rate capability dependence from a direct evaluation of eq. \ref{R_Mat}, 
when recalling the proportionallity of $n_o$ with $h$, but at the price of a reduced signal for 
small $h$. In either one case or the other the implicit assumption that the gain curve is not strongly 
depending on $h$ must be made, and only then an actual $1/h^3$ behavior 
(slightly modified by $C_L$) can be obtained for the rate capability. 

Fig. \ref{compromise} shows
the expected behavior of the rate capability for MWPCs operated in Xe-CO$_2$(85-15)
as a function of the anode-cathode gap $h$ under the former assumptions, 
together with the absorption probability for our benchmark case
in CBM ($\bar{E}_{_{TR}}\simeq 6$ keV). Indeed, a very narrow set of values
can be considered as satisfactory ($h=[2.5-5.5$ mm$]$), with the minimum
value determined by the condition $P_{abs}>80\%$ and the maximum by $\phi_{10}>100$ kHz/cm.

Being the gap dependence so critical for MWPCs in terms of photo-absorption probability, it is 
worth discussing the possibility of using a drift region (of length $D$). For typical pad 
sizes of $1$ cm$^2$ \cite{melanie}, the maximum CBM fluxes would 
imply a particle load of $1/10~\mu$s per channel causing a 10\% pile-up probability within 1 $\mu$s. Taking 
the measured drift times in \cite{AliceTRD}, even in a small drift region corresponding to $D=2h$, the 
electron drift would extend up to a typical time of that order \cite{AliceTRD}, making accurate charge 
measurements very difficult.

On the basis of the above arguments, staying at a MWPC configuration seems to be the most convenient
solution for the CBM TR-detector. 
The possibility of using a mirrored configuration \cite{melanie} with 3 cathode planes deserve
further consideration, since it increases the X-ray detection probability while keeping the rate capability.
If this improvement is worth the extra mechanical complexity or not, will be clarified when the 
first real size prototypes are built.

\section{Conclusions}

We conducted systematic measurements of thin MWPCs 
($h=3$ mm, $s=3,4$ mm) 
filled with Xenon-CO$_2$, Argon-CO$_2$ and Neon-CO$_2$ mixtures under different fractions 
of quencher and at different gains. Employing the Mathieson model 
(including finite beam corrections) and the additive Blanc's law for ion mobilities the 
measured rate capability can be well described theoretically.

When aiming at efficient TR detection at high rates there is a narrow parameter-space
where MWPCs are advantageous. The reason for that is the steep dependence of the rate capability 
with the chamber gap $h$ as $\simeq 1/h^3$ opposed to the exponential behavior of the X-ray absorption
probability  $P_{abs} = 1 - \exp(-2h/\lambda)$. 
We found a good compromise in the value $h=3$ mm where we can comfortably hold 
rates above 100 kHz/cm$^2$ as demanded by the TRD of the CBM experiment, while keeping a typical 
X-ray absorption probability above 80\%.

\acknowledgments

DGD was supported by EU/FP6 contract 515876. 
The authors want to thank Christian Schmidt for discussions and kind help. Special
thanks to M. Ciobanu for his valuable support with analog electronics and to 
J. M. Saa for his help with the bibliography on noble gases properties.
We benefited from the collaboration of A. Battiato, G. Hamar, E. L. Gkougkousis and S. Kohl.


\appendix

\section{Results of the fit}
\label{app_data}
A compilation of data for 36 mixtures (from a total of 44), fitted according to the procedure 
introduced in section \ref{results}, is presented here. The 8 missing plots are not shown 
(but included in the fit) for easier graphical representation. Data for each pitch and 
noble gas mixture (for all measured concentrations of the two gases) 
have been fit with 2 free parameters, namely the
mobilities of the drifting ion in the pure gases. Due to the main scope of this work, data
from Xenon at different `low rate' gains $m_o$ are more abundant than from Argon and, specially,
Neon. For the later, only 1-2 curves are taken per mixture, a choice that relies on the trivial 
scaling of $m_o$ predicted by the Mathieson model (eq. \ref{extrapol}). Each mixture enters
in the fit with the same weight, meaning that the residuals for each curve are divided
by the number of curves per mixture (for instance, the residuals in Xe-CO$_2$(90-10) are divided
by 6). The residuals are additionally divided by the measured value, in order that all data
sets enter with the same weight for a particular mixture.

\newpage

\begin{figure}[ht!!!]
\begin{center}
\includegraphics[width = 7.5 cm]{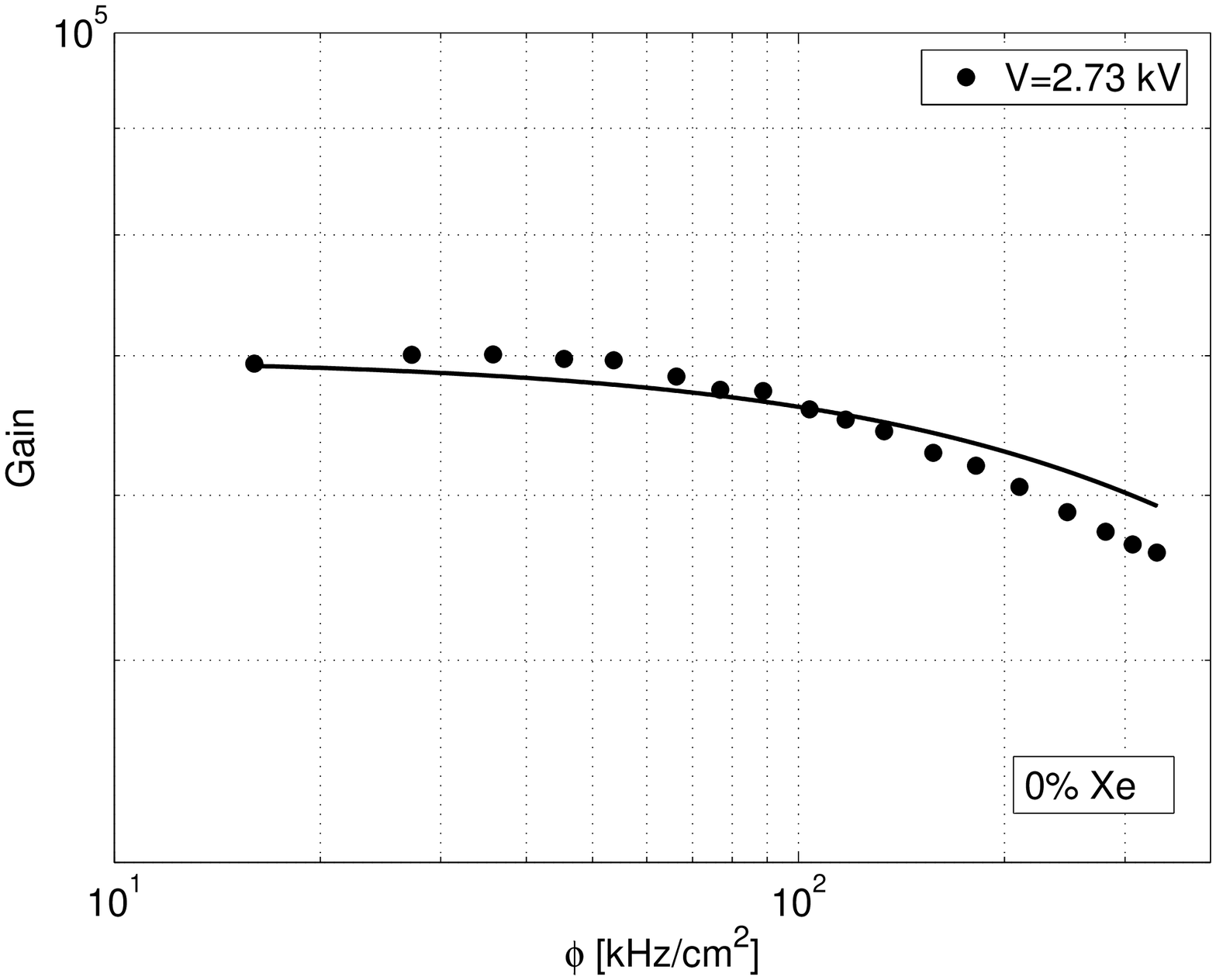}
\includegraphics[width = 7.5 cm]{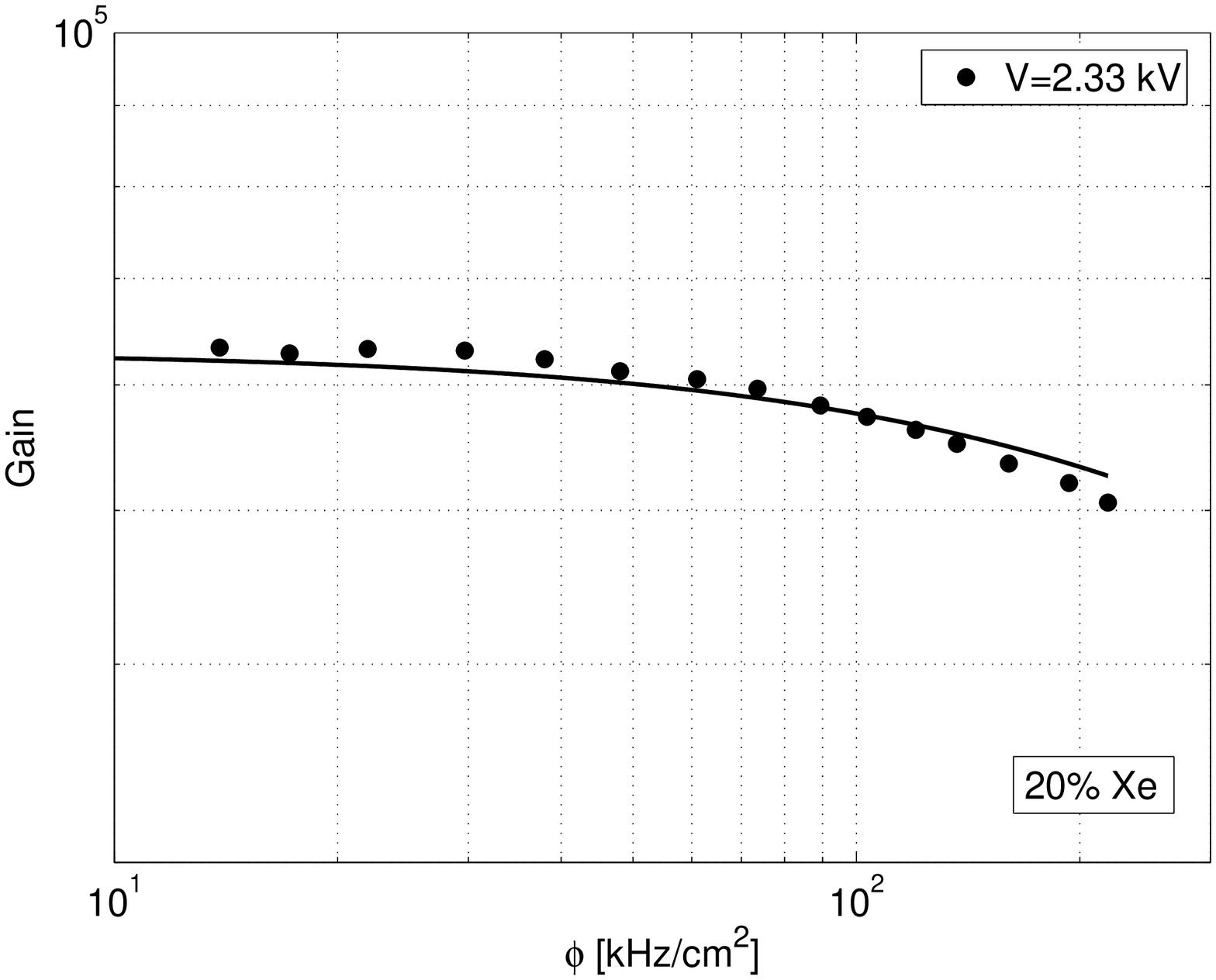}

\includegraphics[width = 7.5 cm]{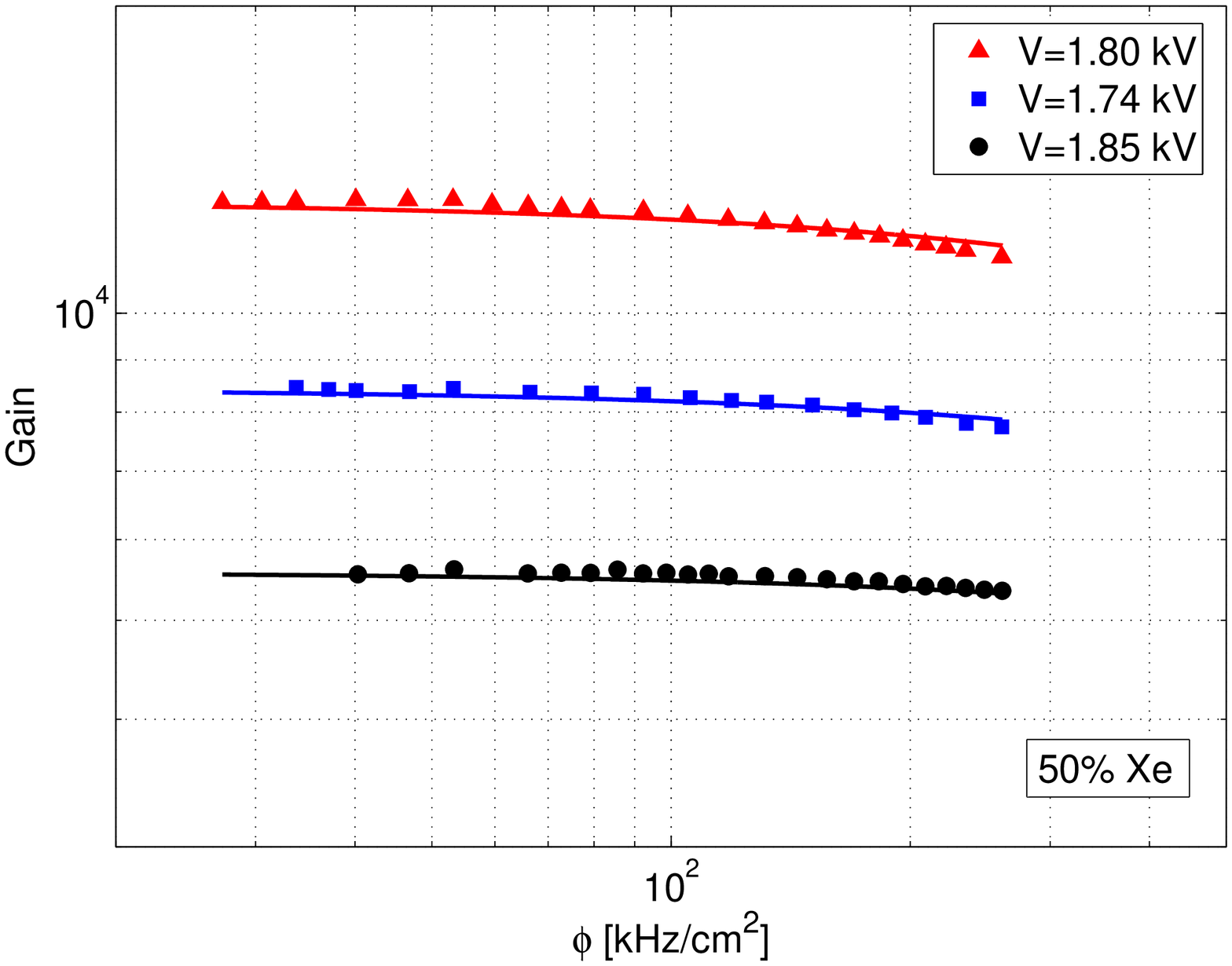}
\includegraphics[width = 7.5 cm]{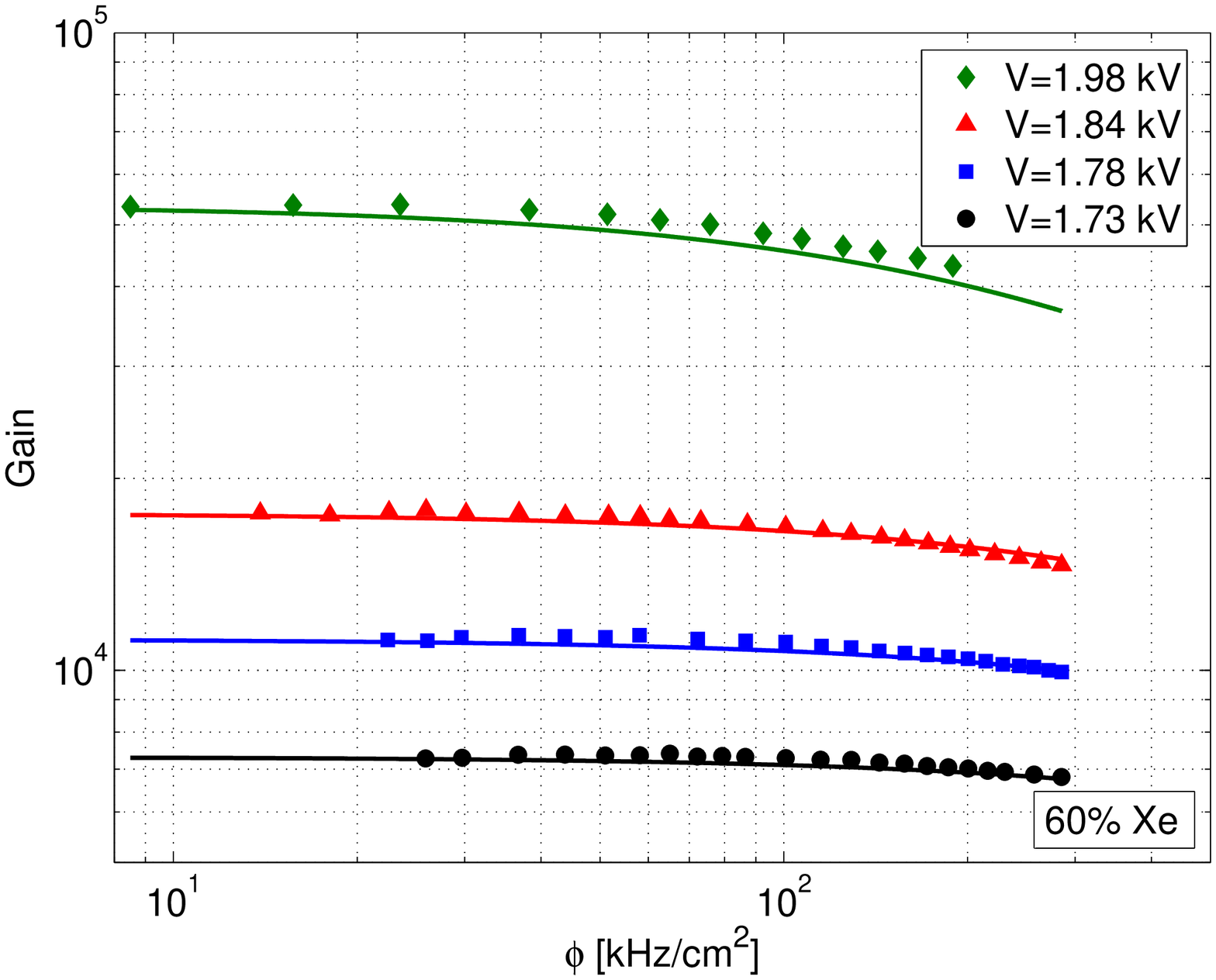}

\includegraphics[width = 7.5 cm]{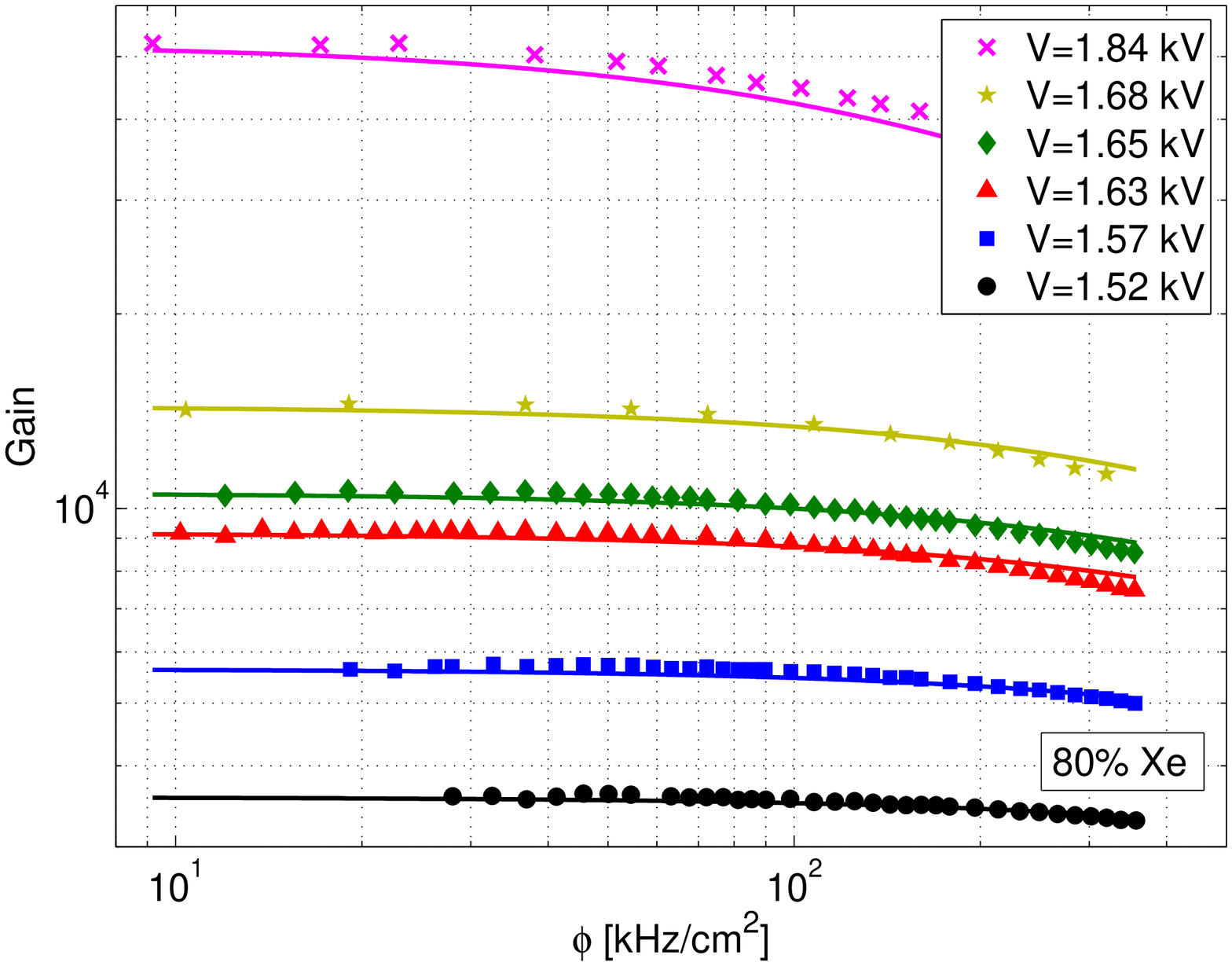}
\includegraphics[width = 7.5 cm]{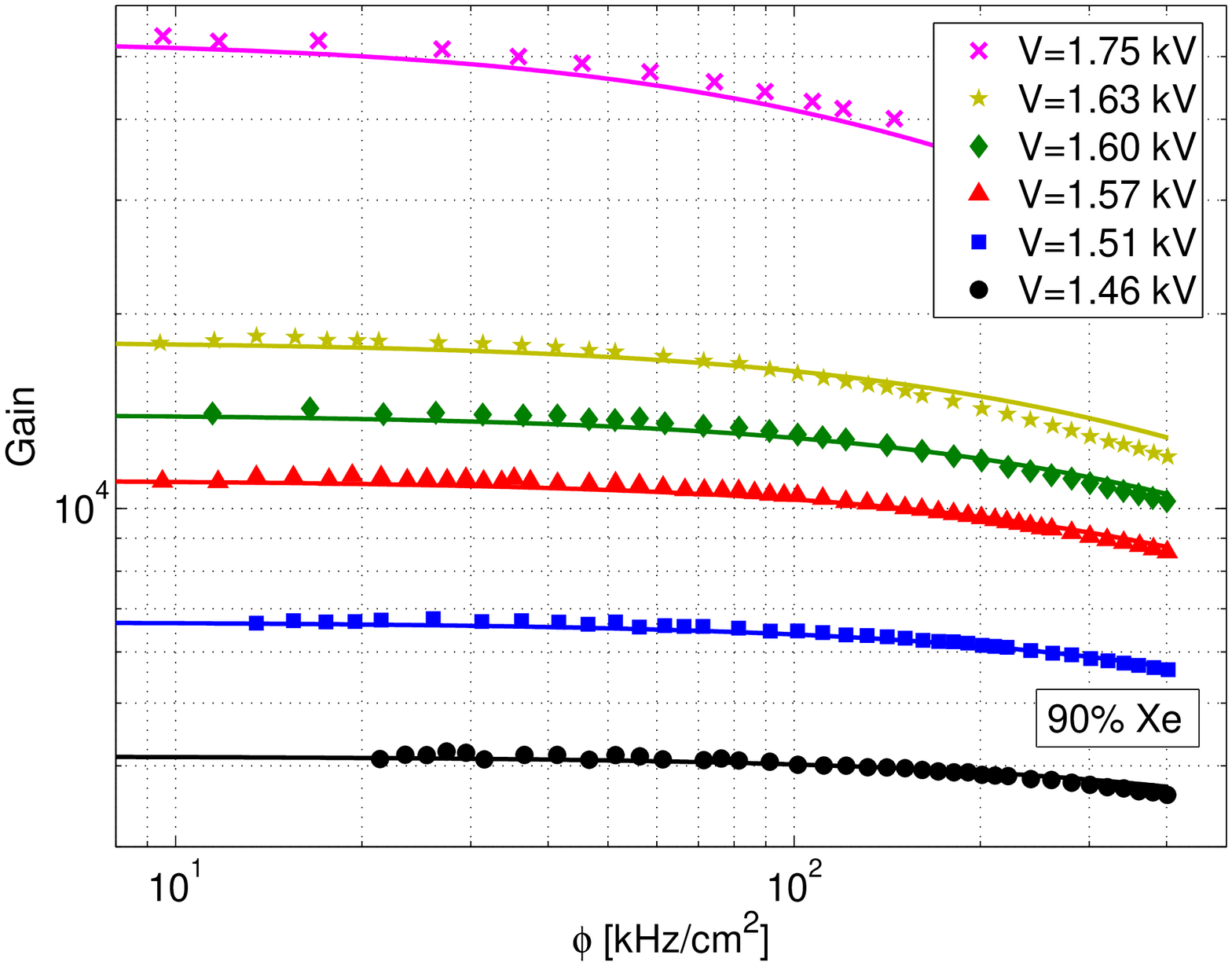}
\caption{\footnotesize Up-left: 2-parameter fit of the rate curves for Xe-CO$_2$ mixtures obtained 
with the $s=4$ mm chamber.}
\label{Xe_4}
\end{center}
\end{figure}

\newpage

\begin{figure}[ht!!!]
\begin{center}
\includegraphics[width = 7.5 cm]{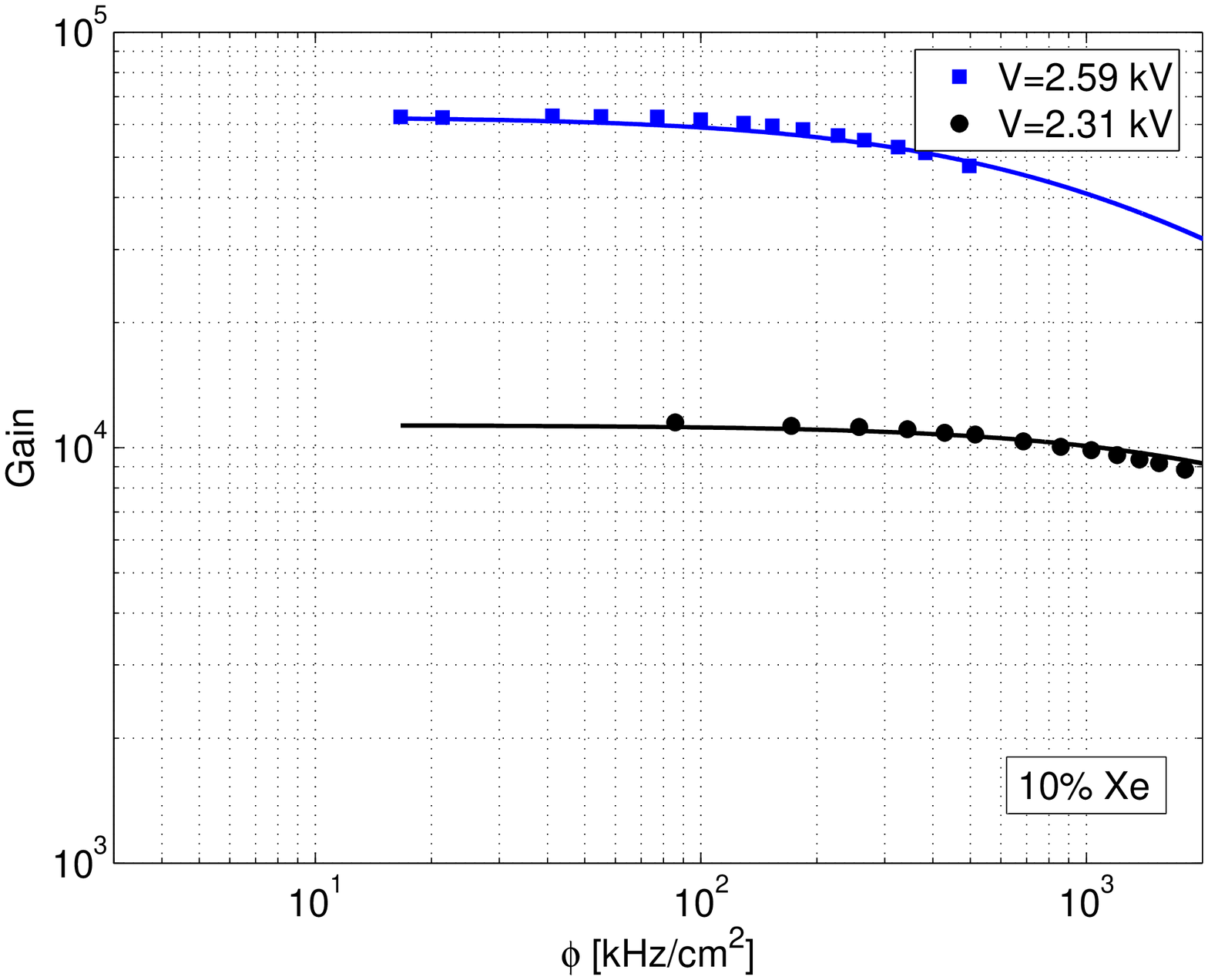}
\includegraphics[width = 7.5 cm]{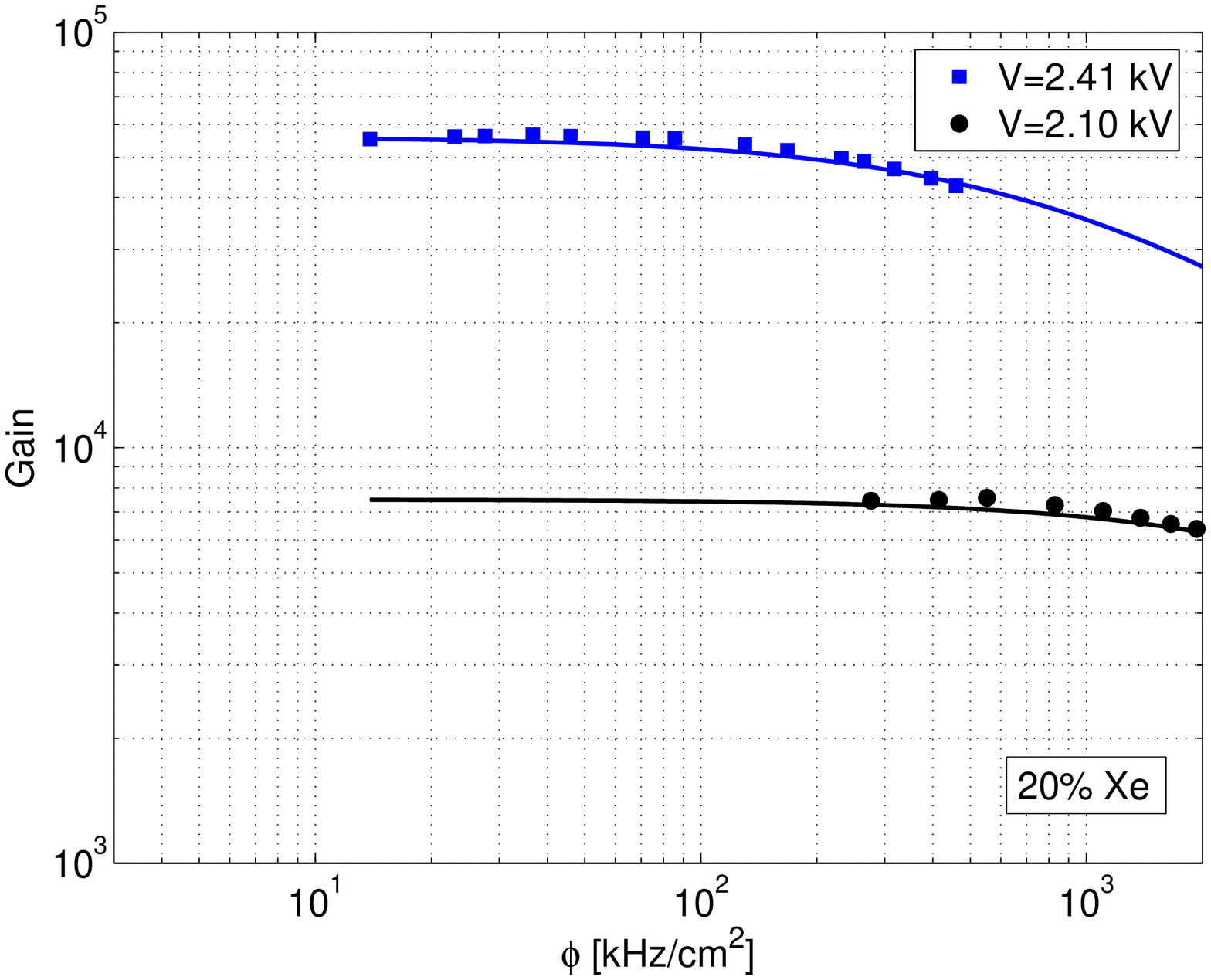}

\includegraphics[width = 7.5 cm]{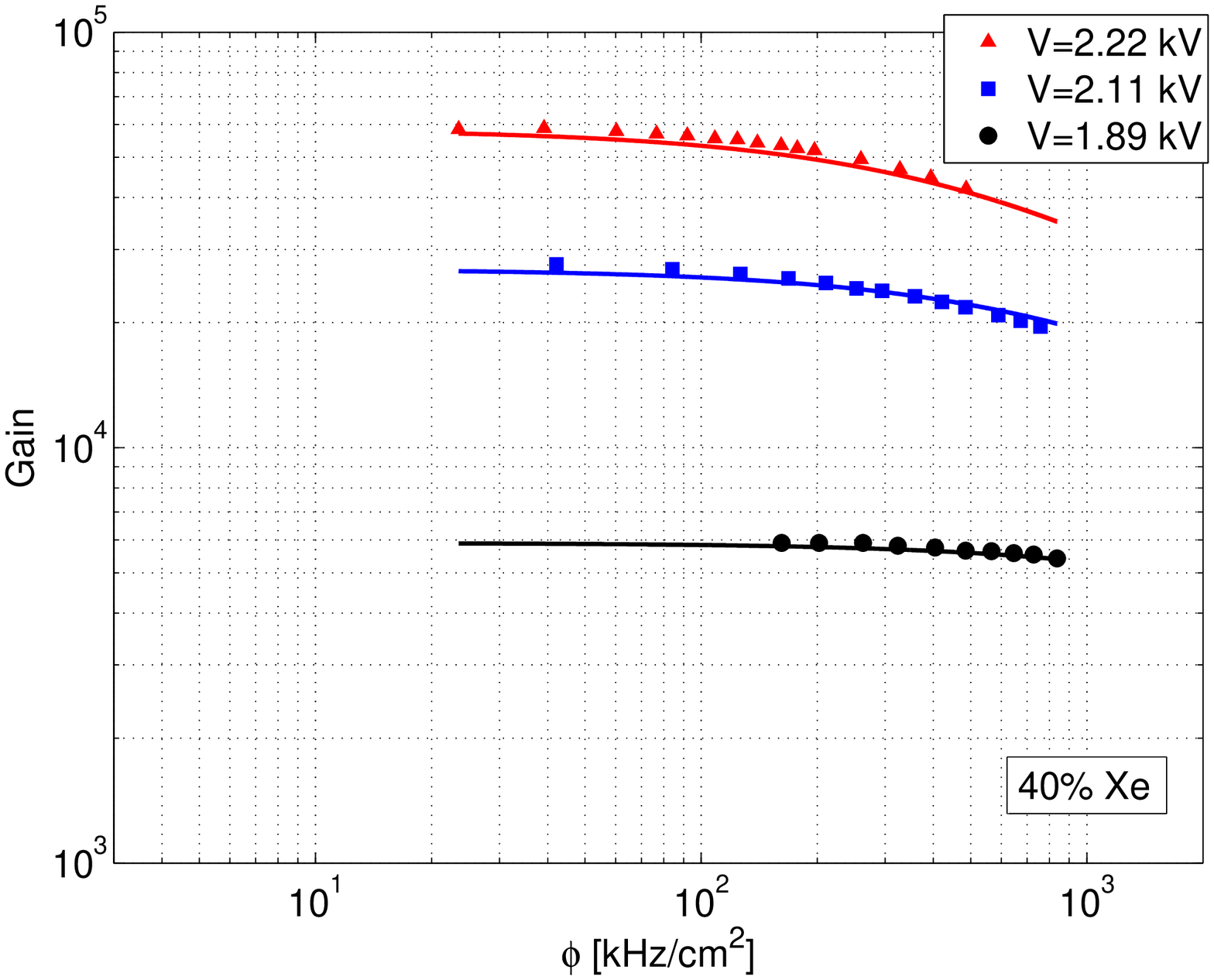}
\includegraphics[width = 7.5 cm]{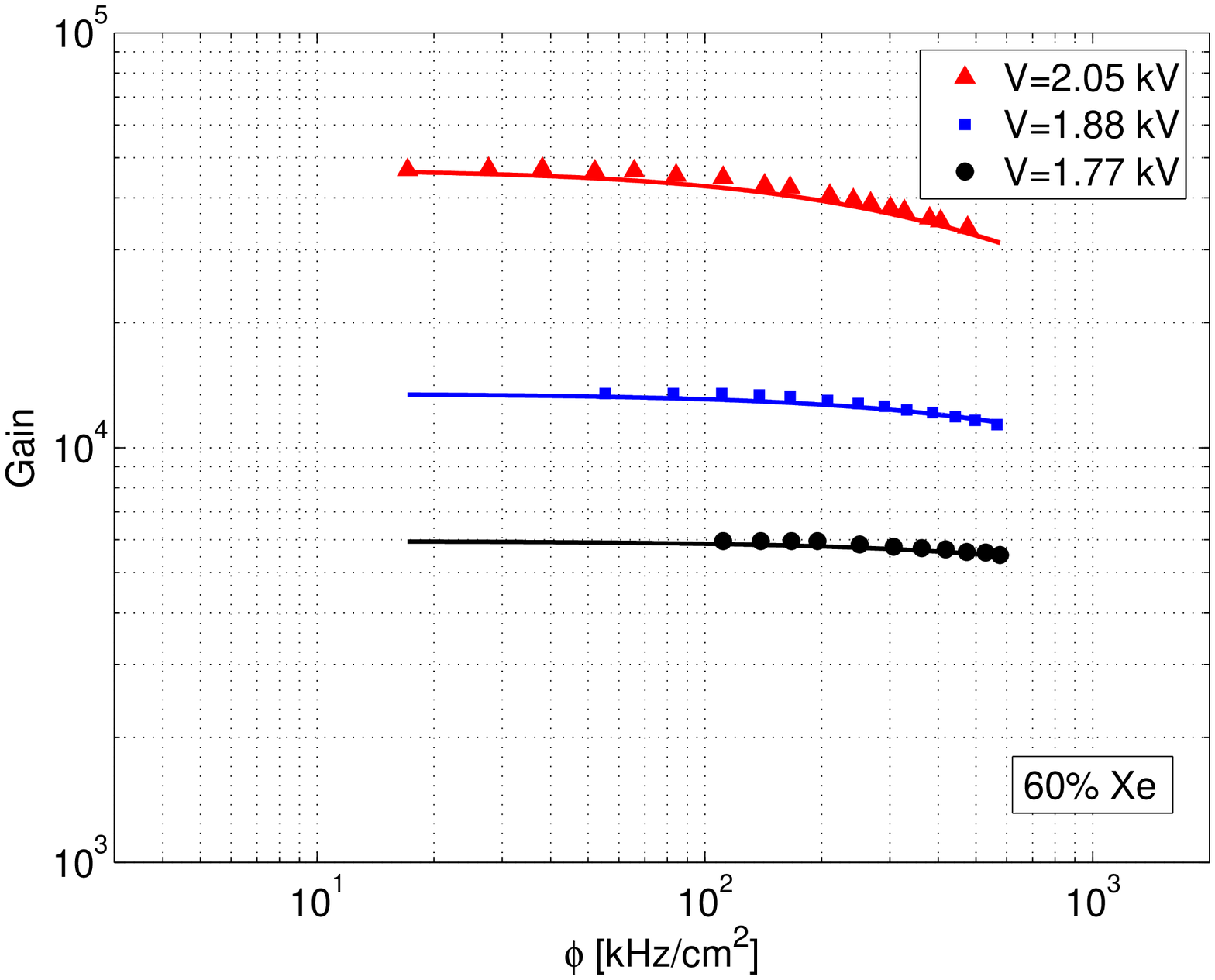}

\includegraphics[width = 7.5 cm]{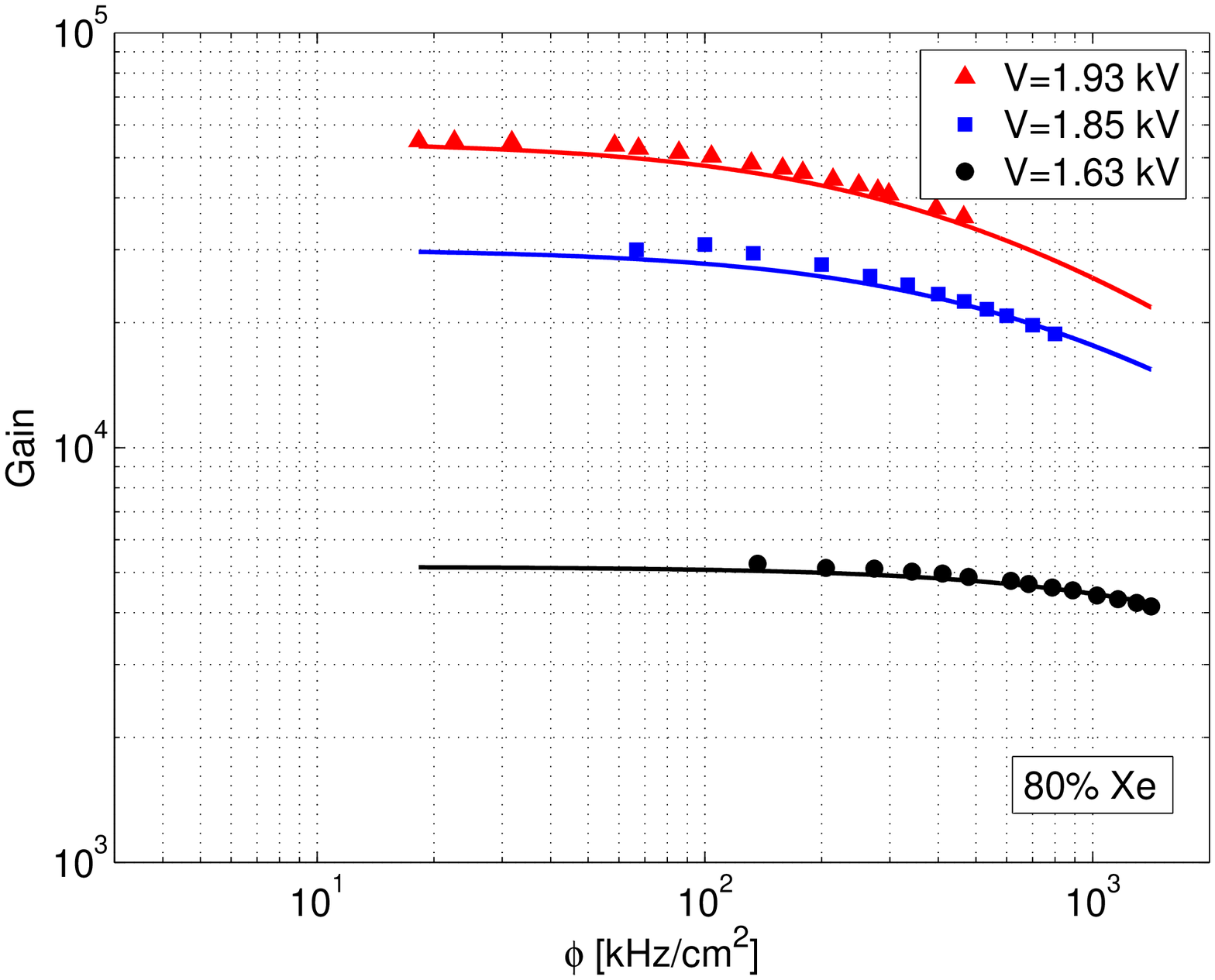}
\includegraphics[width = 7.5 cm]{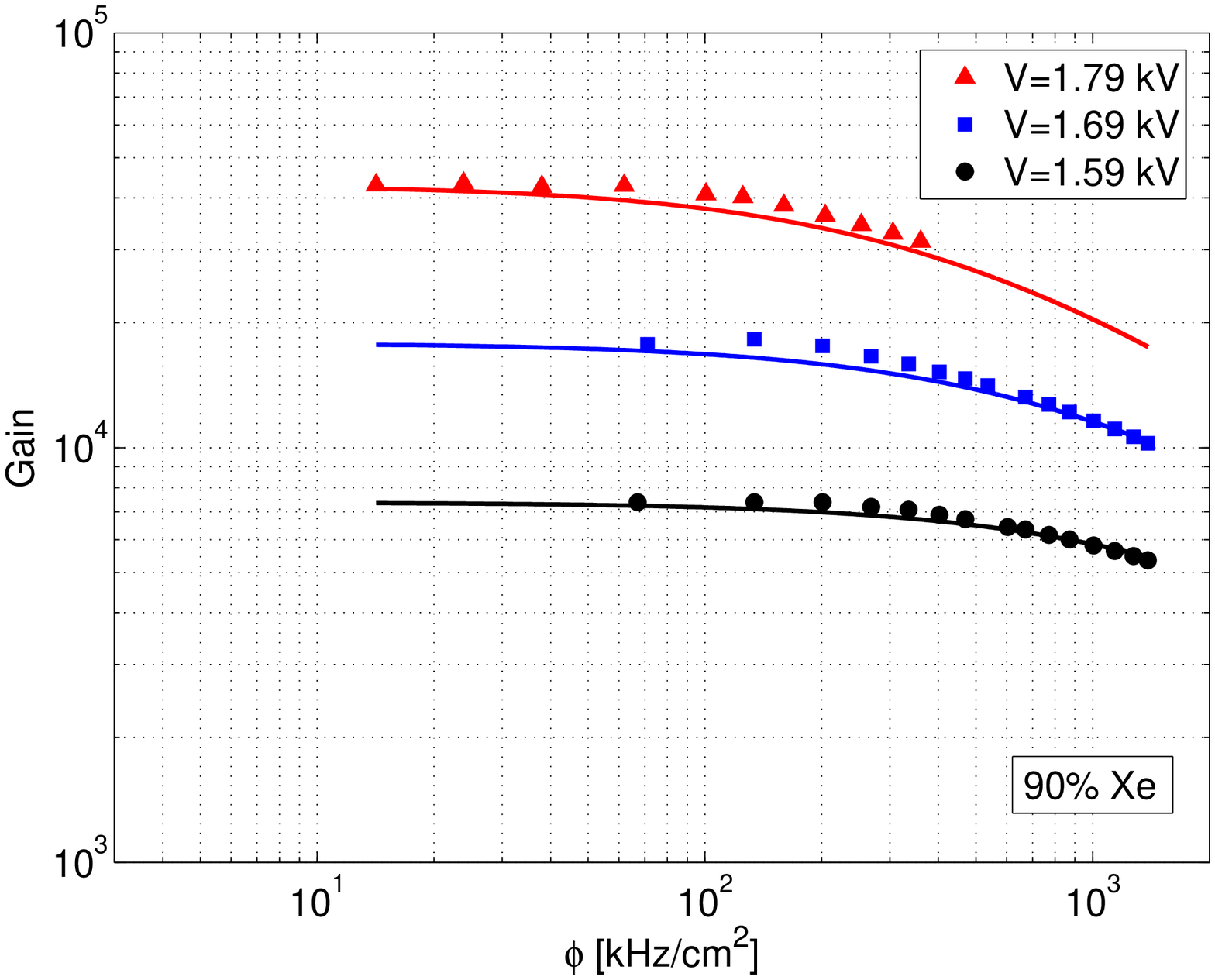}
\caption{\footnotesize Up-left: 2-parameter fit of the rate curves for Xe-CO$_2$ mixtures obtained with the $s=3$ mm chamber.}
\label{Xe_3}
\end{center}
\end{figure}

\newpage

\begin{figure}[ht!!!]
\begin{center}
\includegraphics[width = 7.5 cm]{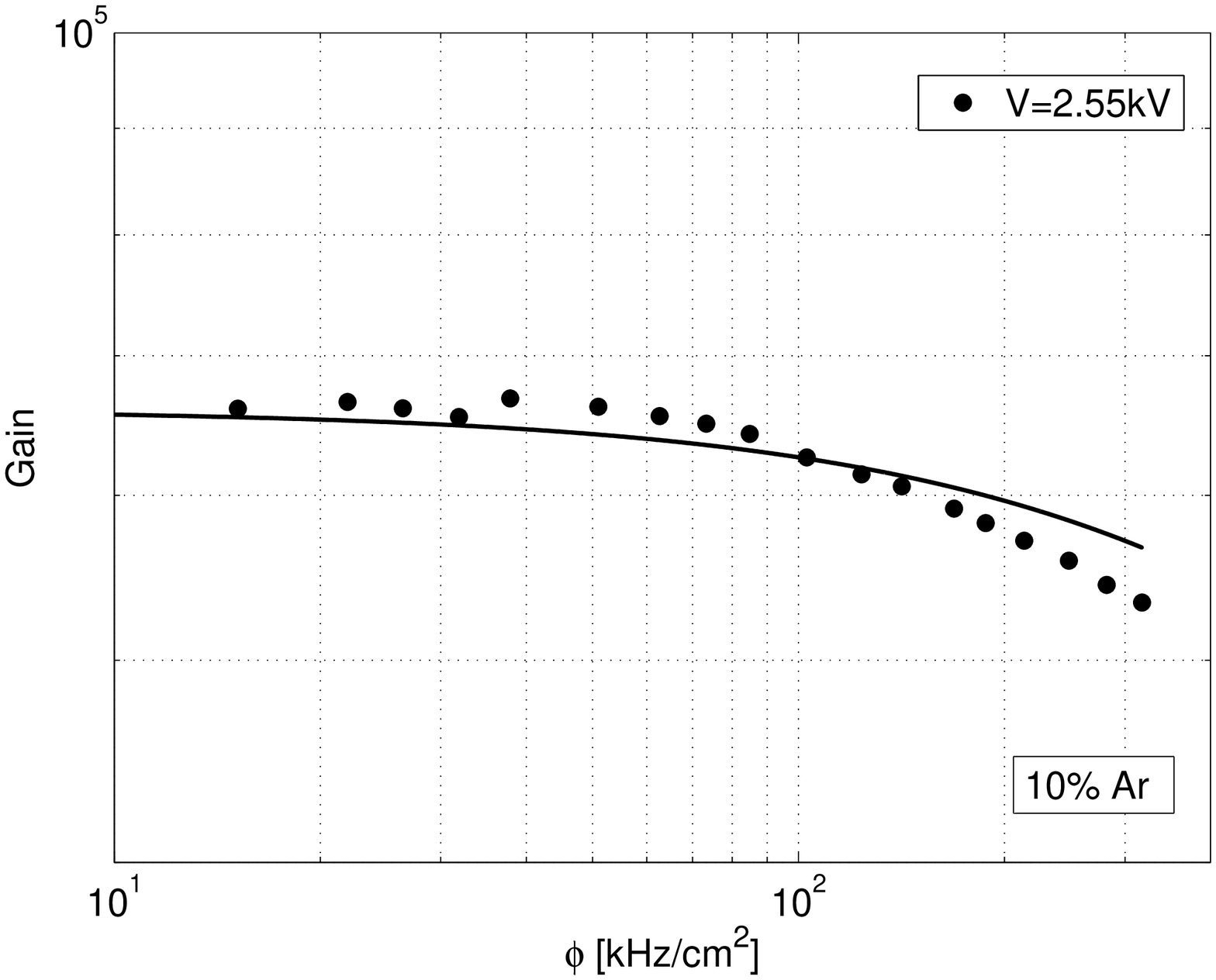}
\includegraphics[width = 7.5 cm]{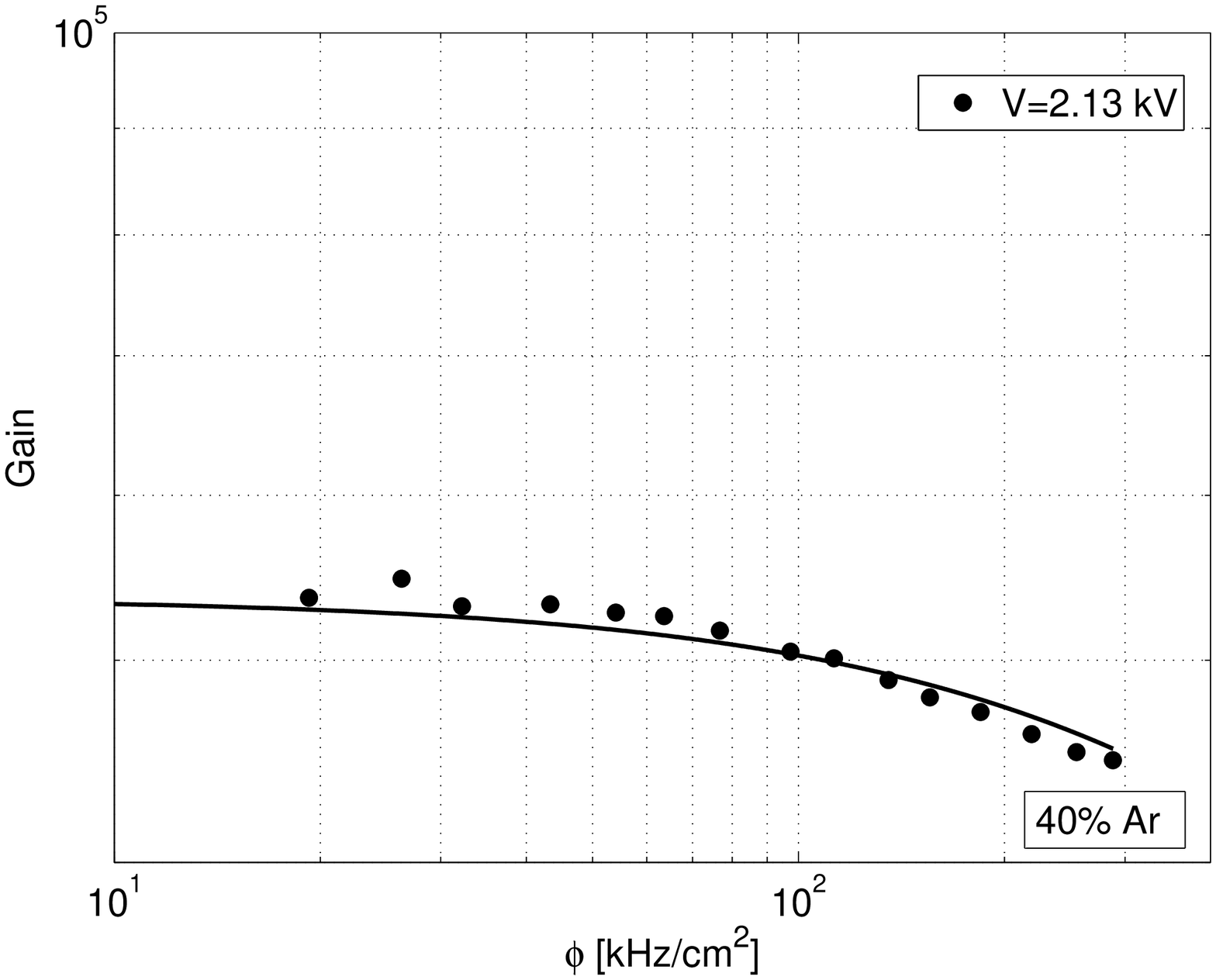}

\includegraphics[width = 7.5 cm]{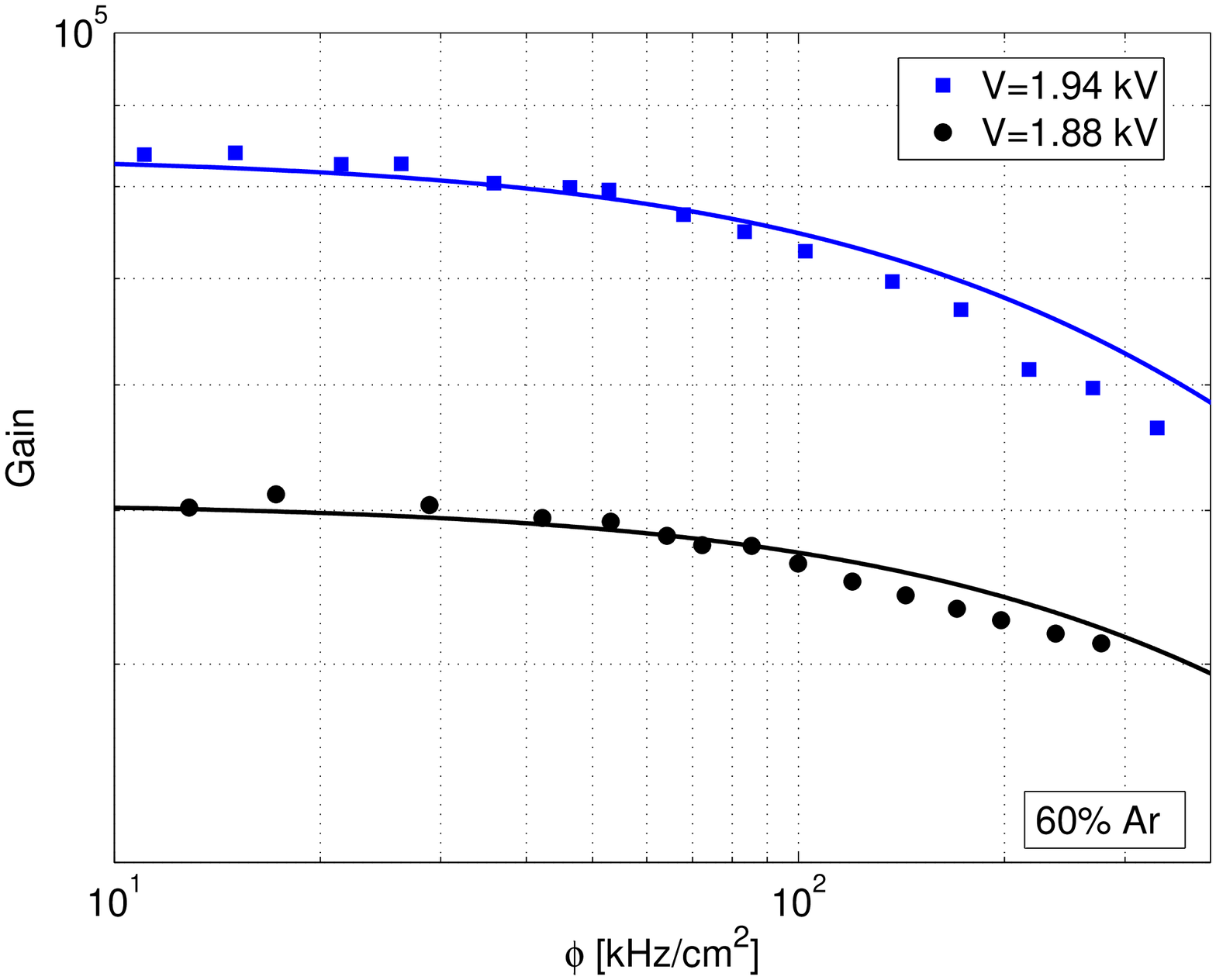}
\includegraphics[width = 7.5 cm]{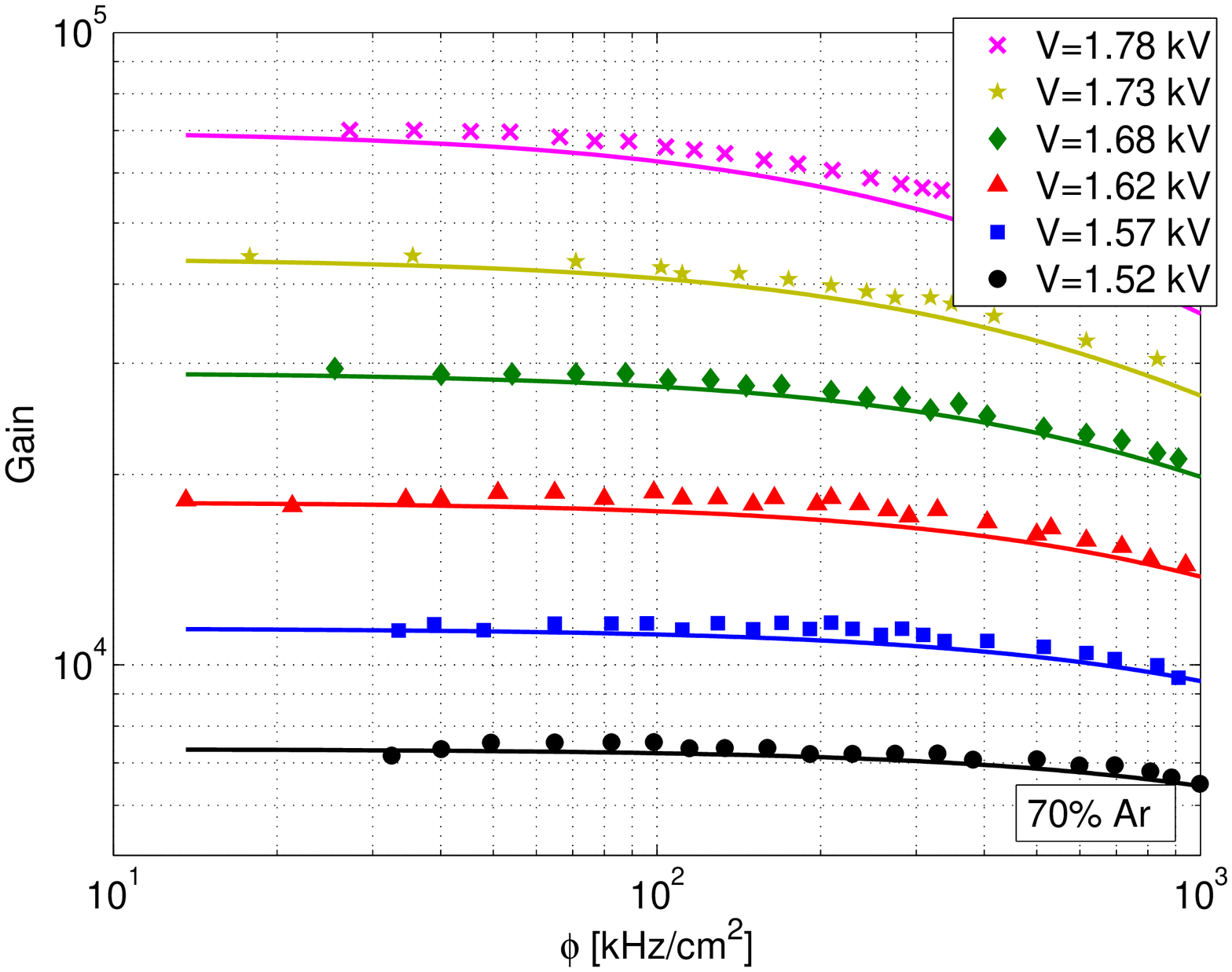}

\includegraphics[width = 7.5 cm]{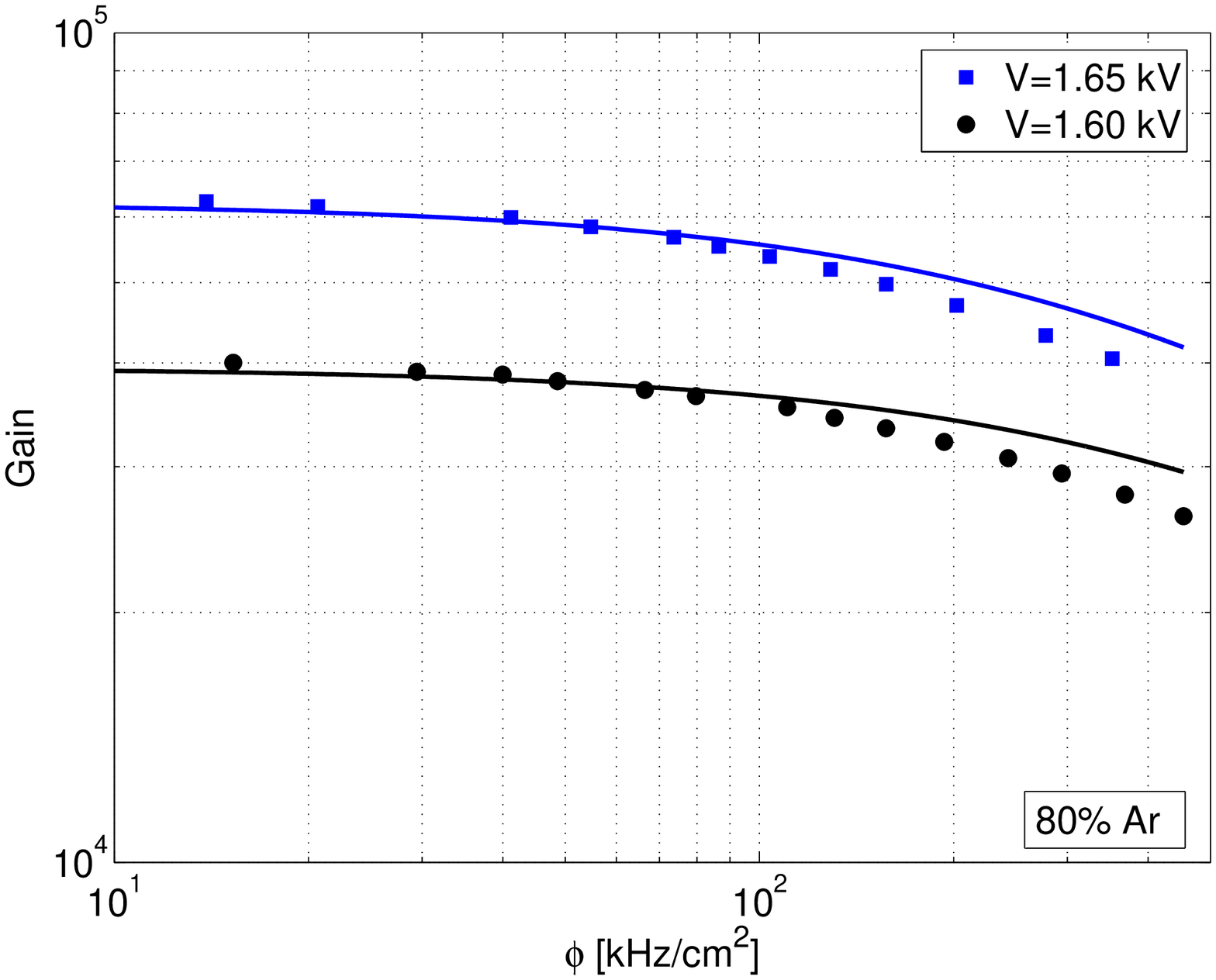}
\includegraphics[width = 7.5 cm]{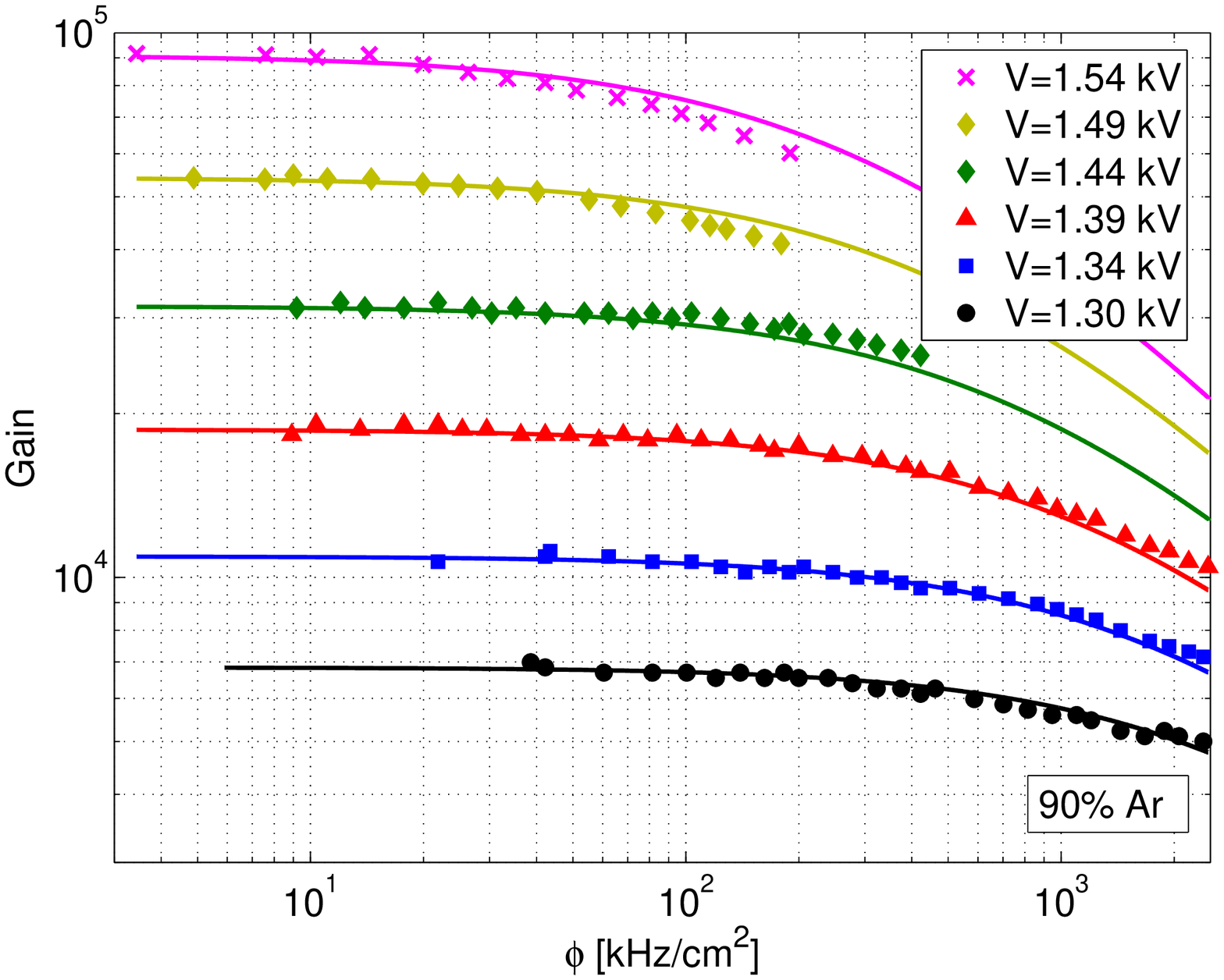}
\caption{\footnotesize Up-left: 2-parameter fit of the rate curves for Ar-CO$_2$ mixtures obtained with the $s=4$ mm chamber.}
\label{Ar_4}
\end{center}
\end{figure}

\newpage

\begin{figure}[ht!!!]
\begin{center}
\includegraphics[width = 7.5 cm]{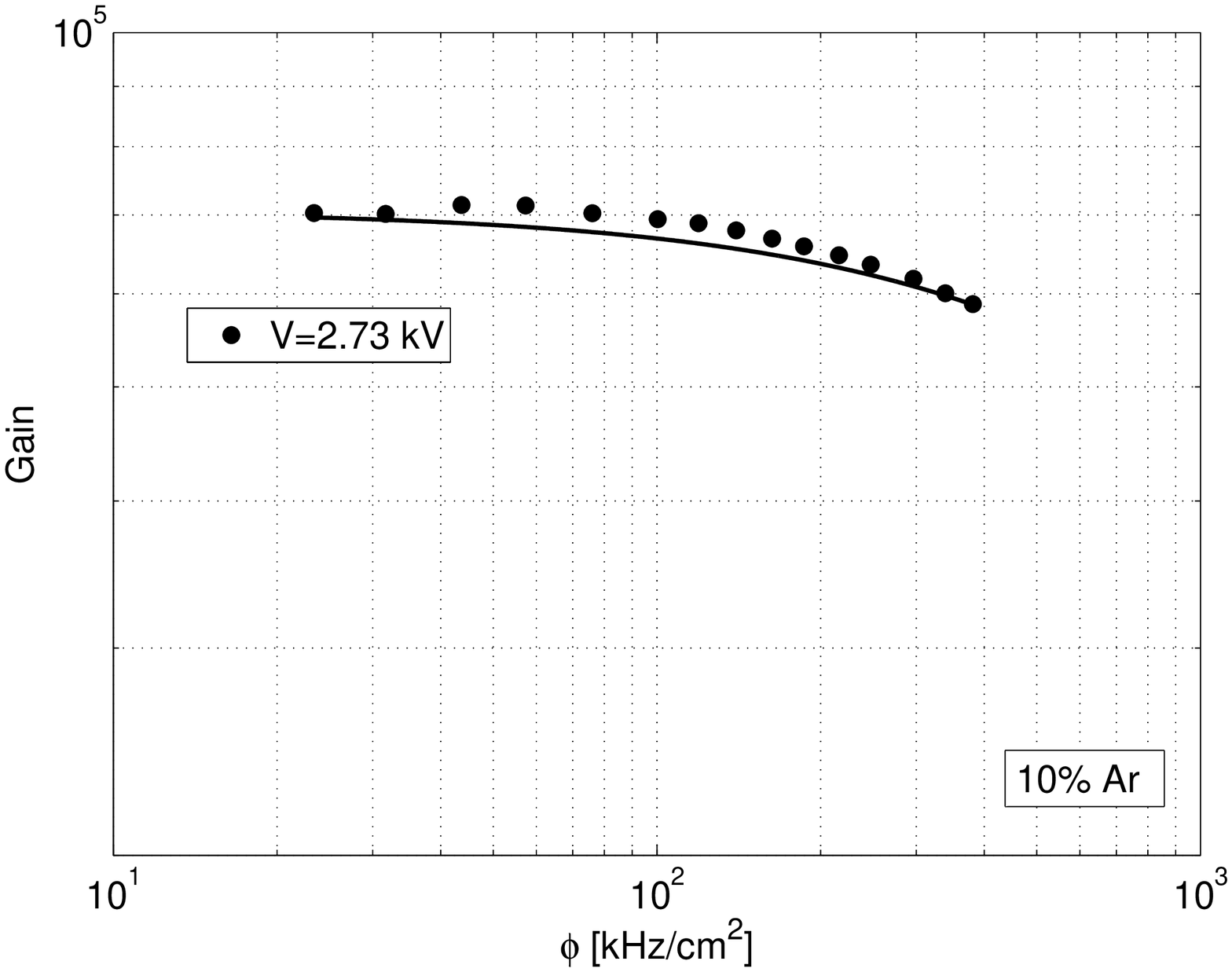}
\includegraphics[width = 7.5 cm]{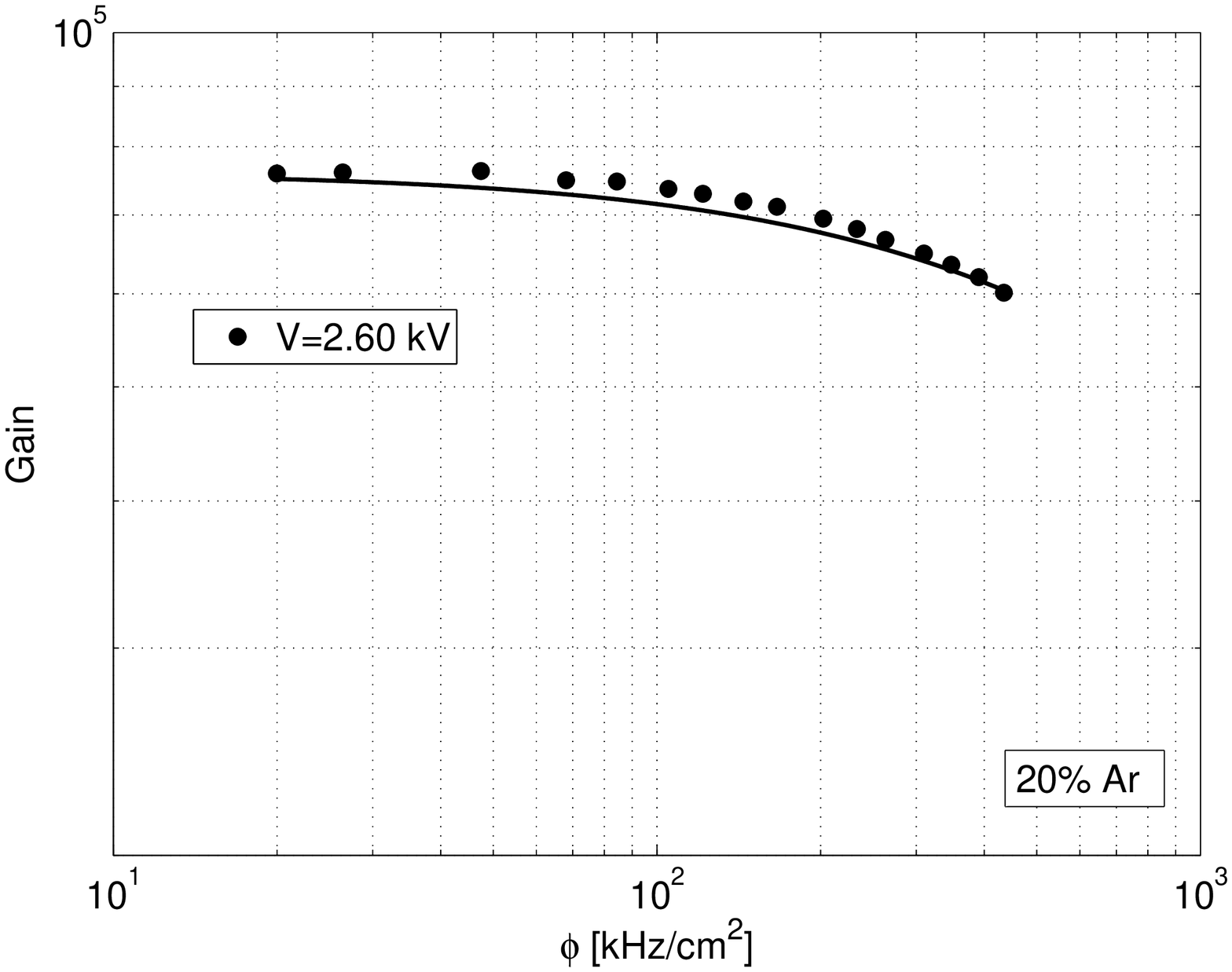}

\includegraphics[width = 7.5 cm]{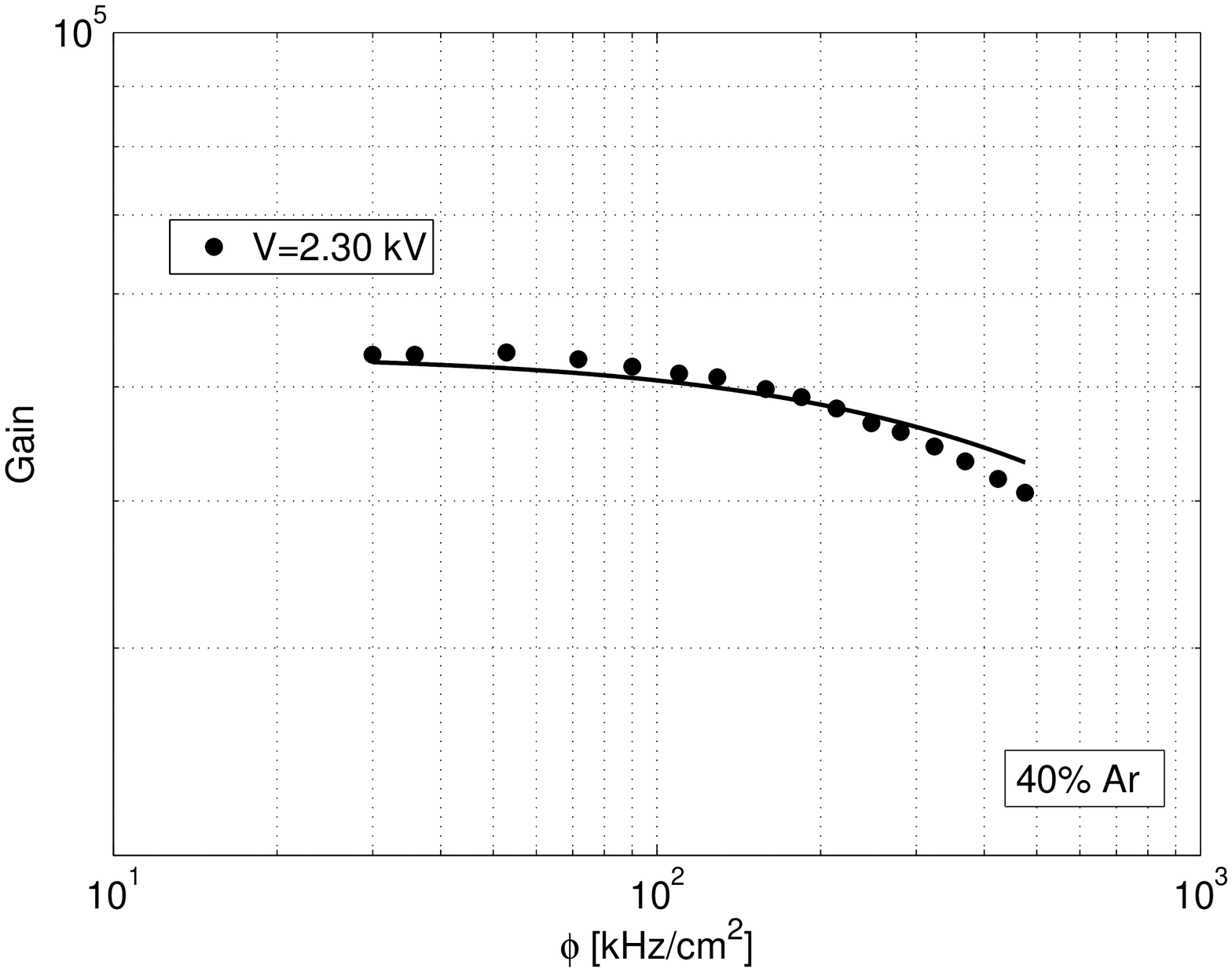}
\includegraphics[width = 7.5 cm]{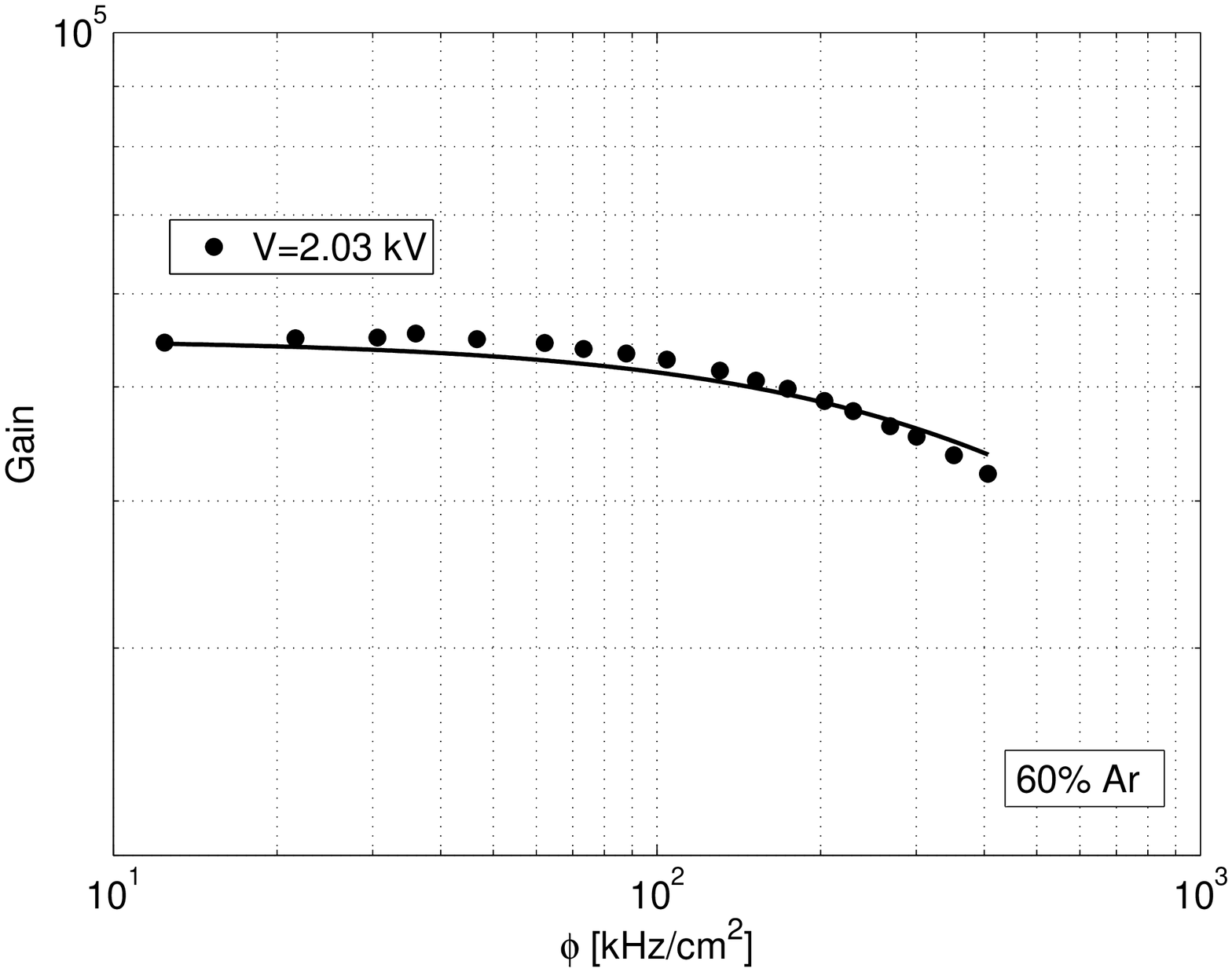}

\includegraphics[width = 7.5 cm]{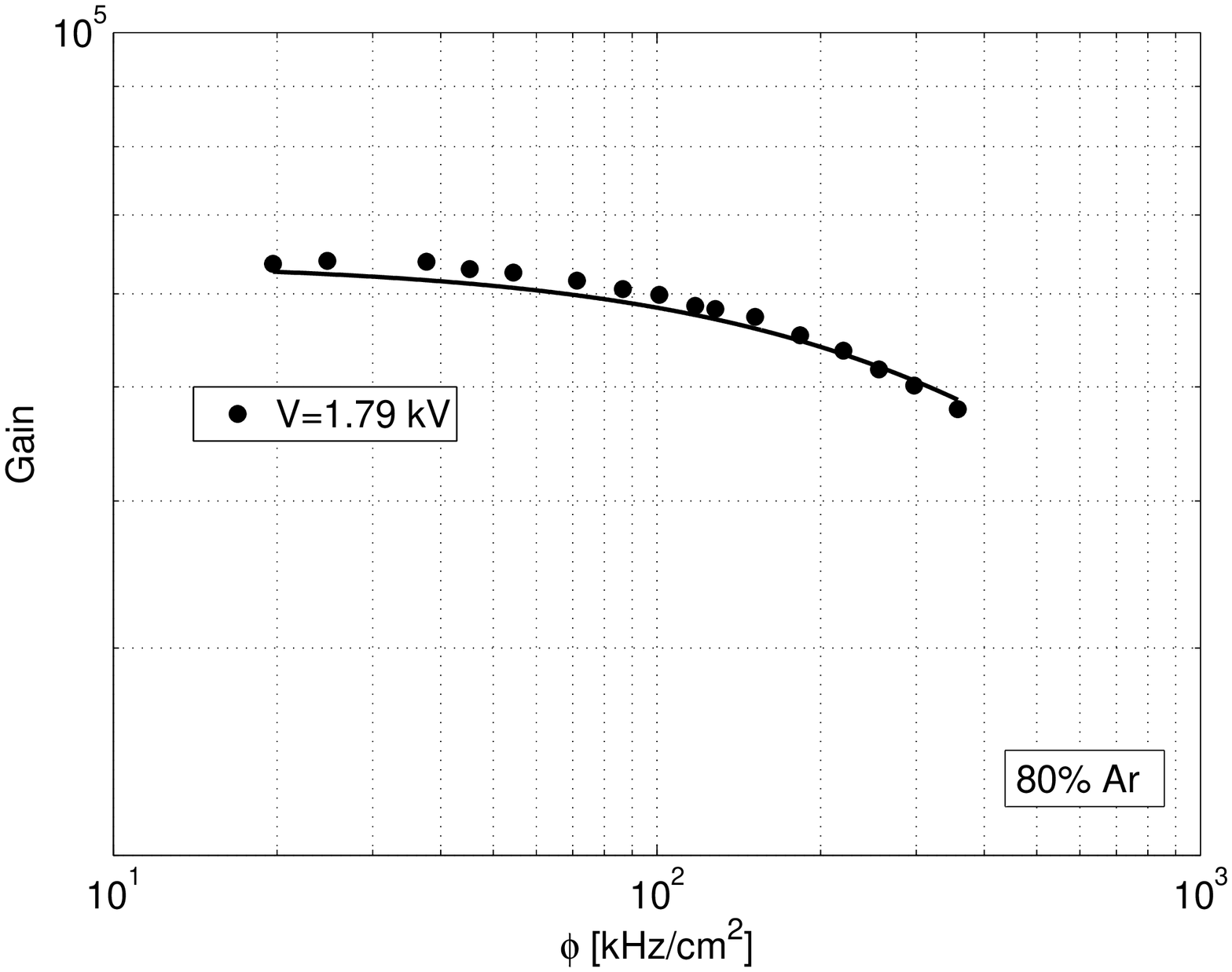}
\includegraphics[width = 7.5 cm]{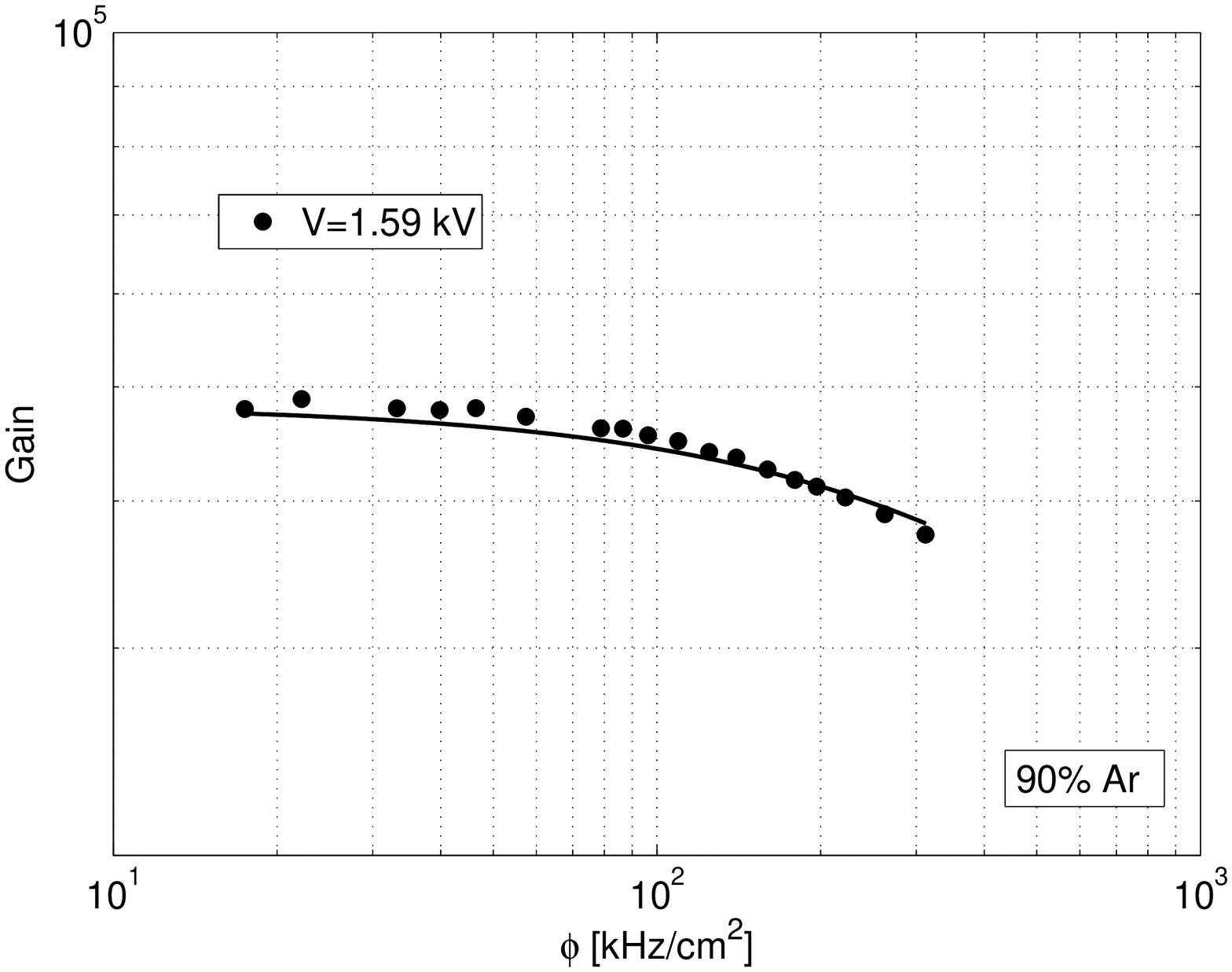}
\caption{\footnotesize Up-left: 2-parameter fit of the rate curves for Ar-CO$_2$ mixtures obtained with the $s=3$ mm chamber.}
\label{Ar_3}
\end{center}
\end{figure}

\newpage

\begin{figure}[ht!!!]
\begin{center}
\includegraphics[width = 7.5 cm]{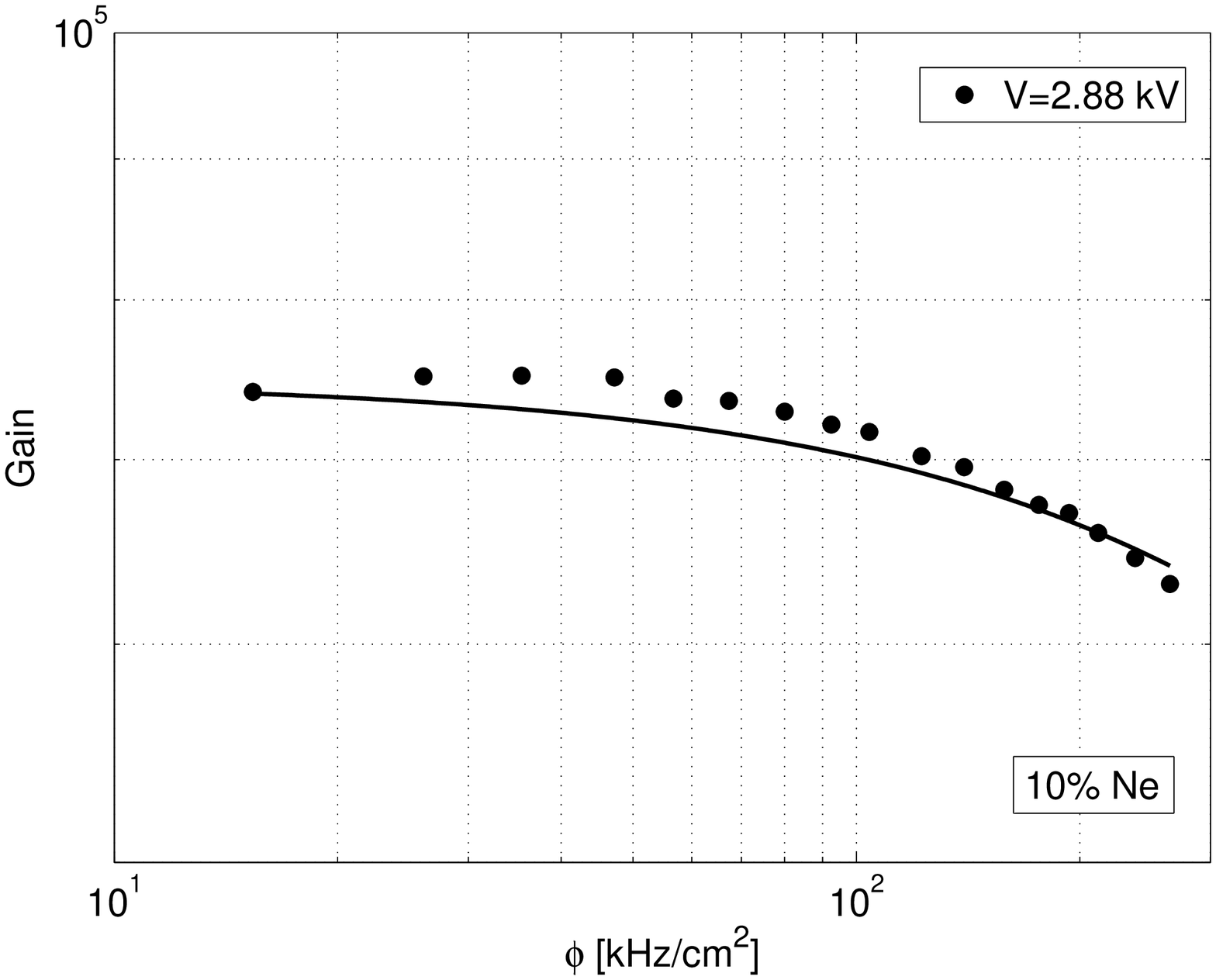}
\includegraphics[width = 7.5 cm]{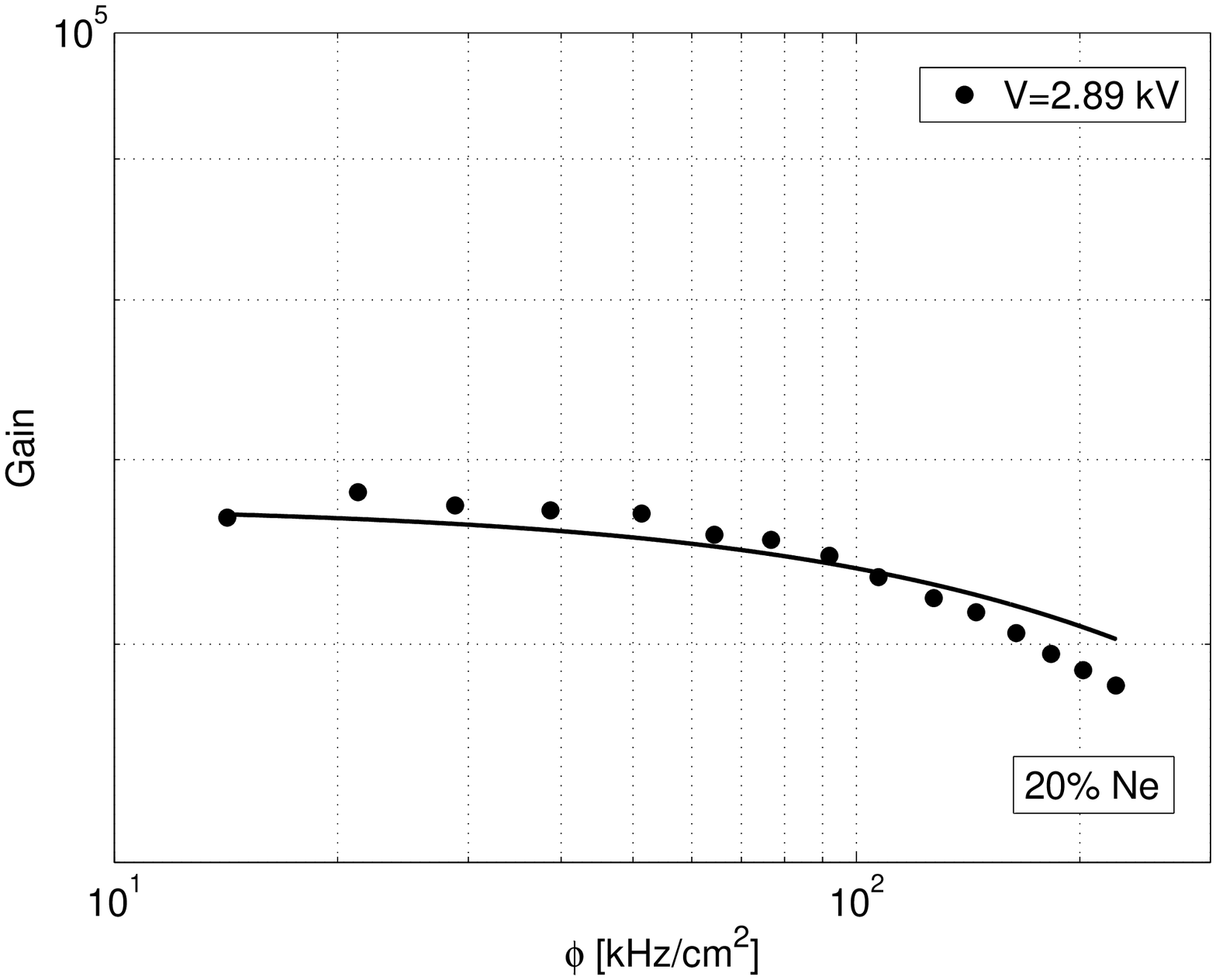}

\includegraphics[width = 7.5 cm]{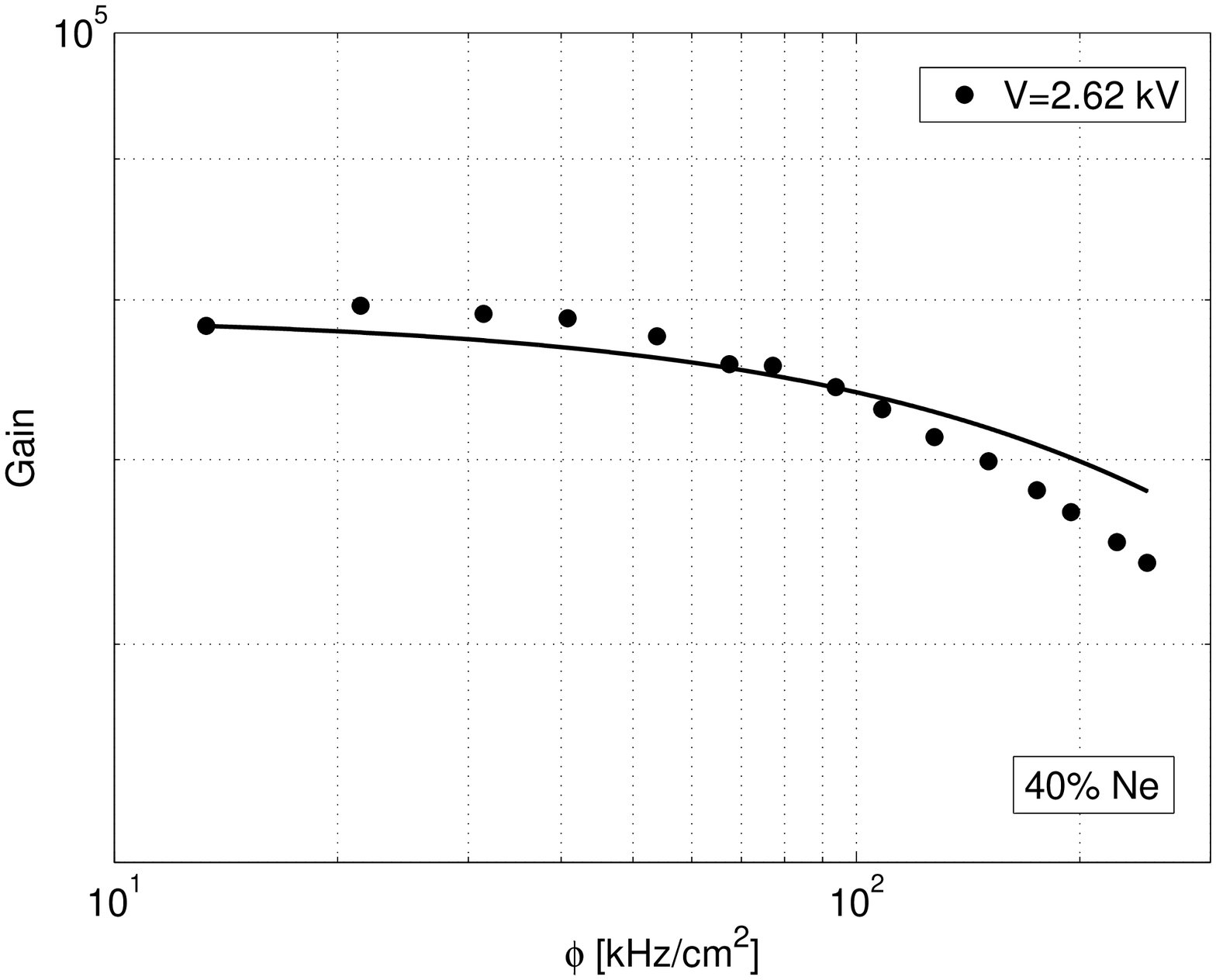}
\includegraphics[width = 7.5 cm]{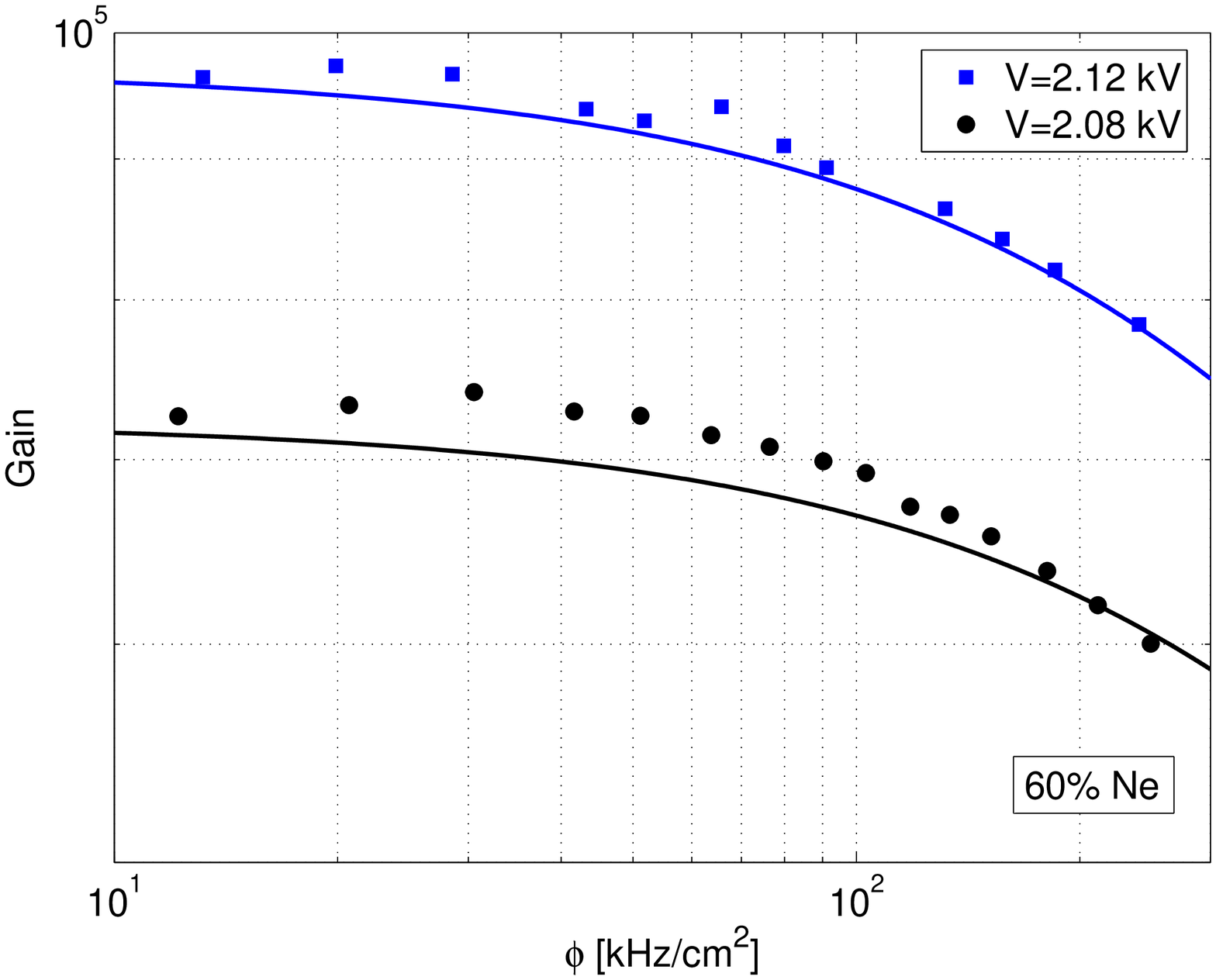}

\includegraphics[width = 7.5 cm]{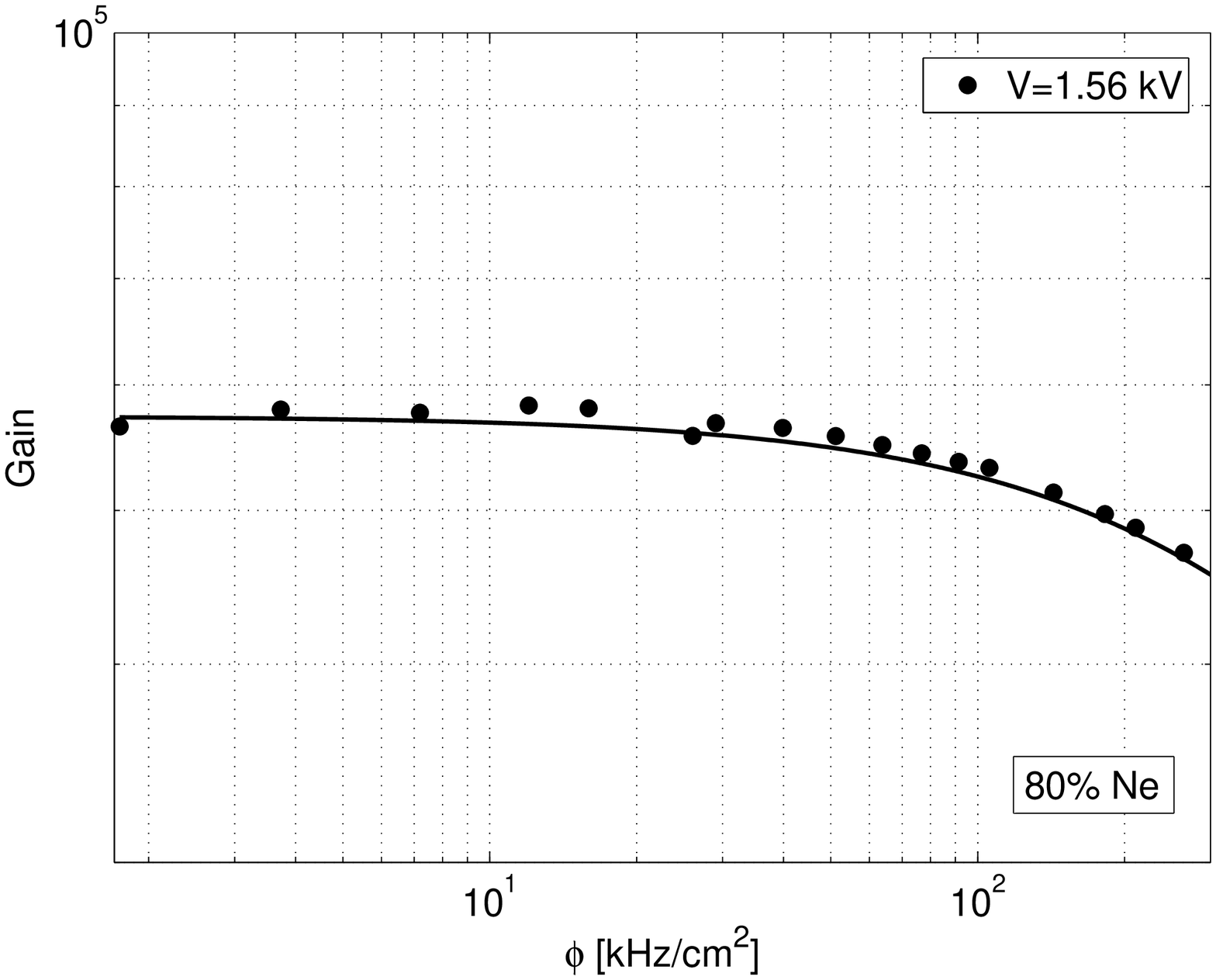}
\includegraphics[width = 7.5 cm]{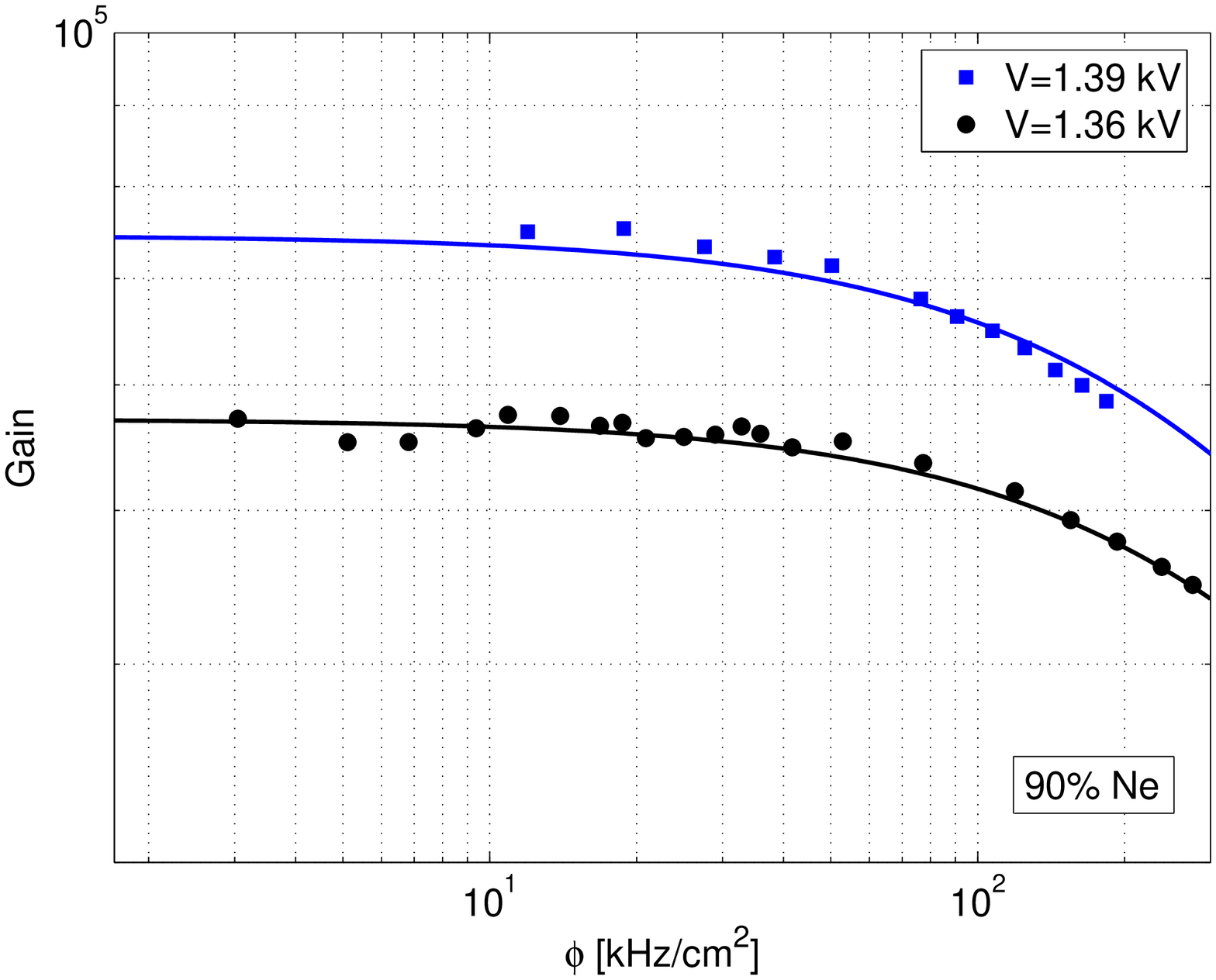}
\caption{\footnotesize Up-left: 2-parameter fit of the rate curves for Ne-CO$_2$ mixtures obtained with the $s=4$ mm chamber.}
\label{Ne_4}
\end{center}
\end{figure}

\newpage

\begin{figure}[ht!!!]
\begin{center}
\includegraphics[width = 7.5 cm]{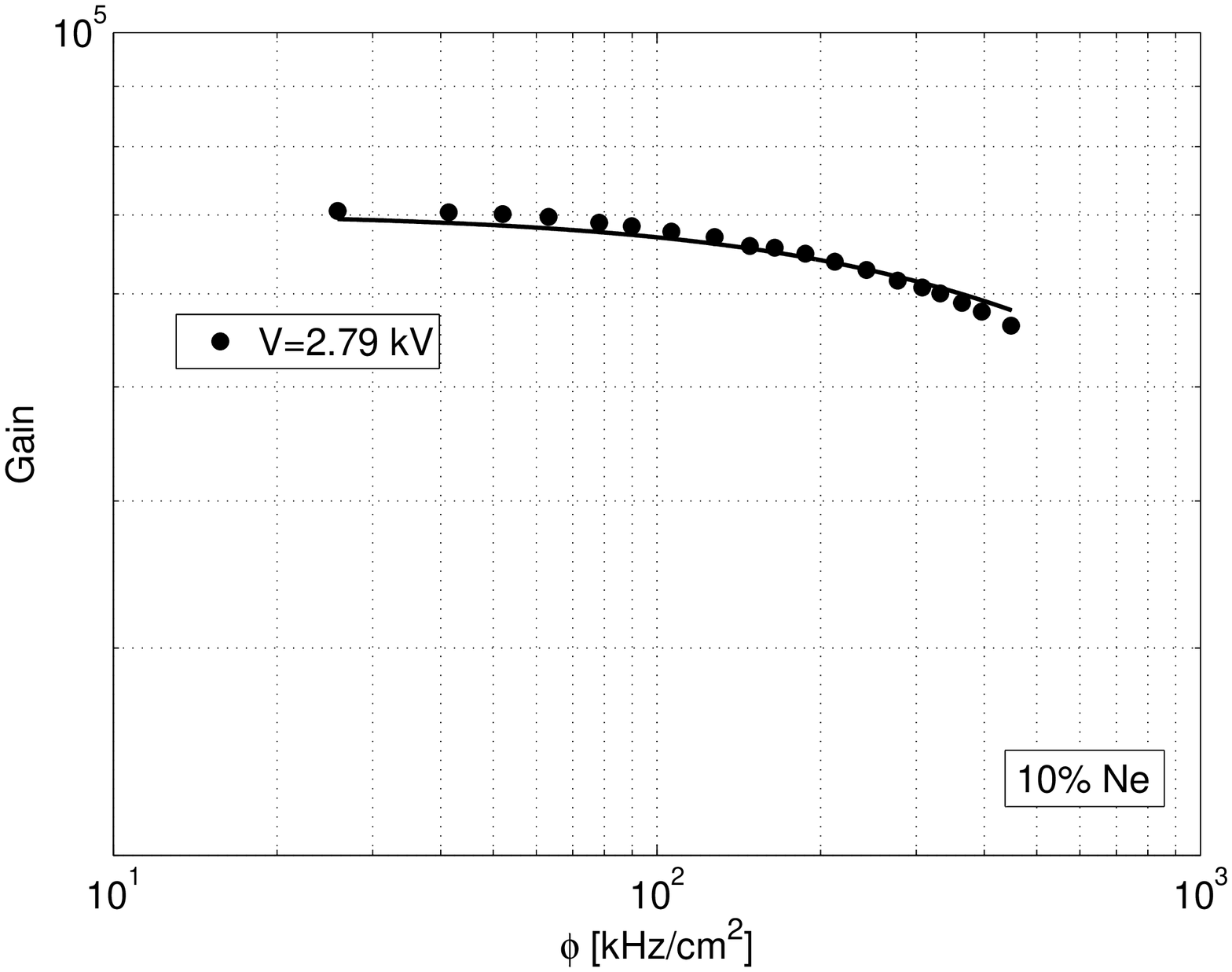}
\includegraphics[width = 7.5 cm]{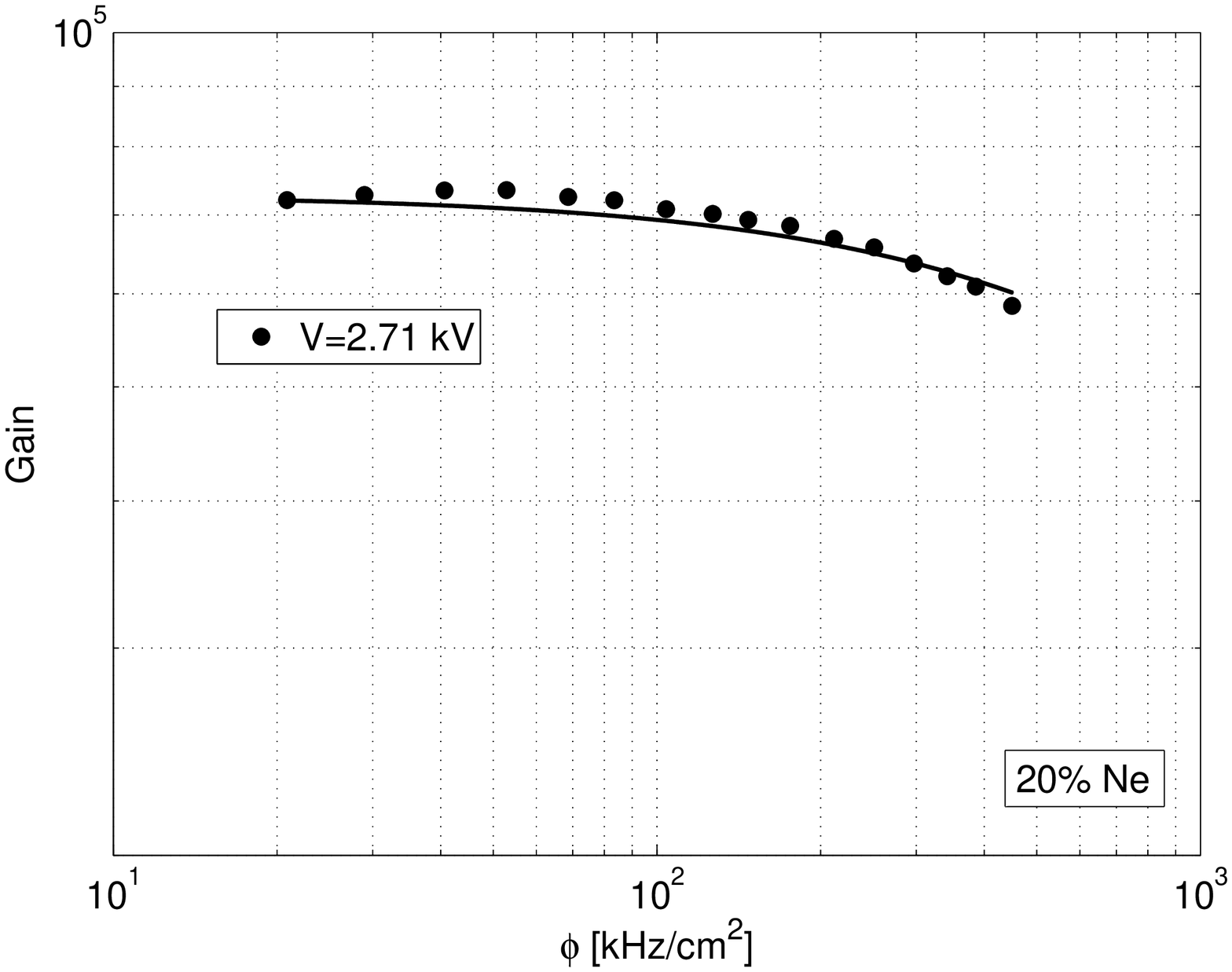}

\includegraphics[width = 7.5 cm]{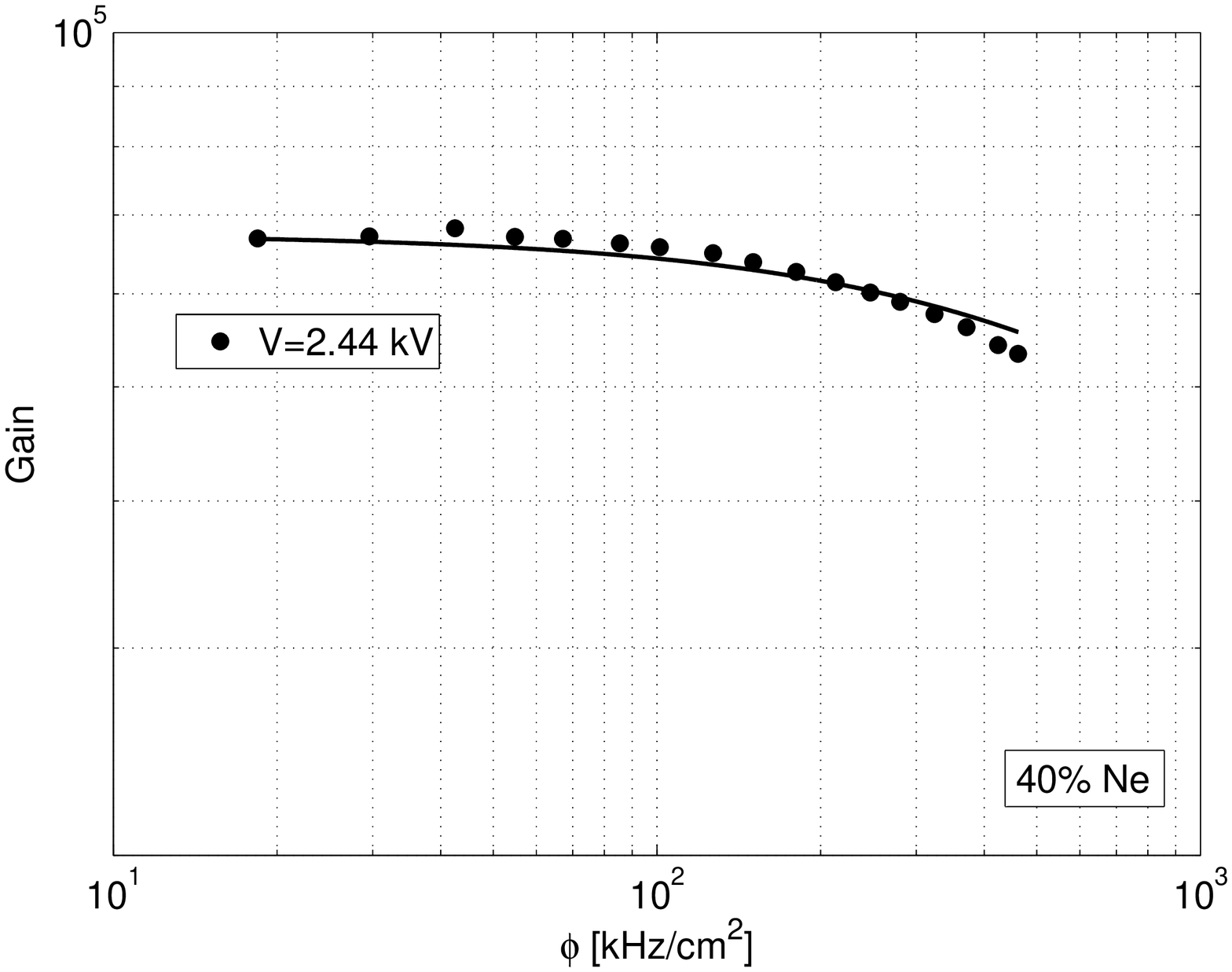}
\includegraphics[width = 7.5 cm]{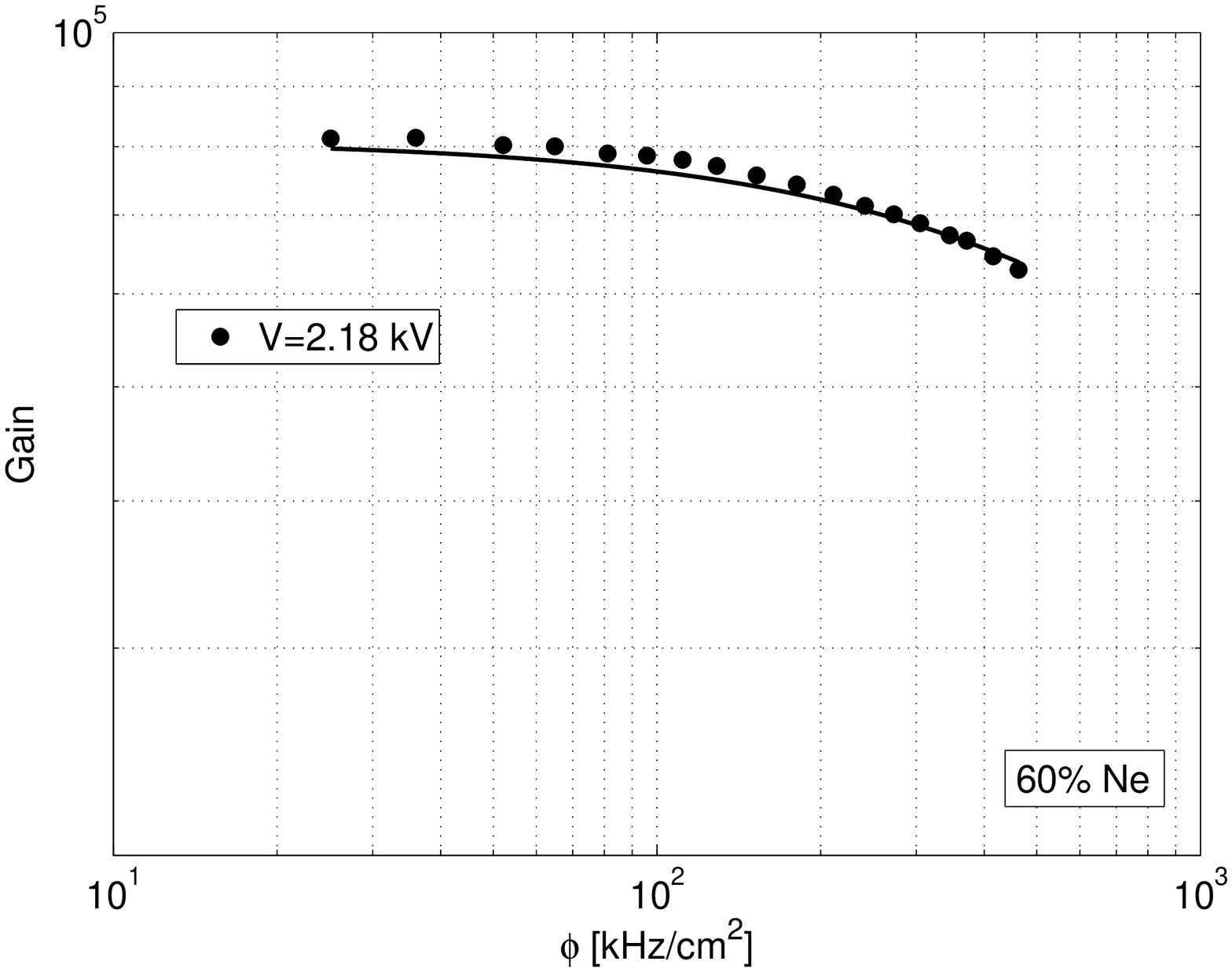}

\includegraphics[width = 7.5 cm]{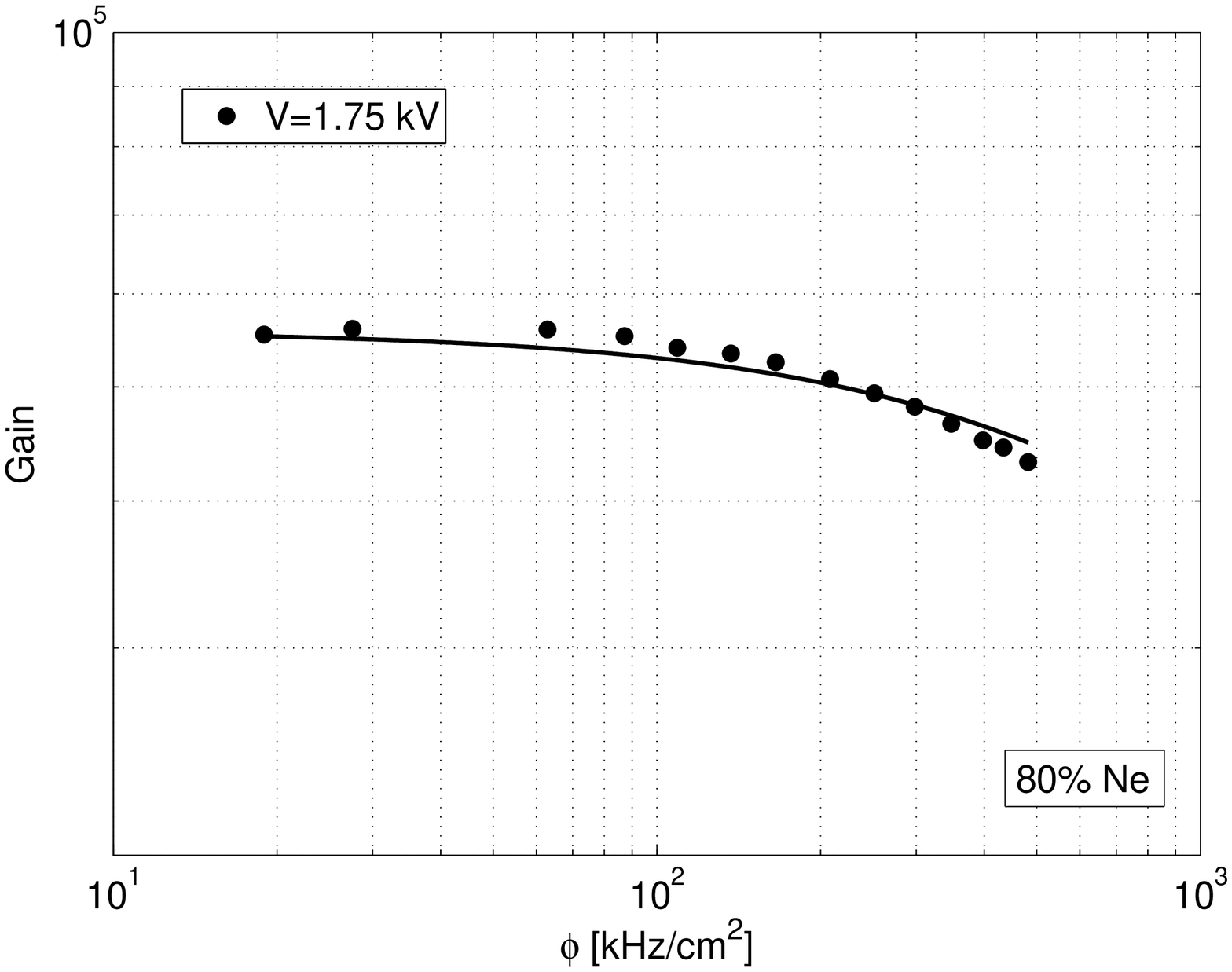}
\includegraphics[width = 7.5 cm]{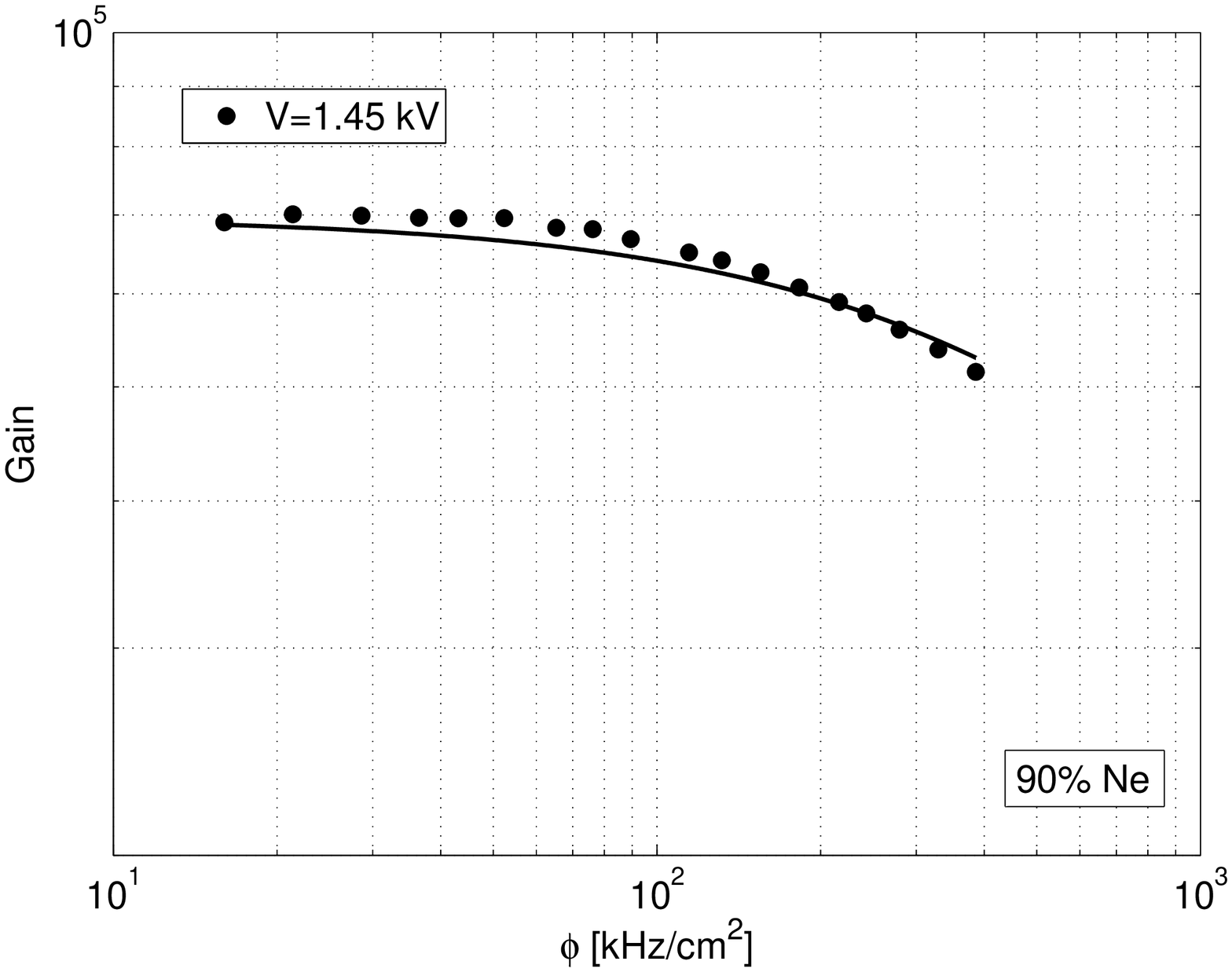}
\caption{\footnotesize Up-left: 2-parameter fit of the rate curves for Ne-CO$_2$ mixtures obtained with the $s=3$ mm chamber.}
\label{Ne_3}
\end{center}
\end{figure}

\end{document}